\documentclass[iicol,pdflatex,sn-mathphys-num]{sn-jnl}% Math and Physical Sciences
\usepackage{multicol}
\usepackage{graphicx}%
\usepackage{multirow}%
\usepackage{amsmath,amssymb,amsfonts}%
\usepackage{amsthm}%
\usepackage{bm}
\usepackage{mathrsfs}%
\usepackage[title]{appendix}%
\usepackage{xcolor}%
\usepackage{textcomp}%
\usepackage{manyfoot}%
\usepackage{booktabs}%
\usepackage{algorithm}%
\usepackage{algorithmicx}%
\usepackage{algpseudocode}%
\usepackage{listings}%
\usepackage{epstopdf}
\usepackage{textcomp}
\usepackage{amsmath}
\usepackage{amssymb}
\usepackage{mathrsfs}
\usepackage{indentfirst}
\usepackage{dsfont}

\newgeometry{
  margin=1in
}
\theoremstyle{thmstyleone}%
\theoremstyle{thmstyletwo}%

\theoremstyle{thmstylethree}%

\begin{document}

\title[Article Title]{Target-Specific \textit{De Novo} Peptide Binder Design with DiffPepBuilder}


\author[1]{\fnm{Fanhao} \sur{Wang}}
\equalcont{These authors contributed equally}

\author[1]{\fnm{Yuzhe} \sur{Wang}}
\equalcont{These authors contributed equally}

\author[3]{\fnm{Laiyi} \sur{Feng}}

\author*[2]{\fnm{Changsheng} \sur{Zhang} \email{changshengzhang@pku.edu.cn}}
%\altaffiliation{A shared footnote}

\author*[1,2,3]{\fnm{Luhua} \sur{Lai} 
\email{lhlai@pku.edu.cn}}

\affil[1]{\orgname{Center for Quantitative Biology, Academy for Advanced Interdisciplinary Studies}, \orgdiv{Peking University}, \orgaddress{\city{Beijing}, \postcode{100871},  \country{China}}}

\affil[2]{\orgname{BNLMS, College of Chemistry and Molecular Engineering}, \orgdiv{Peking University}, \orgaddress{\city{Beijing}, \postcode{100871}, \country{China}}}

\affil[3]{\orgname{Center for Life Sciences, Academy for Advanced Interdisciplinary Studies}, \orgdiv{Peking University}, \orgaddress{\city{Beijing}, \postcode{100871},  \country{China}}}


\abstract{Despite the exciting progress in target-specific \textit{de novo} protein binder design, peptide binder design remains challenging due to the flexibility of peptide structures and the scarcity of protein-peptide complex structure data. In this study, we curated a large synthetic dataset, referred to as PepPC-F, from the abundant protein-protein interface data and developed DiffPepBuilder, a \textit{de novo} target-specific peptide binder generation method that utilizes an SE(3)-equivariant diffusion model trained on PepPC-F to co-design peptide sequences and structures. DiffPepBuilder also introduces disulfide bonds to stabilize the generated peptide structures. We tested DiffPepBuilder on 30 experimentally verified strong peptide binders with available protein-peptide complex structures. DiffPepBuilder was able to effectively recall the native structures and sequences of the peptide ligands and to generate novel peptide binders with improved binding free energy. We subsequently conducted \textit{de novo} generation case studies on three targets. In both the regeneration test and case studies, DiffPepBuilder outperformed AfDesign and RFdiffusion coupled with ProteinMPNN, in terms of sequence and structure recall, interface quality, and structural diversity. Molecular dynamics simulations confirmed that the introduction of disulfide bonds enhanced the structural rigidity and binding performance of the generated peptides. As a general peptide binder \textit{de novo} design tool, DiffPepBuilder can be used to design peptide binders for given protein targets with three dimensional and binding site information.}

%TODO @wyz model detail ESM, IPA




\maketitle

\begin{multicols}{2}

\vspace*{\fill} %
\section{Introduction}\label{sec1}

Peptides have emerged as promising candidates for drug development due to their diverse biological activities and relatively low toxicity\cite{bodanszky1988peptide,craik2013future,fosgerau2015peptide,gomes2018designing,muttenthaler2021trends}. Currently, over 100 peptide-based pharmaceuticals are used to treat various medical conditions including cancer, diabetes, osteoporosis, multiple sclerosis, HIV, and chronic pain\cite{kaspar2013future,henninot2018current,lee2019comprehensive,wang2022therapeutic,chen2024role}. Semaglutide, a notable example recently, is a glucagon-like peptide-1 (GLP-1) receptor agonist used effectively in the management of type 2 diabetes and obesity, offering improved glycemic control and cardiovascular benefits\cite{marso2016semaglutide,husain2019oral,wilding2021once}. The field of peptide drug design has also experienced significant progress, driven by advancements in computational methods\cite{vanhee2011computational,chang2022towards}, structural biology\cite{terada2012recent,miller2016structural}, and synthetic chemistry\cite{stawikowski2012introduction,erak2018peptide,ferrazzano2022sustainability}. Computational tools and algorithms are pivotal in predicting peptide structures, interactions and pharmacokinetic properties, thus facilitating the development of peptide-based therapeutics. Traditional methods such as molecular dynamics simulations and docking studies have enabled researchers to explore the conformational space of peptides and predict their binding affinity to target proteins with improved accuracy\cite{wang2019improved,geng2019applications,bond2007coarse,alam2017high,zhang2019autodock}. For example, Chen et al.\cite{chen2024design} has used Rosetta FlexPepDock\cite{alam2017high} to design peptide scaffolds and sequences from existing peptides. However, traditional design methods encounter challenges in terms of efficiency and accuracy.\\

Peptide binder design remains challenging due to the flexibility of peptide structures and the scarcity of protein-peptide complex structure data. In this study, we curated a large synthetic dataset, referred to as PepPC-F, from the abundant protein-protein interface data and developed DiffPepBuilder, a \textit{de novo} target-specific peptide binder generation method that utilizes an SE(3)-equivariant diffusion model trained on PepPC-F to co-design peptide sequences and structures. DiffPepBuilder also introduces disulfide bonds to stabilize the generated peptide structures. We tested DiffPepBuilder on 30 experimentally verified strong peptide binders with available protein-peptide complex structures. DiffPepBuilder was able to effectively recall the native structures and sequences of the peptide ligands and to generate novel peptide binders with improved binding free energy. We subsequently conducted \textit{de novo} generation case studies on three targets. In both the regeneration test and case studies, DiffPepBuilder outperformed AfDesign and RFdiffusion coupled with ProteinMPNN, in terms of sequence and structure recall, interface quality, and structural diversity. Molecular dynamics simulations confirmed that the introduction of disulfide bonds enhanced the structural rigidity and binding performance of the generated peptides. As a general peptide binder \textit{de novo} design tool, DiffPepBuilder can be used to design peptide binders for given protein targets with three dimensional and binding site information.\\

Over the years, the Protein Data Bank (PDB)\cite{berman2000protein} has compiled a substantial amount of three-dimensional structural information for biomacromolecules and their complexes. Leveraging the expanding volume of available data, deep learning methods have achieved significant success in protein structure prediction\cite{jumper2021highly,baek2021accurate,krishna2024generalized,lin2023evolutionary,evans2021protein} and design\cite{anishchenko2021novo,dauparas2022robust,watson2023novo,ingraham2023illuminating}. However, the number of well-defined protein-peptide complex structures in PDB remains relatively limited. Peptides demonstrate a higher level of conformational plasticity with less secondary structures compared to proteins\cite{fasman2012prediction,brooks1993simulations,abagyan1992optimal,long1998biopolymer,lisanza2023joint}. The flexibility of peptides and the scarcity of data pose challenges for deep learning models. Moreover, co-design of peptide binder backbone structures and sequences is particularly difficult due to the non-conservative nature of peptide backbones. Recently, geometric deep generative models, especially diffusion models\cite{sohl2015deep,ho2020denoising,song2020score} were used to meet these challenges. For instance, Kosugi et al.\cite{kosugi2022solubility} developed the AfDesign\cite{anishchenko2021novo} approach for the generation of backbone conformation and sequence features of peptide binders with good solubility. RFdiffusion\cite{watson2023novo} has demonstrated its capability in generating protein binders, and also suggested potential adaptability for the generation of peptide ligands. Despite these advances, the application of deep generative models in peptide design highlights a pressing yet challenging opportunity, primarily confronted by three substantial hurdles: (1) the need for a comprehensive collection of protein-peptide complex structures for effective model training; (2) the difficult task of implementing co-design of structure and sequence; (3) the requirement of ensuring the stability of binding conformations of the generated peptides.\\

In this study, we first constructed a comprehensive protein-peptide structure dataset from protein-protein interfaces and then developed a diffusion-based generative model to co-design backbone structures and sequences for peptide binders. The SE(3)-equivariant diffusion model architecture was further integrated with a geometric disulfide bond construction module to forge a novel tool referred to as DiffPepBuilder, which excels in peptide ligand sequence-structure co-design tasks. We tested DiffPepBuilder on 30 non-redundant protein-peptide complex systems with strong binding strength and found that it can regenerate natural peptide binders with similar binding positions and high sequence similarity. We then used DiffPepBuilder to design novel linear and cyclic peptide binders for three important targets: activin receptor-like kinase 1 (ALK1), 3C-like protease of SARS-CoV-2 (3CL$^{\rm pro}$), and tumor necrosis factor (TNF-$\alpha$). In both the regeneration tests and case studies, DiffPepBuilder outperformed AfDesign and RFdiffusion coupled with ProteinMPNN (RFd+MPNN).

\begin{figure*}[htb]
\centering
\includegraphics[width=0.95\linewidth]{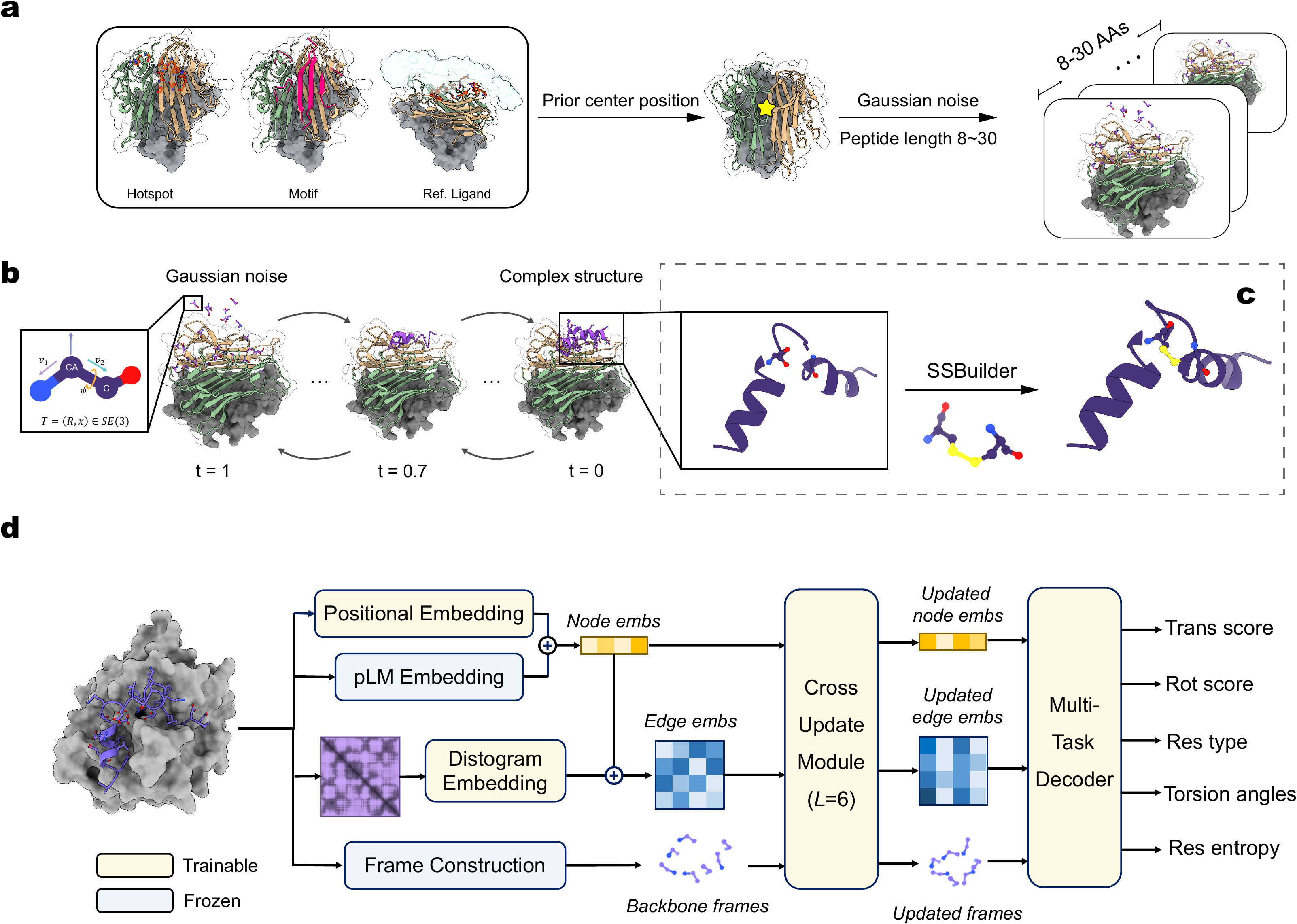}
\caption{\textbf{\textbar\ Overview of DiffPepBuilder.} \textbf{a.} Input preparation. DiffPepBuilder accepts user-specified binding site information formatted as Hotspots (in crimson), Motif (in crimson), or Reference ligand (in transparent, used to define the hotspots). The model then determines the center position (marked by a yellow star) and length range (default 8 to 30) for generated peptide binders and initializes them with Gaussian noise. \textbf{b.} The diffusion process. All residues in the model are represented by local frames that possess SE(3) equivariance. The coordinates of the target protein remain unchanged. The reverse (generative) process involves the peptide starting from Gaussian noise and progressively denoising into a complete binder structure. Conversely, the forward (noising) process entails the addition of Gaussian noise for model training. \textbf{c.} Construction of disulfide bonds with SSBuilder module. After generation, the peptide ligand structure, along with the residue entropy, is fed into SSBuilder. SSBuilder then filters out the residues whose entropy exceeds a threshold and constructs disulfide bonds among them based on geometrically matched fragments. \textbf{d.} The architecture of the diffusion model. It primarily utilizes pLM Embeddings and positional encoding as node information, employs a distogram to additionally encode edge information, and converts three-dimensional coordinates into local frames. After these different modalities of information interact via the Cross Update Module, the Multitask Decoder outputs translational and rotational scores, predicted residue types, predicted torsion angles, and residue entropies.}\label{model_arch}
\end{figure*}

\section{Results}\label{sec2}

\subsection{Overview of the DiffPepBuilder model}\label{result_subsec1}
DiffPepBuilder is an end-to-end \textit{de novo} peptide binder generation model that utilizes a diffusion-based generative procedure to simultaneously design the peptide sequence and structure through gradual structural denoising, coupled with a post-processing procedure designed to enhance the stability of the generated peptide structures by introducing disulfide bonds when appropriate. The \textit{de novo} peptide binder generation process is as follows. DiffPepBuilder takes the full target as input and initializes the protein binding pocket-peptide ligand complex structure based on user-specified binding site information and generation settings (Fig.\ref{model_arch}a, see Sec.\ref{method_subsec2} for details). This initial complex structure is then fed into the diffusion-based generative procedure, as illustrated in Fig.\ref{model_arch}b, where an arbitrary residue is parameterized as an orientation-preserving rigid body frame $\bm{T}_i\in {\rm SE(3)}$ as in AlphaFold 2\cite{jumper2021highly} ($\rm SE(3)$ denotes the special Euclidean group). The full-atom structure of the generated peptide ligand is reconstructed using the final ($t=0$) frame representations. DiffPepBuilder subsequently identifies peptide residues whose amino acid type entropy exceeds a specified threshold, searches the disulfide bond fragment library for matching geometric conformations and replaces the matched residues with disulfide bond-connected cysteine residues (Fig.\ref{model_arch}c, see Sec.\ref{method_subsec4} for details). The structures of the peptide, both before and after the construction of disulfide bonds, are incorporated into the sampling repertoire.\\

The architecture of the denoising network is outlined in Fig.\ref{model_arch}d (see Sec.\ref{method_subsec3} for details). The initial node and edge embeddings, in conjunction with the backbone frames, are processed through a 6-layer Cross Update Module adapted from FrameDiff\cite{yim2023se} that alternately updates the embeddings and frame representations. A Multi-Task Decoder is utilized to predict the translational and rotational scores of the current diffusion step, the residue types of the peptide ligand, the main chain torsion angles $\phi/\psi$ and the side chain torsion angles $\chi_{1}$\(\sim\)$\chi_{4}$. We employed a composite loss scheme (see Sec.\ref{method_subsec5}) and trained the denoising network exclusively on the PepPC-F (Peptide Protein Complexes-Fragment) dataset (see Sec.\ref{result_subsec5} for details).\\

\begin{figure*}[htb]
\centering
\includegraphics[width=1.0\linewidth]{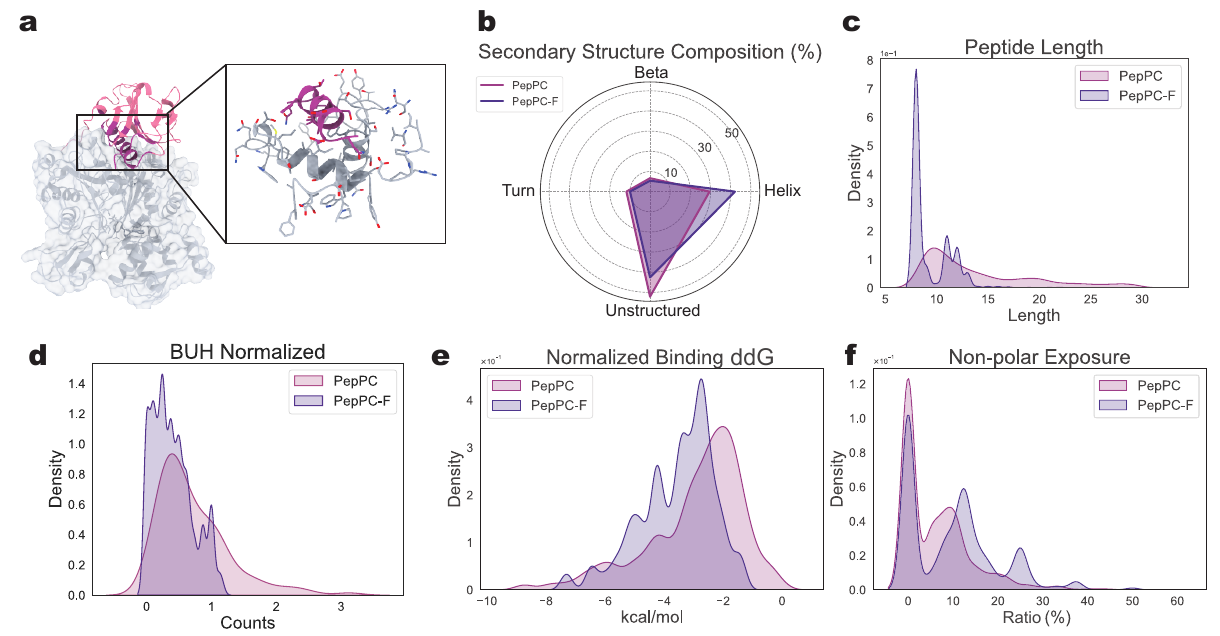}
\caption{\textbf{\textbar\ Statistics of the curated datasets.} \textbf{a.} An illustration of the construction process of protein-fragment complex entries is provided. Interface peptide fragments (in purple) are truncated from the complete binder (in pink), and, together with their corresponding binding proteins (in gray), form the complex structures that are collected in the PepPC-F dataset. \textbf{b.} A comparative diagram of secondary structures across two datasets, PepPC and PepPC-F. \textbf{c.} A graph of length distribution of PepPC and PepPC-F. \textbf{d.} The number of unsatisfied hydrogen bonds at the interface normalized by residue count. \textbf{e.} The distribution graph of Rosetta ddG normalized by residue count. \textbf{f.} The ratio of ligand hydrophobic exposure residues between PepPC-F and PepPC. } \label{database_statistic}
\end{figure*}

\subsection{Protein-peptide complex dataset construction and analysis}\label{result_subsec2}
Currently available protein-peptide complex structures are limited in quantity, and according to experimental data compiled from PDBbind2020\cite{wang2005pdbbind}, peptide binders typically exhibit relatively weak binding affinity, mostly at the micromolar level, which is approximately three orders of magnitude weaker than protein-protein interactions (See Fig.S1b). In order to build a deep learning model to design strong peptide binders, we used the protein-protein complex structures to build a large synthetic protein-peptide structure dataset, PepPC-F. We collected structures of protein-protein complexes from PDB and extracted buried interfacial peptide fragments. The basic xassumption is that peptide fragments that are exposed in the free ligand proteins and become buried by the target proteins represent good binding states of peptides. These peptide fragments, along with their corresponding binding proteins, constitute the complex structures collected in the PepPC-F dataset. At the same time, we also built a dataset of protein-peptide complexes dataset, PepPC for comparison. We  restricted  the peptide  length  in  PepPC  and  PepPC-F dataset between 8 to 30, akin to other databases such as ProPedia\cite{martins2021propedia} and PepBDB\cite{wen2019pepbdb}. We divided the peptides into helical (helix ratio $\geq$ 0.5) and non-helical (including turns and unstructured peptides as shown in Fig.\ref{database_statistic}b). After redundant removal and data cleaning (See Sec.\ref{method_subsec1}), there are 14,897 complexes in PepPC-F, including 4,241 helical peptides and  10,656 non-helical peptides and 3,832 complexes in PepPC, including 232 helical peptides and 3,600 non-helical peptides. The size and diversity of PepPC-F make the training of the diffusion  model  feasible  and  enhance  the  model’s generalization capability on various targets. We compared the characteristics of the peptides in these two datasets. The contents of secondary structures are similar with slightly high helical structures in PepPC-F (Fig.\ref{database_statistic}b). Though both datasets contain peptides with 8-30 residues, short peptides are more populated in PepPC-F(Fig.\ref{database_statistic}c). Peptides in PepPC-F form better interactions with the targeting proteins as indicated by the fewer unsatisfied interface hydrogen bonds (Fig.\ref{database_statistic}d,e) and better normalized binding free energy. One might worry that peptide fragments isolated from proteins may have high proportion of exposed hydrophobic residues since protein interiors often feature tight hydrophobic packing, but our analysis shows that PepPC-F's hydrophobic exposure is not apparently different from that in natural dataset PepPC(Fig.\ref{database_statistic}f). We also analyzed the proportion of hydrophilic and hydrophobic residues in both peptide datasets and observed a similar overall distribution. The PepPC-F dataset is slightly more hydrophobic, but the difference from the natural dataset is not significant (see Fig.S1a). \\

\begin{figure*}[htb]
\centering
\includegraphics[width=1.0\linewidth]{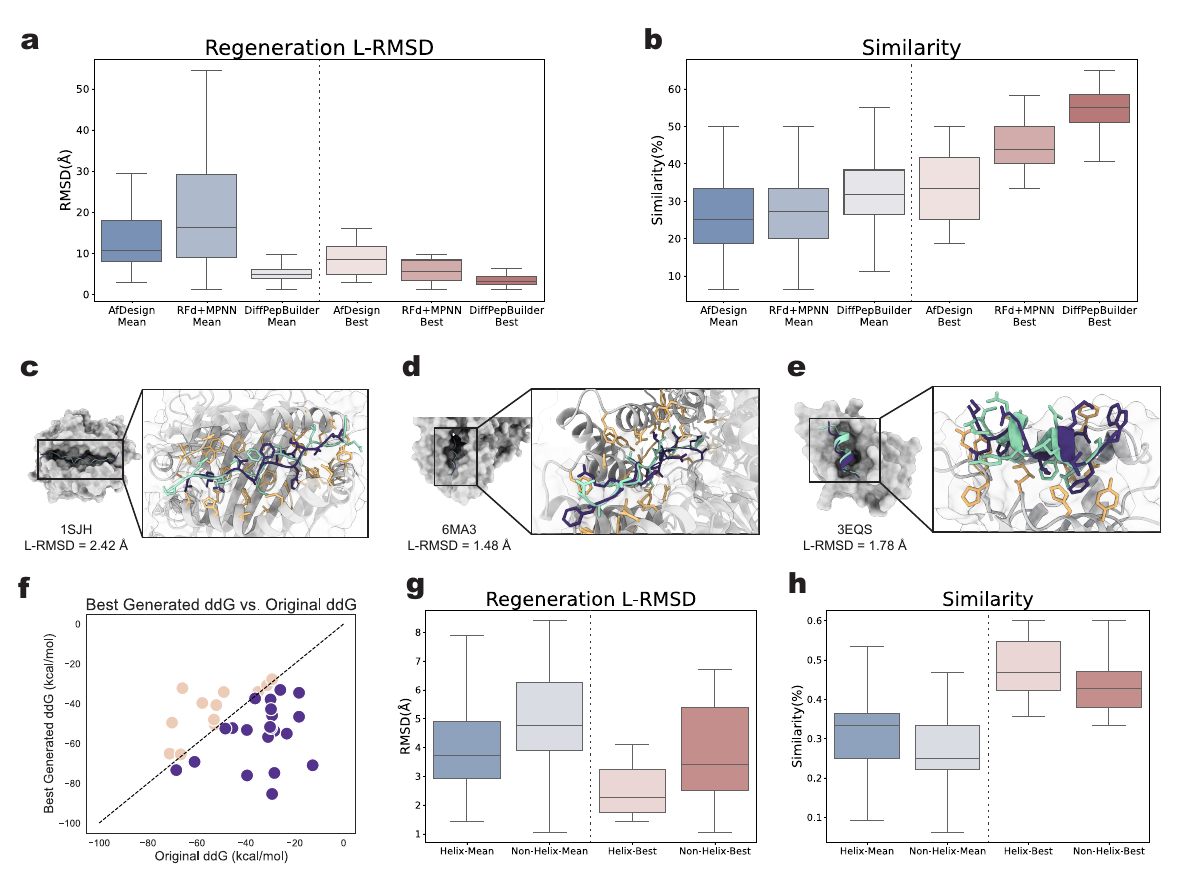}
\caption{\textbf{\textbar\ DiffPepBuilder's performance on regeneration test.} \textbf{a.} Distribution of mean and best L-RMSD values across different methods. \textbf{b.} Distribution of mean and best similarity values across different methods. \textbf{c}-\textbf{e.} Representative regeneration results for various targets. The targets (shown in gray, with interacting side chains in wheat) include human O-GlcNAc transferase, MHC class II and MDM2. The original peptides are depicted in cyan, while the generated peptides are shown in violet. \textbf{f.} Plot comparing the best ddG values of generated peptides with the ddG values of the original peptides in the testing targets. The best ddG refers to the lowest ddG achieved among all peptides generated for the same target. Targets for which the best generated ddG is lower than the original ddG are depicted in violet, while others are shown in wheat. \textbf{g}-\textbf{h.} Distributions of L-RMSD and similarity values for peptides generated by DiffPepBuilder, categorized into helical and non-helical groups.} \label{regeneration}
\end{figure*}

\subsection{DiffPepBuilder regenerates peptide binders in known protein-peptide complexes}\label{result_subsec3}
After training DiffPepBuilder on the PepPC-F dataset, we first tested whether it can regenerate peptide binders in known protein-peptide complexes. For this purpose, we constructed an independent testing dataset, PepPC-HF (PepPC High binding Affinity), that contains 30 non-redundant protein-peptide complexes with high-resolution structures (better than 2.5 \AA) and strong binding potency (See Sec.\ref{method_subsec6} for details). For comparison, we also conducted peptide binder design using RFd+MPNN and AfDesign. We used peptide ligand RMSD (L-RMSD) and peptide sequence similarity(calculated by Biopython's pairwise2 module\cite{cock2009biopython}) to evaluate whether these models can generate ligands that resemble the original peptides in the complexes. \\

For these 30 non-redundant protein-peptide complexes, peptides generated by DiffPepBuilder showed a mean L-RMSD of 4.76 \AA, while that of RFdiffusion and AfDesign were 13.62 \AA\ and 10.76 \AA, respectively. In terms of the average best L-RMSD (the minimum L-RMSD of all generated peptides for the same target), DiffPepBuilder achieved 3.34 \AA, while that of RFdiffusion and AfDesign  were 6.75 \AA \ and 7.28 \AA, respectively. DiffPepBuilder could generate peptide binders with similar sequence as the original binder peptides with the best average sequence similarity of 52.38\% (ranging from 40.71\% to 65.00\%), while that of RFd+MPNN and AfDesign-designed sequences achieved average best similarity of 45.03\% and 33.72\%, respectively. The distribution of L-RMSD and similarity is illustrated in Fig.\ref{regeneration}a,b. These analyses demonstrate that DiffPepBuilder performs well in the regeneration test.\\

Fig.\ref{regeneration}c to Fig.\ref{regeneration}e give examples of regenerated peptide ligands with different types of secondary structures. Fig.\ref{regeneration}c displays an example of major histocompatibility complex class 2 (MHCII, PDB ID: 1SJH\cite{zavala2004hairpin}), where our model generated an extended peptide that aligns well with the original backbone structure, achieving an L-RMSD of 2.42 \AA \ and a sequence similarity of 44.4\%. The second example corresponds to a peptide binder with a loop structure that binds human O-GlcNAc transferase (PDB ID: 6MA3\cite{martin2018structure}), where DiffPepBuilder generated peptide has an L-RMSD of 1.49 \AA\ and sequence similarity of 38.46\%. Fig.\ref{regeneration}e illustrates an example that contains a peptide binder with helical structure, which is a complex structure of MDM2 (mouse double minute 2 homolog) with a 12-mer peptide inhibitor (PDB ID: 3EQS\cite{pazgier2009structural}), where DiffPepBuilder generated a helical peptide with an L-RMSD of 1.78 \AA\ and a sequence similarity of 45.45\%. MDM2 is an important regulator that inhibits the tumor suppressor p53. Peptide inhibitors like ALRN-6924\cite{guerlavais2023discovery} have achieved good therapeutic effects in clinical settings. MDM2 peptide inhibitors share a relatively conserved sequence pattern, ‘PxFxDYWxxL’, where the hydrophobic residues F, W, and L play major roles in hydrophobic interactions with MDM2\cite{bottger1997molecular,klein2004targeting}. We further searched all generated sequences for MDM2 against this pattern and found 3 designs that recover all three of these critical conserved sites. We showcased one of these structures in Fig.S2, from which we can see that the aforementioned conserved residues are closely aligned, demonstrating that our model holds the potential of generating novel inhibitors for MDM2. \\

In general, the best ddG (the lowest ddG of all the generated peptides on the same target) of peptide ligands generated by DiffPepBuilder are lower than those of the original peptides in the complexes, as shown in Fig.\ref{regeneration}f. We further analyzed the performance of DiffPepBuilder on peptide binders with different secondary structures. We differentiated helix from other conformations (mostly loop structures) by setting a criterion where the content of helix secondary structure in the ground truth peptide is at least 40\%, with at least five consecutive residues adopting a helix conformation. As shown in Fig.\ref{regeneration}g and h, DiffPepBuilder performs better on helical structures both for sequence similarity and L-RMSD. In some cases, the loops generated by DiffPepBuilder do not match well with the native ligand, which may stem from the irregular nature of loop structures.\\

\begin{figure*}[htb]
\centering
\includegraphics[width=0.98\linewidth]{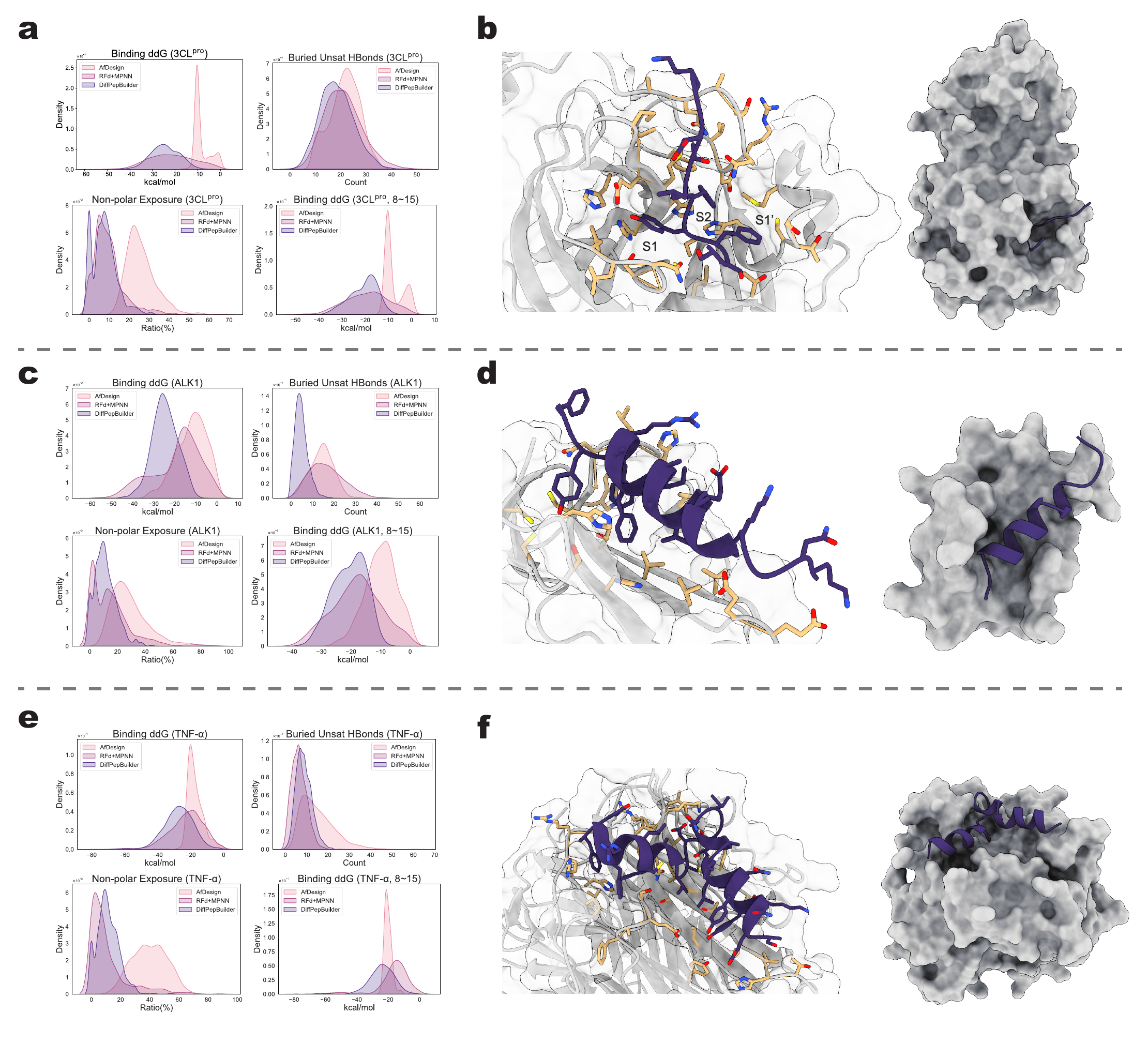}
\caption{\textbf{\textbar\  \textit{De novo} generation performance of DiffPepBuilder.} \textbf{a., c., and e.} Distributions of interface metrics of binding ddG, buried unsatisfied hbond, non-polar exposure ratio, and binding ddG of candidates with 8-15 residues targeting 3CL$^{\rm pro}$, ALK1, and TNF-$\alpha$, respectively. The performance of DiffPepBuilder is compared with that of AfDesign and RfD+MPNN. \textbf{b}, \textbf{d}, and \textbf{f.} Examples of peptides generated by DiffPepBuilder (in violet) targeting three proteins (in gray, with interacting side chains in wheat): 3CL$^{\rm pro}$, ALK1, and TNF-$\alpha$, respectively.}\label{denovo_gen}
\end{figure*}

\begin{table*}[htb]
\caption{\textit{De novo} generation results. The best in-group results are shown in \textbf{bold}.}\label{tab1}
\begin{tabular*}{\textwidth}{@{\extracolsep\fill}cccccc@{}}
\toprule
\multirow{2}{*}{Target} & \multirow{2}{*}{Method} & \multicolumn{4}{c}{Metric (Mean)} \\ \cmidrule(l){3-6} 
                        &                         & ddG (kcal/mol, $\downarrow$)   & ddG$_{\rm 8\sim 15}$ (kcal/mol, $\downarrow$) & pTM-score ($\downarrow$) &  Validity \% ($\uparrow$) \\ \midrule
\multirow{3}{*}{3CL$^{\rm pro}$}  & AfDesign     & -12.33  & -9.77 & \textbf{0.22}        &  47.82          \\
                        & RFd+MPNN             &  \textbf{-23.38} & -21.04 & 0.44     &  68.42     \\
                        & DiffPepBuilder          &  -23.16  & \textbf{-21.17} & 0.27     &  \textbf{96.22}          \\ \midrule
\multirow{3}{*}{ALK1}   & AfDesign                & -7.61  & -7.28 & 0.41         & 30.43      \\
                        & RFd+MPNN             &  -20.31 & -20.18 & 0.67        & 81.89       \\
                        & DiffPepBuilder          &  \textbf{-24.62}    & \textbf{-20.57} & \textbf{0.34} & \textbf{99.63}        \\ \midrule
\multirow{3}{*}{TNF-$\alpha$} & AfDesign       &     -17.16    & -19.98 & 0.30         &      71.36      \\
                        & RFd+MPNN       &      -22.37   & -16.26 & 0.54     &       92.76      \\
                        & DiffPepBuilder   &   \textbf{-26.88} & \textbf{-23.63} & \textbf{0.20}         &   \textbf{96.83}      \\ \bottomrule
\end{tabular*}
\end{table*}

\subsection{\textit{De novo} peptide binder design using DiffPepBuilder}\label{result_subsec4}
We further used DiffPepBuilder to generate novel peptide binders for several important drug targets which do not have homologous proteins in the training dataset to assess its capabilities of generating peptides from scratch and compared its performance to AfDesign and RFd+MPNN. We tested on three targets, 3CL$^{\rm pro}$, ALK1 and TNF-$\alpha$. 3CL$^{\rm pro}$ is critical for SARS-CoV-2 viral maturation and replication, which can be inhibited by drugs like Pfizer’s Paxlovid  containing 3CL$^{\rm pro}$ inhibitor of Nirmatrelvir, yielding significant clinical benefits \cite{owen2021oral}. ALK1 is vital for angiogenesis regulation, which serves as a key target in cancer therapy. Inhibiting the ALK1 signaling pathway that is crucial for tumor vasculature, can substantially reduce tumor growth. Therapies targeting ALK1, such as the monoclonal antibody PF-03446962 and the trap receptor Dalantercept, disrupt its signaling to curb tumor progression\cite{mitchell2010alk1,bendell2014safety}. TNF-$\alpha$, a pivotal cytokine in immune regulation and inflammation, plays a significant role in chronic diseases like rheumatoid arthritis and psoriasis. Its central function in inflammation makes it a key target for drugs such as Infliximab and Adalimumab, which treat autoimmune disorders by neutralizing TNF-$\alpha$ to reduce symptoms and control inflammation\cite{jang2021role,monaco2015anti,rau2002adalimumab}.\\

For 3CL$^{\rm pro}$, we selected the substrate pocket as the peptide targeting site (volume: 681.00 \AA$^3$, surface area: 461.50 \AA$^2$, as measured by CavityPlus\cite{xu2018cavityplus,wang2023cavityplus}), which is suitable to accommodate extended peptides and would be too narrow for helical peptides. The crystal structure of 3CL$^{\rm pro}$ with a PDB ID of 7Z4S\cite{miura2023vitro} was chosen to test the model's proficiency in generating non-helical peptides. Conversely, the interface of ALK1 with BMP10 (Bone Morphogenetic Protein 10, PDB ID: 6SF1\cite{salmon2020molecular}) is more expansive, leading DiffPepBuilder to favor both helical and non-helical peptides. As the interface features a significant proportion of hydrophobic residues (with a hydrophobic interface area of 958.46 \AA$^2$ and a hydrophobic interface ratio of 67\% as calculated by Rosetta\cite{das2008macromolecular}), the peptides need to cover more hydrophobic regions and balance the polar interactions for higher binding affinity. TNF-$\alpha$ forms a stable homo-trimer, and its receptor TNFR1 binds to each of the grooves formed by two TNF-$\alpha$ protomers. This placement demands the generation of peptides capable of simultaneously interacting with two protein chains. Consequently, we selected TNF-$\alpha$ as a case study to evaluate the efficacy of DiffPepBuilder in addressing targets with binding sites formed by more than one chain. Specifically, we utilized the protein structure with PDB ID 7KP7\cite{mcmillan2021structural}, which showcases the TNF-$\alpha$ binding interface with human TNFR1. Given the interface’s large size (2843.77 \AA$^2$), flatness, and hydrophilic nature, the design task needs to generate peptides with a large interaction surface and a sophisticated network of polar interactions.\\

For 3CL$^{\rm pro}$, as detailed in Table \ref{tab1} and Fig.\ref{denovo_gen}a, DiffPepBuilder and RFd+MPNN generated potential peptide binders containing 8 to 30 residues with comparable ddG values. As peptides with about 10 residues are preferable for peptide drugs, we then tested the performance of different models with peptide length between 8 to 15 residues. DiffPepBuilder achieved the best performance with an average ddG of -21.17 kcal/mol. In the case of ALK1, as shown in Fig.\ref{denovo_gen}c and Table \ref{tab1}, DiffPepBuilder achieved an average Rosetta binding free energy of -24.49 kcal/mol, outperforming RFd+MPNN for peptide length of 8 to 30 residues. For peptides with 8 to 15 residues, DiffPepBuilder not only generated structures with best binding energies ($\sim$ -45 kcal/mol) but also demonstrated better average values compared to RFd+MPNN. For TNF-$\alpha$, as illustrated in Fig.\ref{denovo_gen}e and Table \ref{tab1}, DiffPepBuilder achieved the best average ddG of -26.88 kcal/mol, with the lowest value approaching about -60 kcal/mol. In generating peptides with 8 to 15 residues, DiffPepBuilder reached an average ddG of -23.63 kcal/mol, outperforming AfDesign and RFd+MPNN by a large margin. We also normalized the ddG over the number of residues and DiffPepBuilder remains the best (Fig.S5). Further interface analysis revealed that DiffPepBuilder excelled in minimizing buried unsatisfied hydrogen bonds, particularly for ALK1. About 78.4\% of its designs featured fewer than 10 unsaturated hydrogen bonds, with $\leq$ 20 hydrophobic exposure ratio mostly, as indicated in Figure.\ref{denovo_gen}a,c,e. Compared to DiffPepBuilder and RFd+MPNN, AfDesign generated peptides suffer from low calculated binding free energy and high ratio of hydrophobic exposure. We further measured the pTM-score (average TM-score of a generated peptide against all the rest ones) as an metric of structure diversity. The mean pTM-score of DiffPepBuilder for ALK1, 3CL$^{\rm pro}$, and TNF-$\alpha$ were 0.34, 0.27, and 0.20, respectively, indicating a good diversity of the generated peptides. As shown in Table \ref{tab1}, the generation diversity of DiffPepBuilder is remarkably better than that of RFd+MPNN and comparable to AfDesign. Additionally, DiffPepBuilder achieved a higher validity (percentage of ddG $<$ 0 for the generated peptides) than RFd+MPNN and AfDesign. \\ 

Fig.\ref{denovo_gen}b,d,f illustrate examples of generated structures for these three targets, respectively. In Fig.\ref{denovo_gen}b, the S1, S2 and S1' subpockets of the 3CL$^{\rm pro}$ substrate binding pocket are occupied by the F5, Y6 and F7 residues of the generated peptide, forming complementary hydrophobic interactions, as well as hydrogen bonds. In Fig.\ref{denovo_gen}d, the interface of ALK1 is covered with the hydrophobic interactions by a generated helical peptide at the core region, and forms hydrogen bonds with the ligand at the edge. Most residues presented outside the helix are polar or charged. In Fig.\ref{denovo_gen}f, the structure of a generated protein-peptide complex reveals that the peptide occupies the hydrophobic pockets on the groove of the TNF-$\alpha$ homotrimer interface and covers the hydrophobic areas of the $\beta$-sheet motif on the interface. It also forms relatively abundant polar interactions.\\

\subsection{Stabilizing peptide binder conformation by introducing disulfide bond}\label{result_subsec5}

\begin{figure*}[htb]
\centering
\includegraphics[width=\linewidth]{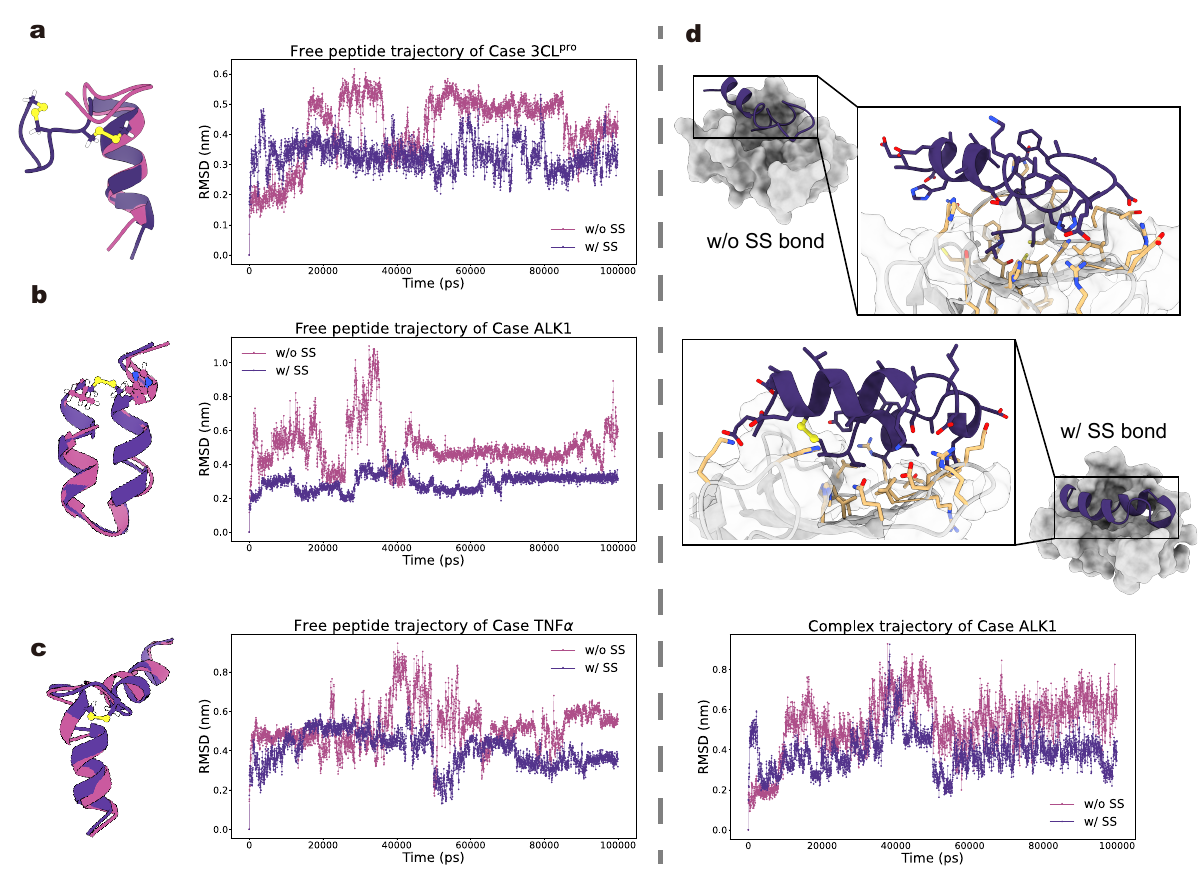}
\caption{\textbf{\textbar\  Case studies of the effect of disulfide bond construction.} \textbf{a}-\textbf{c}. Left: A comparison of free-state peptide structures generated by DiffPepBuilder for 3CL$^{\rm pro}$, ALK1, and TNF-$\alpha$, shown before (in pink) and after (in deep purple) disulfide bond construction at the end of MD simulations. Right: The RMSD trajectories from MD simulations. \textbf{d.} Protein-peptide complex structures of ALK1 at the end of MD simulations. The complex before (above) and after (in the middle) disulfide bond construction are shown. The complex structure is illustrated in gray, with interacting side chains in wheat, highlighting both interfaces featuring hydrogen bonds and hydrophobic interactions. Down: The RMSD trajectories from MD simulations.}\label{md_res}
\end{figure*}

We have shown that DiffPepBuilder performed well in linear peptide binder design. However, linear peptides are often flexible in solution, which may results in large entropy loss upon binding to targets. They are also susceptible to protease degradation, which leads to short half-lives and weak efficacy in vivo\cite{al2022strategies,henninot2018current}. In order to constrict the generated binding conformations and to enhance the binding potency and stability of the designed peptides, we introduced the SSBuilder module into DiffPepBuilder to design disulfide bonds in the generated peptides. After linear peptide generation for the aforementioned three cases, we further used DiffPepBuilder to generate disulfide bond-containing peptides and conducted comparative analyses. \\ 

\begin{table*}[htb]
\caption{Results of MD simulations and MMPBSA assays. The values preceding the "/" symbol represent those w/o disulfide bonds, whereas the values following are w/ disulfide bonds.}\label{tab2}
\begin{tabular*}{\textwidth}{@{\extracolsep\fill}lccc@{}}
\toprule
\multirow{2}{*}{Case name} & \multicolumn{3}{c}{Metric} \\ 
\cmidrule(l){2-4} & Free Peptide RMSD (\AA) & Complex L-RMSD (\AA) & ddG by MMPBSA (kcal/mol) \\
\midrule
3CL$^{\text{pro}}$ & 4.8/3.2 & 7.8/3.2 & -17.32/-47.66 \\
ALK1               & 4.5/3.1 & 5.8/3.3 & -30.98/-56.37 \\
TNF-$\alpha$       & 5.6/3.4 & 4.8/2.7 & -34.45/-54.10 \\
\bottomrule                        
\end{tabular*}
\end{table*}

We first compared the binding strengths of the generated peptide binders with and without disulfide bonds and found that there is no obvious change in Rosetta ddG (See Fig.S6). For each case study, we ranked the genenrated peptides by  Rosetta binding energy, and randomly selected 5 structures from the top 100 for MD simulation studies (details in Sec.S3), including simulations of the free peptides and the protein-peptide complex. We performed an 100 ns MD simulation for each of the generated peptide in free form and calculated the average RMSD (compared to the designed conformation in the complex) of the last 20 ns of the trajectories. The results are shown in Fig.\ref{md_res}a, b, and c. Fig.\ref{md_res}a presents an example for the 3CL$^{\rm pro}$ target, where SSbuilder constructs 2 disulfide bonds, one of which is located at the beginning and end of the peptide loop region, fixing the entire loop region into a cyclic structure and the other is located at the hinge area where the cyclic loop connects with the helix. Fig.\ref{md_res}b illustrates an example peptide generated for the ALK1 target, where the disulfide bond is positioned at the two terminal residues of the peptide, stabilizing the helical hairpin structure that reduced the RMSD from 4.5 \AA\ of the linear peptide to 3.1 \AA. In the example of TNF-$\alpha$ (Fig.\ref{md_res}c), the disulfide bond is also generated in the hinge area between two helical structures, anchoring the two helices together, which causes an RMSD decrease from 5.6 \AA \ to 3.4 \AA. We then filtered the peptides with minimal RMSD and perform 100 ns MD simulation on their complex structures. Fig.\ref{md_res}d displays the structure and L-RMSD trajectory for ALK1 (Fig.\ref{md_res}a). Structures without disulfide bonds exhibit a generally higher flexibility relative to that in the complex compared to those constrained by disulfide bonds. The peptide ligands with disulfide bonds tend to retain or form more stable hydrogen bonds and hydrophobic interactions. Subsequently, we selected the last 20 ns of the complex trajectory for MMPBSA analysis\cite{valdes2021gmx_mmpbsa,miller2012mmpbsa} and found that the peptide with disulfide bonds not only has a significantly lower calculated binding free energy, but also smaller L-RMSD compared with the peptide without disulfide bonds. The other two candidates peptide with disulfied bonds of 3CL$^{\rm pro}$ and TNF-$\alpha$ showed a similar improvement on the binding conformation stability and bindning free  energy (Table \ref{tab2}). The candidate complex structures of ALK1 and 3CL$^{\rm pro}$ after MD are shown in Fig.S8.\\

\section{Conclusion and Discussion}\label{sec3}
Recent advancements in protein binder design have showed promising applications in bio-pharmaceutical and synthetic biology research and development, facilitated by improved computational methods, sophisticated structural biology techniques, and enhanced display technologies\cite{zhang2024revolutionizing,notin2024machine}. However, designing peptide binders remains challenging due to their lower stability, affinity, and specificity compared to proteins. The flexibility and rapid degradation of peptides add complexity to predicting and engineering effective peptide binders, and further innovations are needed to fully harness their potential, possibly involving non-natural amino acids or cyclization strategies to improve their biophysical properties\cite{gupta2022design,wang2024recent}. In this study, we alleviated the scarcity of natural protein-peptide complex data by constructing a diverse, high-quality synthetic dataset and developed DiffPepBuilder, a data-driven diffusion-based generative model. The results in the regeneration tests demonstrated that DiffPepBuilder is capable of generating peptides with conformations closely resembling natural ligands and possesses the potential to explore more optimal binding free energies. Compared to peptides with loop conformations, those with helical conformations can be regenerated better with similar sequences and backbones. This may be due to the inherently non-conservative nature of loop conformations, making it difficult for the model to learn a regular conformational distribution.  In order to stablize the binding conformations, DiffPepBuilder incorporates the  SSBuilder module to construct disulfide bonds in the generated peptides. Molecular  dynamics  simulations and MMPBSA calculations demonstrated that integrating disul- fide bonds into the generated peptide binders can indeed stabilize their binding conformations and enhance binding strength. We further tested the \textit{de novo} peptide binder design ability of DiffPepBuilder on three key drug design targets. DiffPepBuilder outperforms the existing methods in performance, excelling in generating peptides with notable structural diversity and high affinity. \\

A major challenge in deep learning-based, data-driven peptide binder design is the scarcity of high-affinity protein-peptide complex data. While data augmentation is a common practice in representation learning of protein-small molecule interactions\cite{scantlebury2020data,gao2023self}, to the best of our knowledge, similar work in peptide design is absent. In this study, we address the data scarcity issue by utilizing high-quality synthetic data derived from diverse protein-protein complex structures with superior binding performance. DiffPepBuilder demonstrates the ability to capture essential interactions using this synthetic data and successfully regenerate native-like peptide binder structures. This alleviates concerns about a potentially significant gap between natural protein-peptide and synthetic protein-fragment complex data that could mislead the model. The superior \textit{de novo} design performance of DiffPepBuilder on three real-world targets further validates our model’s capability with data augmentation. Additionally, we acknowledge that the model would benefit from training on high-affinity natural protein-peptide data as its availability increases. We also envisage tuning our model on small-quantity data with high affinity via few-shot learning as a future direction.  Moreover, the model's performance could be enhanced by employing a joint diffusion process for categorical and continuous variables, a method that has demonstrated efficacy in the generation of small molecules\cite{huang2023learning,le2023navigating} and proteins\cite{anand2022protein,campbell2024generative}. Additionally, recent advancements in flow matching\cite{lipman2022flow,chen2023riemannian} offer a compelling alternative to diffusion models, presenting a promising avenue for future exploration.\\

As a target-specific \textit{de novo} peptide design method, DiffPepBuilder requires target protein structures, which may be either experimentally determined or \textit{in silico} predicted, along with prior binding site information, as input. In cases where native binding peptides or molecules are already identified, binding site information can be determined either automatically through DiffPepBuilder's Reference Ligand mode or by manually analyzing the interface interactions. Conversely, in the absence of prior binding site data, established pocket searching algorithms such as CavityPlus\cite{xu2018cavityplus,wang2023cavityplus} can be incorporated into the DiffPepBuilder workflow to identify potential binding pockets of target proteins. These pockets are further analyzed to identify hotspot residues via pharmacophore analysis\cite{mannhold2006pharmacophores}. Currently, DiffPepBuilder does not take the flexibility of input target structures into account during model inference. This poses a challenge as conformational changes are frequently observed during the peptide binding process\cite{weikl2014conformational,zarutskie1999conformational,springer1998fast,armstrong2008conformational}, and relying on the rigid structure of target proteins may decrease the model's performance in generating optimal binders. We are actively addressing this challenge by integrating receptor flexibility into DiffPepBuilder to further enhance its generative performance.\\

Turning linear peptide molecules into cyclic ones is an important strategy in peptide drug development\cite{al2022strategies}. Cyclization by introducing disulfide bonds could reduce flexibility and decrease the entropy loss upon target binding\cite{malde2019crystal,nielsen2017orally,gavenonis2014comprehensive}, serving as an important method to enhance the stability of peptide drugs\cite{demmer2009design}. Currently DiffPepBuilder only includes disulfide bonds as conformation restriction methods, in the future, we will further introduce more diverse cyclization methods either with natural or unnatural peptide cyclilization methods. Another issue in peptide drug design is the relatively small sequence space of peptides compared to the chemical space of small molecules or the sequence space of proteins. Further exploration of chemical modification in peptide structures based on increasing experimental data can provide guidance for the design of peptide with unnatural residues, which, incorporating data-driven computational methods, can further broaden the scope of peptide design. \\

\section{Methods}\label{sec4}

\subsection{Dataset construction}\label{method_subsec1}
\textbf{Structure collection} Structures were selected from the Protein Data Bank prior to October 2023, based on the following criteria: composed solely of proteins, comprising more than two chains, and having a resolution lower than 2.5 \AA. Symmetric structures were complemented using biounit information, and missing atoms were rectified with PDBFixer\cite{eastman2017openmm}. A total of 5,358 structures with the shortest chain length between 8 and 30 residues were collected as the peptide-protein dataset, while 11,469 structures with the shortest chain length greater than 30 residues were collected as the protein-protein dataset. For the latter group, the shortest chain of the complex was defined as the ligand, and any chains interacting with the ligand, defined by an interface area (dSASA) of $\geq$ 100 \AA$^2$, were output in pairs. The permanent protein-protein complexes, which has a buried surface area larger than 3,000 \AA$^2$ were not excluded, resulted in 35,338 protein dimer structures.\\

\textbf{Interfacial fragment extraction} For pairwise protein-protein complexes, we first conducted an analysis on the solvent accessible surface area (SASA) of both the ligand and the complex, utilizing the BioPython software package\cite{cock2009biopython}. Then we defined residues with a buried surface area (BSA, $\rm SASA_{ligand} - SASA_{complex}$) greater than 36 \AA$^2$ and a buried proportion (computed by $\rm (SASA_{ligand} - SASA_{complex})/SASA_{ligand}$) exceeding 40\% as buried residues. Within the range of 8 $\sim$ 30 residues, we established a series of truncation windows. For each chain in a  dimer,  we slid  these  windows  across, identifying a segment as a helical structure if it contains a continuous sequence  of $\geq$ 8 residues in a helical conformation. The others are non-helical. For helical segments, the proportion of hotspot residues are $\geq$ 40\% ; for non-helical ones, the requirement was $\geq$ 80\%. For each complex structure, we conducted an initial deduplication check before the overall sequence redundancy reduction, removing redundant entities from the set of fragments associated with the same dimer complex, which was done according to the criteria of L-RMSD $\leq$ 0.5 \AA\ (same length) and ligand sequence similarity $\geq$ 70\%. This process resulted 34,618 entries.\\

\textbf{Redundancy reduction} was implemented on both peptide-protein and protein-protein dataset. It was carried out from two aspects: sequence similarity and structural similarity. The measurement of sequence similarity involved conducting a multi-sequence alignment of the longest chain by CD-HIT\cite{fu2012cd,li2006cd,huang2010cd} with a 90$\%$ similarity cutoff. Then, in each cluster, the structure with the highest X-ray resolution was chosen as the reference, then the L-RMSD of C$\alpha$ atoms and sequence similarity of other ligands (the shortest chain) in the cluster were calculated. Structures with L-RMSD larger than 5 \AA\  and similarity lower than 70$\%$ were preserved for further analysis. This  redundancy  reduction  was implemented for both PepPC (peptide-protein complex) and PepPC-F (peptide-protein complex fragments). After the redundancy reduction process, 117 clusters were generated from 275 helical-ligand complexes and 1,021 clusters of nonhelical-ligand complexes in PepPC; 2,588 clusters were generated from 16,553 helical-ligand complexes and 3,386 clusters of 18,065 loop-ligand complexes in PepPC-F. After further dataset cleaning, including exclusion of disulfide bond and membrane protein (see SI for details), there were 3,832 structures (232 helical ligands) in PepPC, and 14,897 structures (4,241 helical ligands) in PepPC-F.\\

We calculated the \textbf{secondary structure} information for all peptide ligands using the DSSP module in Biopython\cite{cock2009biopython}. In this context, Helix includes A-Helix, 5-Helix and 3-Helix; Beta includes B-sheet; Turn includes B-Bridge, Bend and Turn; Unstructured represents Coil.\\

\subsection{Input preparation}\label{method_subsec2}

In the peptide binder \textit{de novo} generation process, DiffPepBuilder initially takes the full target as input and truncates it to the binding pocket according to the user provided binding site information. Specifically, the binding site information can be formatted into three types, each corresponding to one of the three pattern diagrams shown in Fig.\ref{model_arch}a. The first type, Motif, requires the user to provide information about the target binding motif (can be a structurally truncated PDB file or sequence information). The second type, Hotspots, requires the hotspot information (often provided as strings of hotspot residue IDs). The third type, Reference Ligand, allows for specifying a reference ligand by inputting a PDB file and specify the receptor. The model will delineate residues on the receptor that are a certain distance from the ligand C$\alpha$ (default is 8 \AA) as the hotspots. We We tested several cutoff values and found that 8 \AA cutoff maintains an optimal protein motif size (See Fig.S4 and Table S2). For all three types of information, residues whose C$\alpha$ atom are within 10 \AA \ of the C$\alpha$ atoms from these binding motif or hotspots compose the binding pocket. The model then calculates the geometric center of the pocket's C$\alpha$ atoms and adds a $\sigma_0=2$ \AA\  Gaussian noise to the center position to enhance the generation diversity:

\begin{equation}
    \bm{x}_{\mathrm{center}} = \sum_{i=1}^{N_{\mathrm{rec}}} \bm{x}_{i}^{\mathrm{rec}} + \bm{\varepsilon}, \  \ \bm{\varepsilon} \sim  \mathcal{N}(0,\sigma_0 \bm{I})\label{eq_1}
\end{equation}

\noindent
where $\bm{x}_i^{\mathrm{rec}}$ represents the coordinates of the C$\alpha$ atom of the $i$-th pocket residue, and $N_{\mathrm{rec}}$ is the number of pocket residues. For each peptide length $N_{\rm lig}$ within the user-defined range $[N_{\rm min}, N_{\rm max}]$, the model independently samples $N_{\rm lig}$ positions, $\{\bm{x}_{i}^{\mathrm{lig}}\}_{i=1}^{N_{\mathrm{lig}}}$, from an isotropic Gaussian distribution $\mathcal{N}(\bm{x}_{\mathrm{center}},\sigma\bm{I})$, with $\sigma$ acting as an adjustable hyper-parameter that controls the structural diversity of the \textit{de novo} generated peptides. The model subsequently recenters the target pocket-peptide ligand complex to the noise-adjusted centroid $\bm{x}_{\mathrm{center}}$ to ensure the (approximate) SE(3)-equivariance in the diffusion process, aligning with methodologies suggested by previous works\cite{kohler2020equivariant,xu2022geodiff,yim2023se}:

\begin{equation}
    \bm{x}_i\leftarrow\bm{x}_i-\bm{x}_{\mathrm{center}} \label{eq_2}
\end{equation}

\noindent 
here $\bm{x}_i$ denotes the C$\alpha$ coordinates of an arbitrary residue $i$ within the pocket-peptide ligand complex. Following this, the model constructs a fully-connected graph of the complex, which is then processed through the diffusion-based generative procedure. The model additionally generates a binary mask,  $\bm{m}\in\{0,1\}^{N}$, that indicates the positions of the pocket residues which are to be fixed during the diffusion process.\\

Prior to the diffusion-based generative procedure, the backbone of the target pocket-peptide ligand complex is parameterized as a collection of rigid-body frames, $\mathbf{T}=[\bm{T}_1,\ldots,\mathbf{T}_N]$, akin to the approach in AlphaFold 2\cite{jumper2021highly}, where $N=N_{\mathrm{rec}}+N_{\mathrm{lig}}$. Note that each frame $\bm{T}_i=(\bm{R}_i, \bm{x}_i)$ consists of a 3-dimensional rotation $\bm{R}_i\in \mathrm{SO(3)}$ and a translation $\bm{x}_{n}\in \mathcal{R}^3$. Initial frames of the peptide to be generated are independently sampled from a uniform distribution on SO(3) i.e. $\mathcal{U}^{\mathrm{SO(3)}}$.\\

\subsection{Architecture of the denoising network}\label{method_subsec3}

The diffusion model utilized in DiffPepBuilder essentially introduces multi-level noises to the ground-truth peptide ligand conformation and learns to recover the original structure through a denoising network. During the inference process, it reverses this noising process and \textit{de novo} generates the peptide ligand conditioned on the pocket (Fig.\ref{model_arch}b). Mathematically, the forward noising process and the reverse denoising process on $\mathrm{SE(3)}^N$ are characterized by:

\begin{align}
    \mathrm{d}\mathbf{T}^{(t)}&=\big[0,-\textstyle\frac{1}{2}\mathrm{P}\mathbf{X}^{(t)}\big]\mathrm{d}t+\big[\mathrm{d}\mathbf{B}^{(t)}_{\mathrm{SO(3)}^N},\mathrm{d}\mathbf{B}^{(t)}_{\mathcal{R}^{3N}}\big]\label{eq_3}
\end{align}

\noindent
and

\begin{align}
    \mathrm{d}\overleftarrow{\mathbf{R}}^{(t)} &= \nabla_R\log{p_t}(\overleftarrow{\mathbf{T}}^{(t)})\mathrm{d}t+\mathrm{d}\mathbf{B}^{(t)}_{\mathrm{SO(3)}^N}\\
    \mathrm{d}\overleftarrow{\mathbf{X}}^{(t)} &= \mathrm{P}\big(\textstyle\frac{1}{2}\mathbf{X}^{(t)}+\nabla_x\log{p_t}(\overleftarrow{\mathbf{T}}^{(t)})\big)\mathrm{d}t+\mathrm{Pd}\mathbf{B}^{(t)}_{\mathcal{R}^{3N}}\label{eq_4}
\end{align}

\noindent
respectively\cite{de2022riemannian,yim2023se}, where $\mathrm{P}$ is the projection matrix removing center of mass. Yim et al.\cite{yim2023se} proved that the score i.e. $[\nabla_R\log{p_t},\nabla_x\log{p_t}]$ can be computed seperately for the rotation and translation of each residue. The denoising network is trained to approximate the score with $[s^R_{\theta},s^x_{\theta}]$, a process known as denoising score matching (DSM)\cite{ho2020denoising,song2020score}.\\

As depicted in Fig.\ref{model_arch}d, the denoising network of DiffPepBuilder is composed of three main components: the Embedding Module, the Cross Update Module, and the Multi-Task Decoder. In the \textbf{embedding stage}, initial node embeddings $\mathbf{h}_0=[h_1,\ldots,h_N]$ are generated by concatenating diffusion timestep embeddings, residue index embeddings, residue type embeddings, and pLM embeddings. We employed sinusoidal positional embeddings\cite{vaswani2017attention} for residue index embeddings and utilized a frozen 650M parameter ESM-2\cite{lin2022language} encoder to extract the pLM embeddings from the sequence of the untruncated target. Peptide ligand residues are assigned a special residue type \texttt{<mask>} in the residue type embedding process, and the average pLM embeddings of pocket residues are used as their pLM embeddings. To indicate breaks between chains, we incorporated a 100-residue gap in the residue indices between each chain, following the approach in RoseTTAFold\cite{baek2021accurate}. Node embeddings $\mathbf{h}_0$ are subject to cross-concatenation and are subsequently concatenated with self-conditioning distogram embeddings\cite{chen2022analog,watson2023novo} and sequence distance embeddings, resulting in the preliminary edge embeddings $\mathbf{z}_0=(z_{ij})_{i,j=1}^N$. Following RFdiffusion, we performed self-conditioning of predicted C$\alpha$ pairwise distances, $\hat{\mathbf{d}}$, in 50\% of the examples in the training process.\\

The initial node embeddings $\mathbf{h}_0$, edge embeddings $\mathbf{z}_0$, and the backbone frames $\mathbf{T}_0$ are subsequently processed through the \textbf{Cross Update Module} (Fig.S3). Specifically, within an arbitrary layer $l$, the node update is achieved through a combination of Invariant Point Attention (IPA)\cite{jumper2021highly}, Transformer encoder layers\cite{vaswani2017attention}, Multi-Layer Perceptron (MLP), Layer Normalization (LayerNorm)\cite{ba2016layer}, and Linear layers:

\begin{align}
    \mathbf{h}_{\mathrm{ipa}} &= \mathrm{LayerNorm}(\mathrm{IPA}(\mathbf{h}_l,\mathbf{z}_l,\mathbf{T}_l)+\mathbf{h}_l) \\
    \mathbf{h}_{\mathrm{trans}} &= \mathrm{Transformer}(\mathbf{h}_{\mathrm{ipa}}\,\|\, \mathrm{Linear}(\mathbf{h}_0))\\
    \mathbf{h}_{l+1} &= \mathrm{MLP}(\mathbf{h}_\mathrm{ipa}+\mathrm{Linear}(\mathbf{h}_\mathrm{trans}))\label{eq_7}
\end{align}

\noindent
here $\|$ denotes concatenation process. Subsequently, the edge update is performed as:

\begin{align}
    \mathbf{h}_\mathrm{proj} &= \mathrm{Linear}(\mathbf{h}_{l+1})\\
    \mathbf{z}_\mathrm{concat} &= \mathbf{z}_l \,\|\, \mathrm{CrossConcat}(\mathbf{h}_\mathrm{proj})\\
    \mathbf{z}_{l+1} &= \mathrm{LayerNorm}(\mathrm{MLP}(\mathbf{z}_\mathrm{concat}))\label{eq_8}
\end{align}

\noindent
here \texttt{proj} is an abbreviation for projection, which is implemented by a Linear layer. Lastly, the frame update is executed following the \texttt{BackboneUpdate} algorithm in AlphaFold 2:

\begin{align}
    \mathbf{T}_\mathrm{update} &= \mathrm{CalcRot}(\mathrm{Linear}(\mathbf{h}_l))\\
    \mathbf{T}_\mathrm{masked} &= \mathrm{MaskRec}(\mathbf{T}_\mathrm{update},\bm{m})\\
    \mathbf{T}_{l+1} &= \mathbf{T}_{l}\cdot \mathbf{T}_\mathrm{masked}\label{eq_9}
\end{align}

\noindent
where the \texttt{CalcRot} function transforms the input non-unit quaternion into a rotation matrix $\bm{R}_{i,\mathrm{update}}$ and outputs it along with the coordinate update $\bm{x}_{i,\mathrm{update}}$. The \texttt{MaskRec} function modifies the frame update of target pocket residues to $\bm{T}_i^\mathrm{rec}=(\bm{R}_i^\mathrm{rec},\bm{x}_i^\mathrm{rec})=(\bm{I},\bm{0})$, ensuring the target remains fixed during the diffusion process. We stacked $L=6$ layers with no shared parameters to form the Cross Update Module.\\

The updated node embeddings $\mathbf{h}_L$, edge embeddings $\mathbf{z}_L$, and the backbone frames $\mathbf{T}_L$ are then fed into the \textbf{Multi-Task Decoder}. This decoder predicts the translational and rotational scores of the current diffusion step, the residue types of the peptide ligand, the main chain torsion angles $\psi$ and the side chain torsion angles $\chi_{1}$\(\sim\)$\chi_{4}$. We take the final frame representation $\mathbf{T}_L\equiv\hat{\mathbf{T}}^{(0)}=(\hat{\mathbf{R}}^{(0)},\hat{\mathbf{X}}^{(0)})$ to calculate the translational score $s^x_{\theta}$ and rotational score $s^r_{\theta}$ following Yim et al.\cite{yim2023se}. Torsion angle prediction is implemented as (using $\psi$ prediction as an example):

\begin{align}
    \mathbf{h}_\mathrm{psi} &= \mathrm{MLP}(\mathbf{h}_L)\\
    \bm{\psi}_\mathrm{unnorm} &= \mathrm{Linear}(\mathbf{h}_\mathrm{psi}+\mathbf{h}_L)\\
    \bm{\psi}_\mathrm{pred} &= \mathrm{Normalize}(\bm{\psi}_\mathrm{unnorm})\label{eq_10}
\end{align}

\noindent
where the \texttt{Normalize} function is used to normalize the output torsion angles to be $-180^\circ\sim 180^\circ$. Residue type prediction is achieved in a one-shot decoding manner:

\begin{align}
    \mathbf{d}_\mathrm{pred} &= \mathrm{CalcPairDist}(\mathbf{X}_L)\\
    \mathbf{z}_\mathrm{mean} &= \mathrm{RowMean}(\mathrm{DistMask}(\mathbf{z}_L,\mathbf{d}_\mathrm{pred}))\\
    \mathbf{z}_\mathrm{logits} &= \mathrm{MLP}(\mathbf{h}_L\,\|\, \mathbf{z}_\mathrm{mean})\\
    \mathbf{s}_\mathrm{prob} &= \mathrm{SoftMax}(\mathbf{z}_\mathrm{logits}/\tau)\\
    \bm{s}_\mathrm{pred} &= \mathrm{Sample}(\mathbf{s}_\mathrm{prob})\label{eq_11}
\end{align}

\noindent
where the \texttt{CalcPairDist} function calculates pairwise distances of C$\alpha$ atoms based on given C$\alpha$ coordinates. The \texttt{DistMask} function masks residue pairs whose C$\alpha$ distances exceed a specified threshold, $d_\mathrm{max}=12$ \AA. The sampling temperature $\tau$ is a hyper-parameter that controls the randomness of the sampling process. We set $\tau=0.1$ throughout the generation process. The \texttt{Sample} function samples from the multinomial distribution $\mathbf{s}_\mathrm{prob}$ to determine the final amino acid type $\bm{s}_\mathrm{pred}$. In the sampling process, the peptide backbone frames are first iteratively denoised, and the residue types are subsequently predicted leverageing the final ($t=0$) backbone frames. During the sampling of the residue types, we additionally calculate per-residue entropy as:

\begin{align}
    S_i &= -\sum_{k=1}^{20}s_{\mathrm{prob},k}\log{s_{\mathrm{prob},k}}\label{eq_12} 
\end{align}

\noindent
where $s_{\mathrm{prob},k}$ represents the probability of the $k$-th amino acid type for residue $i$ of an arbitrary peptide ligand.\\

After the denoising process, DiffPepBuilder reconstructs the backbone of the generated peptide ligand using the final frame representations and the predicted $\psi$ torsion angles as that in AlphaFold 2\cite{jumper2021highly}, and ultimately rebuilds the side chain conformations with PDBFixer\cite{eastman2017openmm} according to the predicted C$\beta$ orientation to obtain an all-atom structure. The predicted complex structures undergo optimization using Rosetta's FastRelax\cite{conway2014relaxation} with a Fixbb (Fixed backbone) protocol to eliminate clashes of the side chains.\\

\subsection{Details of SSBuilder}\label{method_subsec4}

\textbf{SSBuilder} method is based on a structure library of disulfide-bonded pairs, which is extracted from the PDB entries with a X-ray resolution $<$ 2.5 \AA \ and prior to December 2023. we extracted all the structures of disulfide-bonded two cysteine residues. We then calculated the values of the dihedral angles formed by C$\alpha$1-C$\beta$1-C$\beta$2-C$\alpha$2 (Dihedral), the values of the two bond angles C$\alpha$1-C$\beta$1-C$\beta$2 (angle1) and C$\beta$1-C$\beta$2-C$\alpha$2 (angle2), as well as the value of the C$\beta$1-C$\beta$2 (distance), to serve as geometric matching criteria (See Fig.S9). Angles were binned every $5^\circ$, and distance bins' width were set to 0.1 \AA.\\

Residue entropy in our deep learning model output was used to identify potential disulfide bond site. This serves as an additional filter, excluding residues critical to binding interactions based on the model's confidence in residue type, thereby ensuring that only non-essential residues are considered for cysteine substitution. Residues with an entropy greater than 0.01 will be selected and paired. The angles and distance parameters of each residue pair are quickly matched to corresponding bins(those corresponding bins cannot be found will be skipped). This is followed by a comparison with the same geometric parameters of the disulfide bond building blocks within those bins. The residue pair that achieves the best parameter matching (where the sum of the absolute differences of all parameters is minimized) is then spliced with its corresponding disulfide bond unit to create a cyclized structure. \\
 
\subsection{Training details}\label{method_subsec5}

In the training of the denoising network, we employed a composite loss scheme that includes denoising score matching (DSM) loss $\mathcal{L}_{\rm dsm}$, peptide amino acid type loss $\mathcal{L}_{\rm aa}$, and a series of auxiliary losses comprising peptide backbone position loss $\mathcal{L}_{\rm bb}$, pairwise atomic distance loss $\mathcal{L}_{\rm dist}$, side chain torsion angle loss $\mathcal{L}_{\rm chi}$, and C$\alpha$ atom clash loss $\mathcal{L}_{\rm clash}$. The SE(3) DSM loss is given by:

\begin{align}
    \mathcal{L}_\mathrm{dsm} &= \mathbb{E}_{i,t}\big[\lambda^{x}_t\|\nabla_x\log{p_{t|0}}(\bm{x}^{(t)}_i|\hat{\bm{x}}^{(0)}_i)-s^x_\theta(t,\mathbf{T}_i^{(t)})\|\big]\nonumber\\
    +\mathbb{E}_{i,t}&\big[\lambda^{R}_t\|\nabla_R\log{p_{t|0}}(\bm{R}^{(t)}_i|\hat{\bm{R}}^{(0)}_i)-s^R_\theta(t,\mathbf{T}_i^{(t)})\|\big]\label{eq_13}
\end{align}

\noindent
where $i\in \{1,\ldots,N_\mathrm{lig}\}$, $t\sim\mathcal{U}[0,1]$. We utilized the weight schedule following previous works\cite{song2020score,yim2023se}:

\begin{align}
    \lambda_t^R &= 1/\mathbb{E}(\big[\|\nabla\log{p_{t|0}}(\bm{R}^{(t)}_i|\bm{R}^{(0)}_i)\|^2_{\mathrm{SO(3)}}\big])\\
    \lambda_t^x &= e^{t/2}(1-e^{-t})\label{eq_14}
\end{align}

\noindent
The peptide amino acid type loss $\mathcal{L}_\mathrm{aa}$ is calculated as a residue-averaged cross entropy loss for all peptide residues. For the auxiliary losses, the peptide backbone position loss is formulated as a mean squared error (MSE) loss on backbone atom positions:

\begin{align}
    \mathcal{L}_{\rm bb} &= \frac{1}{4N_\mathrm{lig}}\sum_{i=1}^{N_\mathrm{lig}}\sum_{n}\|\bm{x}^{(0)}_{i,n}-\hat{\bm{x}}^{(0)}_{i,n}\|^2\label{eq_15}
\end{align}

\noindent
where $n\in\{\texttt{N},\texttt{C},\texttt{C}\alpha,\texttt{O}\}$. The pairwise atomic distance loss is also a MSE loss on pairwise atomic distances:

\begin{align}
    \mathcal{L}_\mathrm{dist} &= \frac{1}{Z_\mathrm{dist}}\sum_{i=1}^{N}\sum_{j=1}^{N_\mathrm{lig}}\sum_{n,m}\mathds{1}(d_{ij}^{nm}<d_\mathrm{dist})\|d_{ij}^{nm}-\hat{d}_{ij}^{nm}\|^2\label{eq_16}
\end{align}

\noindent
where $d_\mathrm{dist}=6$ \AA, $n,m\in\{\texttt{N},\texttt{C},\texttt{C}\alpha,\texttt{O}\}$, and the normalizing factor $Z_\mathrm{dist}$ is given by:

\begin{align}
    Z_\mathrm{dist} &= \sum_{i=1}^{N}\sum_{j=1}^{N_\mathrm{lig}}\sum_{n,m}\mathds{1}(d_{ij}^{nm}<d_\mathrm{dist})-N_\mathrm{lig}\label{eq_17}
\end{align}

\noindent
The side chain torsion angle loss is a MSE loss as computed in AlphaFold 2\cite{jumper2021highly}. We only considered $\chi_1$ and $\chi_2$ angles because $\chi_3$ and $\chi_4$ are relatively less informative and more challenging to predict accurately\cite{dauparas2023atomic}. We empirically found that this side chain torsion angle loss helps the model better capture inter-residue interactions. The C$\alpha$ atom clash loss $\mathcal{L}_{\rm clash}$ is defined as:

\begin{align}
    \mathcal{L}_\mathrm{clash} &= \frac{\rm 1e3}{Z_\mathrm{clash}} \sum_{i=1}^{N_\mathrm{rec}}\sum_{i=1}^{N_\mathrm{lig}}\mathds{1}(d_{ij}^{\mathrm{C}\alpha}<d_\mathrm{clash})\label{eq_18}
\end{align}

\noindent
where we set the clash threshold $d_\mathrm{clash}=4$ \AA, and the normalizing factor $Z_\mathrm{clash}$ is computed by $Z_\mathrm{clash} = N_\mathrm{rec}\times N_\mathrm{lig}$. The C$\alpha$ atom clash loss is utilized to minimize steric clashes between the target and the generated peptide ligand. The full training loss is formulated as:

\begin{align}
    \mathcal{L} &= \mathcal{L}_\mathrm{dsm}\nonumber\\
    &+\mathds{1}(t<0.25)w_1\mathcal{L}_\mathrm{aa}\nonumber\\
    &+\mathds{1}(t<0.25)w_2\big(\mathcal{L}_\mathrm{bb}+\mathcal{L}_\mathrm{dist}+\mathcal{L}_\mathrm{chi}\big)\nonumber\\
    &+\mathds{1}(t<0.5)w_3\mathcal{L}_\mathrm{clash}\label{eq_19}
\end{align}

\noindent
where we set the loss weights $w_1=2,\, w_2=0.25,\, w_3=0.25$. We applied the amino acid type loss $\mathcal{L}_{aa}$ and auxiliary losses primarily near $t=0$ to encourage the model to learn fine-grained characteristics.\\

The denoising network consists of $\sim$104M parameters and was trained exclusively on the PepPC-F dataset using 8 NVIDIA A800 80GB GPUs. The training lasted $\sim$5 days. We employed the AdamW optimizer\cite{loshchilov2017decoupled} with a learning rate of 1e-5. For multi-GPU training and inference, we used \texttt{DistributedDataParallel} implemented by PyTorch\cite{paszke2019pytorch}.\\

\subsection{Regeneration details}\label{method_subsec6}

For the regeneration test set PepPC-HF, we first collected a total of 822 structures from the PepPC dataset that overlapped with the PDBbind2020 database\cite{wang2005pdbbind}. Redundant entries were removed with a 40\% sequence similarity threshold using CD-HIT\cite{fu2012cd,li2006cd,huang2010cd}. We then filtered out entries with an activity value ($K_i$, $K_d$, $\rm IC_{50}$) $\leq$ 0.1 $\mu$M and conducted further deduplication against the PepPC-F data using a maximum 60\% sequence similarity criterion for target proteins (mostly $\leq$ 30\%). Ultimately, these processes yielded 30 complex structures featuring high-activity peptide ligands (See Table S1 for details).\\

For \textbf{DiffPepBuilder}, we used the Reference Ligand mode and set the peptide length to that of the original peptide. We set the number of denoising steps, the noise scale, and the sampling temperature to 500, 1.0, and 0.1, respectively. For \textbf{AfDesign}, we utilized its fine-tuned version for generating peptide binders\cite{kosugi2022solubility}. We set the weights for various parameters as follows: solubility at 0.5, msa\_ent (multiple sequence alignment entropy) at 0.01, pLDDT (predicted Local Distance Difference Test) at 0.1, pae\_intra (predicted aligned error for intra-chain) at 0.1, and pae\_inter (predicted aligned error for inter-chain) at 1.0. The weights for con\_intra (contact predictions within chains) and con\_inter (contact predictions between chains) were set to 0.1 and 0.5, respectively. The cycles for the three stages were 100, 100, and 10. For \textbf{RFdiffusion} used for peptide backbone generation, we set noise scale to 0. We employed the dlbinder design method\cite{bennett2023improving} for the subsequent ProteinMPNN\cite{dauparas2022robust} sequence design and Rosetta side-chain assembly and FastRelax. Pocket residues within 5 \AA \ of the reference peptide were selected as the hotspots for all of the three methods, and each method performed 128 samplings.\\

\subsection{\textit{De novo} generation settings}\label{method_subsec7}

For \textbf{DiffPepBuilder}, we sampled 128 times for each peptide length. The number of the denoising steps is set to 500. We set the noise scale to 1.0, 0.5, and 2.0 for ALK1, 3CL$^{\rm pro}$ and TNF-$\alpha$, respectively. The impact of the noise scale settings is illustrated in Fig.S7. We note that adjusting the noise scale provides versatile control over the average binding strength and generation diversity. For \textbf{AfDesign}, we adopted the parameters in Sec.\ref{method_subsec6} and executed 128 sampling for each length. For \textbf{RFdiffusion}, we adhered to the default noise scale settings and employed the dlbinder design method as in Sec.\ref{method_subsec6}. We sampled 128 times for each specified length. For sequence design, we generated one sequence for each backbone. Peptide length was set to 8$\sim$30 for generation for each method, and the structures generated by all programs were optimized using Rosetta FastRelax\cite{conway2014relaxation} before evaluation. The interface parameters, including ddG and buried unsatisfied hydrogen bonds, were also calculated by Rosetta.\\

As there are no reference peptide ligands available in de novo generation studies, we used hotpots to define peptide binding sites. Hotspots for 3CL$^{\rm pro}$, ALK1 and TNF-$\alpha$ are  "B24-B45-B46-B49-B140-B142-B143-B144-B145-B163-B164-B165-B166-B168-B188-B189-B191-B192","B40-B58-B59-B71-B72-B87","C23-C25-C138-C139-C140-C141-D67-D68-D69-D71-D73-D79-D80-D81-D82-D83-D84-D87-D88-D89-D125-D127-D128-D129", respectively. For 3CL$^{\rm pro}$, based on complex structure in 7Z4S, we initially selected all residues on the 3CL$^{\rm pro}$ receptor that are within 8 \AA\ of the C$\alpha$ atoms of the cyclic peptide residues. We then performed an alanine scan using Rosetta on these residues. Those with $\rm ddG_{mutant}- ddG_{wild-type} \geq\ 1\, $kcal/mol are identified as hotspots, For ALK1, in the protein-protein complex structure of 6SF1, we used BMP10 as a reference and selected hotspots with the same criteria of 3CL$^{\rm pro}$. For the TNF-$\alpha$, on of three TNFR1s in 7KP7 was selected as reference ligand, hotspots were selected from the TNF-$\alpha$ trimer structure according to the same criteria. All models use the same hotspot information as input for peptide generation.\\

\backmatter

\bmhead{Code Availability}
The source code is available at \url{https://github.com/YuzheWangPKU/DiffPepBuilder}.

\bmhead{Data Availability}
The PepPC and PepPC-F datasets are available at \url{https://github.com/YuzheWangPKU/DiffPepBuilder/tree/main/datasets}.

\bmhead{Acknowledgements}

This work was supported in part by the National Key R\&D Program of China (2022YFA1303700), and the National Natural Science Foundation of China (21977007, T2321001), and the Chinese Academy of Medical Science (2021-I2M-5-014).

\bmhead{Supporting Information}
Detailed descriptions of dataset processing and division, specifics of the regeneration task, illustration of the model's cross update module, \textit{de novo} generation task, molecular dynamics simulation parameter settings and statistical details of SSbuilder parameters.\\

\end{multicols}

%%======================================================
\newpage
\bibliography{sn-bibliography}% common bib file


\begin{thebibliography}{108}
\ifx \bisbn   \undefined \def \bisbn  #1{ISBN #1}\fi
\ifx \binits  \undefined \def \binits#1{#1}\fi
\ifx \bauthor  \undefined \def \bauthor#1{#1}\fi
\ifx \batitle  \undefined \def \batitle#1{#1}\fi
\ifx \bjtitle  \undefined \def \bjtitle#1{#1}\fi
\ifx \bvolume  \undefined \def \bvolume#1{\textit{#1}}\fi
\ifx \byear  \undefined \def \byear#1{#1}\fi
\ifx \bissue  \undefined \def \bissue#1{#1}\fi
\ifx \bfpage  \undefined \def \bfpage#1{#1}\fi
\ifx \blpage  \undefined \def \blpage #1{#1}\fi
\ifx \burl  \undefined \def \burl#1{\textsf{#1}}\fi
\ifx \doiurl  \undefined \def \doiurl#1{\url{https://doi.org/#1}}\fi
\ifx \betal  \undefined \def \betal{\textit{et al.}}\fi
\ifx \binstitute  \undefined \def \binstitute#1{#1}\fi
\ifx \binstitutionaled  \undefined \def \binstitutionaled#1{#1}\fi
\ifx \bctitle  \undefined \def \bctitle#1{#1}\fi
\ifx \beditor  \undefined \def \beditor#1{#1}\fi
\ifx \bpublisher  \undefined \def \bpublisher#1{#1}\fi
\ifx \bbtitle  \undefined \def \bbtitle#1{#1}\fi
\ifx \bedition  \undefined \def \bedition#1{#1}\fi
\ifx \bseriesno  \undefined \def \bseriesno#1{#1}\fi
\ifx \blocation  \undefined \def \blocation#1{#1}\fi
\ifx \bsertitle  \undefined \def \bsertitle#1{#1}\fi
\ifx \bsnm \undefined \def \bsnm#1{#1}\fi
\ifx \bsuffix \undefined \def \bsuffix#1{#1}\fi
\ifx \bparticle \undefined \def \bparticle#1{#1}\fi
\ifx \barticle \undefined \def \barticle#1{#1}\fi
\bibcommenthead
\ifx \bconfdate \undefined \def \bconfdate #1{#1}\fi
\ifx \botherref \undefined \def \botherref #1{#1}\fi
\ifx \url \undefined \def \url#1{\textsf{#1}}\fi
\ifx \bchapter \undefined \def \bchapter#1{#1}\fi
\ifx \bbook \undefined \def \bbook#1{#1}\fi
\ifx \bcomment \undefined \def \bcomment#1{#1}\fi
\ifx \oauthor \undefined \def \oauthor#1{#1}\fi
\ifx \citeauthoryear \undefined \def \citeauthoryear#1{#1}\fi
\ifx \endbibitem  \undefined \def \endbibitem {}\fi
\ifx \bconflocation  \undefined \def \bconflocation#1{#1}\fi
\ifx \arxivurl  \undefined \def \arxivurl#1{\textsf{#1}}\fi
\csname PreBibitemsHook\endcsname

%%% 1
\bibitem[\protect\citeauthoryear{Bodanszky}{1988}]{bodanszky1988peptide}
\begin{botherref}
\oauthor{\bsnm{Bodanszky}, \binits{M.}}:
Peptide chemistry.
A Practical Textbook,
(1988)
\end{botherref}
\endbibitem

%%% 2
\bibitem[\protect\citeauthoryear{Craik et~al.}{2013}]{craik2013future}
\begin{barticle}
\bauthor{\bsnm{Craik}, \binits{D.J.}},
\bauthor{\bsnm{Fairlie}, \binits{D.P.}},
\bauthor{\bsnm{Liras}, \binits{S.}},
\bauthor{\bsnm{Price}, \binits{D.}}:
\batitle{The future of peptide-based drugs}.
\bjtitle{Chemical biology \& drug design}
\bvolume{81},
\bfpage{136}--\blpage{147}
(\byear{2013})
\end{barticle}
\endbibitem

%%% 3
\bibitem[\protect\citeauthoryear{Fosgerau and Hoffmann}{2015}]{fosgerau2015peptide}
\begin{barticle}
\bauthor{\bsnm{Fosgerau}, \binits{K.}},
\bauthor{\bsnm{Hoffmann}, \binits{T.}}:
\batitle{Peptide therapeutics: current status and future directions}.
\bjtitle{Drug discovery today}
\bvolume{20},
\bfpage{122}--\blpage{128}
(\byear{2015})
\end{barticle}
\endbibitem

%%% 4
\bibitem[\protect\citeauthoryear{Gomes et~al.}{2018}]{gomes2018designing}
\begin{barticle}
\bauthor{\bsnm{Gomes}, \binits{B.}},
\bauthor{\bsnm{Augusto}, \binits{M.T.}},
\bauthor{\bsnm{Fel{\'\i}cio}, \binits{M.R.}},
\bauthor{\bsnm{Hollmann}, \binits{A.}},
\bauthor{\bsnm{Franco}, \binits{O.L.}},
\bauthor{\bsnm{Gon{\c{c}}alves}, \binits{S.}},
\bauthor{\bsnm{Santos}, \binits{N.C.}}:
\batitle{Designing improved active peptides for therapeutic approaches against infectious diseases}.
\bjtitle{Biotechnology advances}
\bvolume{36},
\bfpage{415}--\blpage{429}
(\byear{2018})
\end{barticle}
\endbibitem

%%% 5
\bibitem[\protect\citeauthoryear{Muttenthaler et~al.}{2021}]{muttenthaler2021trends}
\begin{barticle}
\bauthor{\bsnm{Muttenthaler}, \binits{M.}},
\bauthor{\bsnm{King}, \binits{G.F.}},
\bauthor{\bsnm{Adams}, \binits{D.J.}},
\bauthor{\bsnm{Alewood}, \binits{P.F.}}:
\batitle{Trends in peptide drug discovery}.
\bjtitle{Nature reviews Drug discovery}
\bvolume{20},
\bfpage{309}--\blpage{325}
(\byear{2021})
\end{barticle}
\endbibitem

%%% 6
\bibitem[\protect\citeauthoryear{Kaspar and Reichert}{2013}]{kaspar2013future}
\begin{barticle}
\bauthor{\bsnm{Kaspar}, \binits{A.A.}},
\bauthor{\bsnm{Reichert}, \binits{J.M.}}:
\batitle{Future directions for peptide therapeutics development}.
\bjtitle{Drug discovery today}
\bvolume{18},
\bfpage{807}--\blpage{817}
(\byear{2013})
\end{barticle}
\endbibitem

%%% 7
\bibitem[\protect\citeauthoryear{Henninot et~al.}{2018}]{henninot2018current}
\begin{barticle}
\bauthor{\bsnm{Henninot}, \binits{A.}},
\bauthor{\bsnm{Collins}, \binits{J.C.}},
\bauthor{\bsnm{Nuss}, \binits{J.M.}}:
\batitle{The current state of peptide drug discovery: back to the future?}
\bjtitle{Journal of medicinal chemistry}
\bvolume{61},
\bfpage{1382}--\blpage{1414}
(\byear{2018})
\end{barticle}
\endbibitem

%%% 8
\bibitem[\protect\citeauthoryear{Lee et~al.}{2019}]{lee2019comprehensive}
\begin{barticle}
\bauthor{\bsnm{Lee}, \binits{A.C.-L.}},
\bauthor{\bsnm{Harris}, \binits{J.L.}},
\bauthor{\bsnm{Khanna}, \binits{K.K.}},
\bauthor{\bsnm{Hong}, \binits{J.-H.}}:
\batitle{A comprehensive review on current advances in peptide drug development and design}.
\bjtitle{International journal of molecular sciences}
\bvolume{20},
\bfpage{2383}
(\byear{2019})
\end{barticle}
\endbibitem

%%% 9
\bibitem[\protect\citeauthoryear{Wang et~al.}{2022}]{wang2022therapeutic}
\begin{barticle}
\bauthor{\bsnm{Wang}, \binits{L.}},
\bauthor{\bsnm{Wang}, \binits{N.}},
\bauthor{\bsnm{Zhang}, \binits{W.}},
\bauthor{\bsnm{Cheng}, \binits{X.}},
\bauthor{\bsnm{Yan}, \binits{Z.}},
\bauthor{\bsnm{Shao}, \binits{G.}},
\bauthor{\bsnm{Wang}, \binits{X.}},
\bauthor{\bsnm{Wang}, \binits{R.}},
\bauthor{\bsnm{Fu}, \binits{C.}}:
\batitle{Therapeutic peptides: current applications and future directions}.
\bjtitle{Signal Transduction and Targeted Therapy}
\bvolume{7},
\bfpage{48}
(\byear{2022})
\end{barticle}
\endbibitem

%%% 10
\bibitem[\protect\citeauthoryear{Chen et~al.}{2024}]{chen2024role}
\begin{barticle}
\bauthor{\bsnm{Chen}, \binits{Z.}},
\bauthor{\bsnm{Wang}, \binits{R.}},
\bauthor{\bsnm{Guo}, \binits{J.}},
\bauthor{\bsnm{Wang}, \binits{X.}}:
\batitle{The role and future prospects of artificial intelligence algorithms in peptide drug development}.
\bjtitle{Biomedicine \& Pharmacotherapy}
\bvolume{175},
\bfpage{116709}
(\byear{2024})
\end{barticle}
\endbibitem

%%% 11
\bibitem[\protect\citeauthoryear{Marso et~al.}{2016}]{marso2016semaglutide}
\begin{barticle}
\bauthor{\bsnm{Marso}, \binits{S.P.}},
\bauthor{\bsnm{Bain}, \binits{S.C.}},
\bauthor{\bsnm{Consoli}, \binits{A.}},
\bauthor{\bsnm{Eliaschewitz}, \binits{F.G.}},
\bauthor{\bsnm{J{\'o}dar}, \binits{E.}},
\bauthor{\bsnm{Leiter}, \binits{L.A.}},
\bauthor{\bsnm{Lingvay}, \binits{I.}},
\bauthor{\bsnm{Rosenstock}, \binits{J.}},
\bauthor{\bsnm{Seufert}, \binits{J.}},
\bauthor{\bsnm{Warren}, \binits{M.L.}}, \betal:
\batitle{Semaglutide and cardiovascular outcomes in patients with type 2 diabetes}.
\bjtitle{New England Journal of Medicine}
\bvolume{375},
\bfpage{1834}--\blpage{1844}
(\byear{2016})
\end{barticle}
\endbibitem

%%% 12
\bibitem[\protect\citeauthoryear{Husain et~al.}{2019}]{husain2019oral}
\begin{barticle}
\bauthor{\bsnm{Husain}, \binits{M.}},
\bauthor{\bsnm{Birkenfeld}, \binits{A.L.}},
\bauthor{\bsnm{Donsmark}, \binits{M.}},
\bauthor{\bsnm{Dungan}, \binits{K.}},
\bauthor{\bsnm{Eliaschewitz}, \binits{F.G.}},
\bauthor{\bsnm{Franco}, \binits{D.R.}},
\bauthor{\bsnm{Jeppesen}, \binits{O.K.}},
\bauthor{\bsnm{Lingvay}, \binits{I.}},
\bauthor{\bsnm{Mosenzon}, \binits{O.}},
\bauthor{\bsnm{Pedersen}, \binits{S.D.}}, \betal:
\batitle{Oral semaglutide and cardiovascular outcomes in patients with type 2 diabetes}.
\bjtitle{New England Journal of Medicine}
\bvolume{381},
\bfpage{841}--\blpage{851}
(\byear{2019})
\end{barticle}
\endbibitem

%%% 13
\bibitem[\protect\citeauthoryear{Wilding et~al.}{2021}]{wilding2021once}
\begin{barticle}
\bauthor{\bsnm{Wilding}, \binits{J.P.}},
\bauthor{\bsnm{Batterham}, \binits{R.L.}},
\bauthor{\bsnm{Calanna}, \binits{S.}},
\bauthor{\bsnm{Davies}, \binits{M.}},
\bauthor{\bsnm{Van~Gaal}, \binits{L.F.}},
\bauthor{\bsnm{Lingvay}, \binits{I.}},
\bauthor{\bsnm{McGowan}, \binits{B.M.}},
\bauthor{\bsnm{Rosenstock}, \binits{J.}},
\bauthor{\bsnm{Tran}, \binits{M.T.}},
\bauthor{\bsnm{Wadden}, \binits{T.A.}}, \betal:
\batitle{Once-weekly semaglutide in adults with overweight or obesity}.
\bjtitle{New England Journal of Medicine}
\bvolume{384},
\bfpage{989}--\blpage{1002}
(\byear{2021})
\end{barticle}
\endbibitem

%%% 14
\bibitem[\protect\citeauthoryear{Vanhee et~al.}{2011}]{vanhee2011computational}
\begin{barticle}
\bauthor{\bsnm{Vanhee}, \binits{P.}},
\bauthor{\bsnm{Sloot}, \binits{A.M.}},
\bauthor{\bsnm{Verschueren}, \binits{E.}},
\bauthor{\bsnm{Serrano}, \binits{L.}},
\bauthor{\bsnm{Rousseau}, \binits{F.}},
\bauthor{\bsnm{Schymkowitz}, \binits{J.}}:
\batitle{Computational design of peptide ligands}.
\bjtitle{Trends in biotechnology}
\bvolume{29},
\bfpage{231}--\blpage{239}
(\byear{2011})
\end{barticle}
\endbibitem

%%% 15
\bibitem[\protect\citeauthoryear{Chang et~al.}{2022}]{chang2022towards}
\begin{barticle}
\bauthor{\bsnm{Chang}, \binits{L.}},
\bauthor{\bsnm{Mondal}, \binits{A.}},
\bauthor{\bsnm{Perez}, \binits{A.}}:
\batitle{Towards rational computational peptide design}.
\bjtitle{Frontiers in Bioinformatics}
\bvolume{2},
\bfpage{1046493}
(\byear{2022})
\end{barticle}
\endbibitem

%%% 16
\bibitem[\protect\citeauthoryear{Terada and Inui}{2012}]{terada2012recent}
\begin{barticle}
\bauthor{\bsnm{Terada}, \binits{T.}},
\bauthor{\bsnm{Inui}, \binits{K.-i.}}:
\batitle{Recent advances in structural biology of peptide transporters}.
\bjtitle{Current topics in membranes}
\bvolume{70},
\bfpage{257}--\blpage{274}
(\byear{2012})
\end{barticle}
\endbibitem

%%% 17
\bibitem[\protect\citeauthoryear{Miller and Gulick}{2016}]{miller2016structural}
\begin{botherref}
\oauthor{\bsnm{Miller}, \binits{B.R.}},
\oauthor{\bsnm{Gulick}, \binits{A.M.}}:
Structural biology of nonribosomal peptide synthetases.
Nonribosomal Peptide and Polyketide Biosynthesis: Methods and Protocols,
3--29
(2016)
\end{botherref}
\endbibitem

%%% 18
\bibitem[\protect\citeauthoryear{Stawikowski and Fields}{2012}]{stawikowski2012introduction}
\begin{barticle}
\bauthor{\bsnm{Stawikowski}, \binits{M.}},
\bauthor{\bsnm{Fields}, \binits{G.B.}}:
\batitle{Introduction to peptide synthesis}.
\bjtitle{Current protocols in protein science}
\bvolume{69},
\bfpage{18}--\blpage{1}
(\byear{2012})
\end{barticle}
\endbibitem

%%% 19
\bibitem[\protect\citeauthoryear{Erak et~al.}{2018}]{erak2018peptide}
\begin{barticle}
\bauthor{\bsnm{Erak}, \binits{M.}},
\bauthor{\bsnm{Bellmann-Sickert}, \binits{K.}},
\bauthor{\bsnm{Els-Heindl}, \binits{S.}},
\bauthor{\bsnm{Beck-Sickinger}, \binits{A.G.}}:
\batitle{Peptide chemistry toolbox--transforming natural peptides into peptide therapeutics}.
\bjtitle{Bioorganic \& medicinal chemistry}
\bvolume{26},
\bfpage{2759}--\blpage{2765}
(\byear{2018})
\end{barticle}
\endbibitem

%%% 20
\bibitem[\protect\citeauthoryear{Ferrazzano et~al.}{2022}]{ferrazzano2022sustainability}
\begin{barticle}
\bauthor{\bsnm{Ferrazzano}, \binits{L.}},
\bauthor{\bsnm{Catani}, \binits{M.}},
\bauthor{\bsnm{Cavazzini}, \binits{A.}},
\bauthor{\bsnm{Martelli}, \binits{G.}},
\bauthor{\bsnm{Corbisiero}, \binits{D.}},
\bauthor{\bsnm{Cantelmi}, \binits{P.}},
\bauthor{\bsnm{Fantoni}, \binits{T.}},
\bauthor{\bsnm{Mattellone}, \binits{A.}},
\bauthor{\bsnm{De~Luca}, \binits{C.}},
\bauthor{\bsnm{Felletti}, \binits{S.}}, \betal:
\batitle{Sustainability in peptide chemistry: current synthesis and purification technologies and future challenges}.
\bjtitle{Green Chemistry}
\bvolume{24},
\bfpage{975}--\blpage{1020}
(\byear{2022})
\end{barticle}
\endbibitem

%%% 21
\bibitem[\protect\citeauthoryear{Wang et~al.}{2019}]{wang2019improved}
\begin{barticle}
\bauthor{\bsnm{Wang}, \binits{J.}},
\bauthor{\bsnm{Alekseenko}, \binits{A.}},
\bauthor{\bsnm{Kozakov}, \binits{D.}},
\bauthor{\bsnm{Miao}, \binits{Y.}}:
\batitle{Improved modeling of peptide-protein binding through global docking and accelerated molecular dynamics simulations}.
\bjtitle{Frontiers in molecular biosciences}
\bvolume{6},
\bfpage{112}
(\byear{2019})
\end{barticle}
\endbibitem

%%% 22
\bibitem[\protect\citeauthoryear{Geng et~al.}{2019}]{geng2019applications}
\begin{barticle}
\bauthor{\bsnm{Geng}, \binits{H.}},
\bauthor{\bsnm{Chen}, \binits{F.}},
\bauthor{\bsnm{Ye}, \binits{J.}},
\bauthor{\bsnm{Jiang}, \binits{F.}}:
\batitle{Applications of molecular dynamics simulation in structure prediction of peptides and proteins}.
\bjtitle{Computational and structural biotechnology journal}
\bvolume{17},
\bfpage{1162}--\blpage{1170}
(\byear{2019})
\end{barticle}
\endbibitem

%%% 23
\bibitem[\protect\citeauthoryear{Bond et~al.}{2007}]{bond2007coarse}
\begin{barticle}
\bauthor{\bsnm{Bond}, \binits{P.J.}},
\bauthor{\bsnm{Holyoake}, \binits{J.}},
\bauthor{\bsnm{Ivetac}, \binits{A.}},
\bauthor{\bsnm{Khalid}, \binits{S.}},
\bauthor{\bsnm{Sansom}, \binits{M.S.}}:
\batitle{Coarse-grained molecular dynamics simulations of membrane proteins and peptides}.
\bjtitle{Journal of structural biology}
\bvolume{157},
\bfpage{593}--\blpage{605}
(\byear{2007})
\end{barticle}
\endbibitem

%%% 24
\bibitem[\protect\citeauthoryear{Alam et~al.}{2017}]{alam2017high}
\begin{barticle}
\bauthor{\bsnm{Alam}, \binits{N.}},
\bauthor{\bsnm{Goldstein}, \binits{O.}},
\bauthor{\bsnm{Xia}, \binits{B.}},
\bauthor{\bsnm{Porter}, \binits{K.A.}},
\bauthor{\bsnm{Kozakov}, \binits{D.}},
\bauthor{\bsnm{Schueler-Furman}, \binits{O.}}:
\batitle{High-resolution global peptide-protein docking using fragments-based piper-flexpepdock}.
\bjtitle{PLoS computation al biology}
\bvolume{13},
\bfpage{1005905}
(\byear{2017})
\end{barticle}
\endbibitem

%%% 25
\bibitem[\protect\citeauthoryear{Zhang and Sanner}{2019}]{zhang2019autodock}
\begin{barticle}
\bauthor{\bsnm{Zhang}, \binits{Y.}},
\bauthor{\bsnm{Sanner}, \binits{M.F.}}:
\batitle{Autodock crankpep: combining folding and docking to predict protein--peptide complexes}.
\bjtitle{Bioinformatics}
\bvolume{35},
\bfpage{5121}--\blpage{5127}
(\byear{2019})
\end{barticle}
\endbibitem

%%% 26
\bibitem[\protect\citeauthoryear{Chen et~al.}{2024}]{chen2024design}
\begin{barticle}
\bauthor{\bsnm{Chen}, \binits{S.}},
\bauthor{\bsnm{Lin}, \binits{T.}},
\bauthor{\bsnm{Basu}, \binits{R.}},
\bauthor{\bsnm{Ritchey}, \binits{J.}},
\bauthor{\bsnm{Wang}, \binits{S.}},
\bauthor{\bsnm{Luo}, \binits{Y.}},
\bauthor{\bsnm{Li}, \binits{X.}},
\bauthor{\bsnm{Pei}, \binits{D.}},
\bauthor{\bsnm{Kara}, \binits{L.B.}},
\bauthor{\bsnm{Cheng}, \binits{X.}}:
\batitle{Design of target specific peptide inhibitors using generative deep learning and molecular dynamics simulations}.
\bjtitle{Nature Communications}
\bvolume{15},
\bfpage{1611}
(\byear{2024})
\end{barticle}
\endbibitem

%%% 27
\bibitem[\protect\citeauthoryear{Berman et~al.}{2000}]{berman2000protein}
\begin{barticle}
\bauthor{\bsnm{Berman}, \binits{H.M.}},
\bauthor{\bsnm{Westbrook}, \binits{J.}},
\bauthor{\bsnm{Feng}, \binits{Z.}},
\bauthor{\bsnm{Gilliland}, \binits{G.}},
\bauthor{\bsnm{Bhat}, \binits{T.N.}},
\bauthor{\bsnm{Weissig}, \binits{H.}},
\bauthor{\bsnm{Shindyalov}, \binits{I.N.}},
\bauthor{\bsnm{Bourne}, \binits{P.E.}}:
\batitle{The protein data bank}.
\bjtitle{Nucleic acids research}
\bvolume{28},
\bfpage{235}--\blpage{242}
(\byear{2000})
\end{barticle}
\endbibitem

%%% 28
\bibitem[\protect\citeauthoryear{Jumper et~al.}{2021}]{jumper2021highly}
\begin{barticle}
\bauthor{\bsnm{Jumper}, \binits{J.}},
\bauthor{\bsnm{Evans}, \binits{R.}},
\bauthor{\bsnm{Pritzel}, \binits{A.}},
\bauthor{\bsnm{Green}, \binits{T.}},
\bauthor{\bsnm{Figurnov}, \binits{M.}},
\bauthor{\bsnm{Ronneberger}, \binits{O.}},
\bauthor{\bsnm{Tunyasuvunakool}, \binits{K.}},
\bauthor{\bsnm{Bates}, \binits{R.}},
\bauthor{\bsnm{{\v{Z}}{\'\i}dek}, \binits{A.}},
\bauthor{\bsnm{Potapenko}, \binits{A.}}, \betal:
\batitle{Highly accurate protein structure prediction with alphafold}.
\bjtitle{Nature}
\bvolume{596},
\bfpage{583}--\blpage{589}
(\byear{2021})
\end{barticle}
\endbibitem

%%% 29
\bibitem[\protect\citeauthoryear{Baek et~al.}{2021}]{baek2021accurate}
\begin{barticle}
\bauthor{\bsnm{Baek}, \binits{M.}},
\bauthor{\bsnm{DiMaio}, \binits{F.}},
\bauthor{\bsnm{Anishchenko}, \binits{I.}},
\bauthor{\bsnm{Dauparas}, \binits{J.}},
\bauthor{\bsnm{Ovchinnikov}, \binits{S.}},
\bauthor{\bsnm{Lee}, \binits{G.R.}},
\bauthor{\bsnm{Wang}, \binits{J.}},
\bauthor{\bsnm{Cong}, \binits{Q.}},
\bauthor{\bsnm{Kinch}, \binits{L.N.}},
\bauthor{\bsnm{Schaeffer}, \binits{R.D.}}, \betal:
\batitle{Accurate prediction of protein structures and interactions using a three-track neural network}.
\bjtitle{Science}
\bvolume{373},
\bfpage{871}--\blpage{876}
(\byear{2021})
\end{barticle}
\endbibitem

%%% 30
\bibitem[\protect\citeauthoryear{Krishna et~al.}{2024}]{krishna2024generalized}
\begin{botherref}
\oauthor{\bsnm{Krishna}, \binits{R.}},
\oauthor{\bsnm{Wang}, \binits{J.}},
\oauthor{\bsnm{Ahern}, \binits{W.}},
\oauthor{\bsnm{Sturmfels}, \binits{P.}},
\oauthor{\bsnm{Venkatesh}, \binits{P.}},
\oauthor{\bsnm{Kalvet}, \binits{I.}},
\oauthor{\bsnm{Lee}, \binits{G.R.}},
\oauthor{\bsnm{Morey-Burrows}, \binits{F.S.}},
\oauthor{\bsnm{Anishchenko}, \binits{I.}},
\oauthor{\bsnm{Humphreys}, \binits{I.R.}}, et al.:
Generalized biomolecular modeling and design with rosettafold all-atom.
Science,
2528
(2024)
\end{botherref}
\endbibitem

%%% 31
\bibitem[\protect\citeauthoryear{Lin et~al.}{2023}]{lin2023evolutionary}
\begin{barticle}
\bauthor{\bsnm{Lin}, \binits{Z.}},
\bauthor{\bsnm{Akin}, \binits{H.}},
\bauthor{\bsnm{Rao}, \binits{R.}},
\bauthor{\bsnm{Hie}, \binits{B.}},
\bauthor{\bsnm{Zhu}, \binits{Z.}},
\bauthor{\bsnm{Lu}, \binits{W.}},
\bauthor{\bsnm{Smetanin}, \binits{N.}},
\bauthor{\bsnm{Verkuil}, \binits{R.}},
\bauthor{\bsnm{Kabeli}, \binits{O.}},
\bauthor{\bsnm{Shmueli}, \binits{Y.}}, \betal:
\batitle{Evolutionary-scale prediction of atomic-level protein structure with a language model}.
\bjtitle{Science}
\bvolume{379},
\bfpage{1123}--\blpage{1130}
(\byear{2023})
\end{barticle}
\endbibitem

%%% 32
\bibitem[\protect\citeauthoryear{Evans et~al.}{2021}]{evans2021protein}
\begin{botherref}
\oauthor{\bsnm{Evans}, \binits{R.}},
\oauthor{\bsnm{O’Neill}, \binits{M.}},
\oauthor{\bsnm{Pritzel}, \binits{A.}},
\oauthor{\bsnm{Antropova}, \binits{N.}},
\oauthor{\bsnm{Senior}, \binits{A.}},
\oauthor{\bsnm{Green}, \binits{T.}},
\oauthor{\bsnm{{\v{Z}}{\'\i}dek}, \binits{A.}},
\oauthor{\bsnm{Bates}, \binits{R.}},
\oauthor{\bsnm{Blackwell}, \binits{S.}},
\oauthor{\bsnm{Yim}, \binits{J.}}, et al.:
Protein complex prediction with alphafold-multimer.
biorxiv,
2021--10
(2021)
\end{botherref}
\endbibitem

%%% 33
\bibitem[\protect\citeauthoryear{Anishchenko et~al.}{2021}]{anishchenko2021novo}
\begin{barticle}
\bauthor{\bsnm{Anishchenko}, \binits{I.}},
\bauthor{\bsnm{Pellock}, \binits{S.J.}},
\bauthor{\bsnm{Chidyausiku}, \binits{T.M.}},
\bauthor{\bsnm{Ramelot}, \binits{T.A.}},
\bauthor{\bsnm{Ovchinnikov}, \binits{S.}},
\bauthor{\bsnm{Hao}, \binits{J.}},
\bauthor{\bsnm{Bafna}, \binits{K.}},
\bauthor{\bsnm{Norn}, \binits{C.}},
\bauthor{\bsnm{Kang}, \binits{A.}},
\bauthor{\bsnm{Bera}, \binits{A.K.}}, \betal:
\batitle{De novo protein design by deep network hallucination}.
\bjtitle{Nature}
\bvolume{600},
\bfpage{547}--\blpage{552}
(\byear{2021})
\end{barticle}
\endbibitem

%%% 34
\bibitem[\protect\citeauthoryear{Dauparas et~al.}{2022}]{dauparas2022robust}
\begin{barticle}
\bauthor{\bsnm{Dauparas}, \binits{J.}},
\bauthor{\bsnm{Anishchenko}, \binits{I.}},
\bauthor{\bsnm{Bennett}, \binits{N.}},
\bauthor{\bsnm{Bai}, \binits{H.}},
\bauthor{\bsnm{Ragotte}, \binits{R.J.}},
\bauthor{\bsnm{Milles}, \binits{L.F.}},
\bauthor{\bsnm{Wicky}, \binits{B.I.}},
\bauthor{\bsnm{Courbet}, \binits{A.}},
\bauthor{\bsnm{Haas}, \binits{R.J.}},
\bauthor{\bsnm{Bethel}, \binits{N.}}, \betal:
\batitle{Robust deep learning--based protein sequence design using proteinmpnn}.
\bjtitle{Science}
\bvolume{378},
\bfpage{49}--\blpage{56}
(\byear{2022})
\end{barticle}
\endbibitem

%%% 35
\bibitem[\protect\citeauthoryear{Watson et~al.}{2023}]{watson2023novo}
\begin{barticle}
\bauthor{\bsnm{Watson}, \binits{J.L.}},
\bauthor{\bsnm{Juergens}, \binits{D.}},
\bauthor{\bsnm{Bennett}, \binits{N.R.}},
\bauthor{\bsnm{Trippe}, \binits{B.L.}},
\bauthor{\bsnm{Yim}, \binits{J.}},
\bauthor{\bsnm{Eisenach}, \binits{H.E.}},
\bauthor{\bsnm{Ahern}, \binits{W.}},
\bauthor{\bsnm{Borst}, \binits{A.J.}},
\bauthor{\bsnm{Ragotte}, \binits{R.J.}},
\bauthor{\bsnm{Milles}, \binits{L.F.}}, \betal:
\batitle{De novo design of protein structure and function with rfdiffusion}.
\bjtitle{Nature}
\bvolume{620},
\bfpage{1089}--\blpage{1100}
(\byear{2023})
\end{barticle}
\endbibitem

%%% 36
\bibitem[\protect\citeauthoryear{Ingraham et~al.}{2023}]{ingraham2023illuminating}
\begin{barticle}
\bauthor{\bsnm{Ingraham}, \binits{J.B.}},
\bauthor{\bsnm{Baranov}, \binits{M.}},
\bauthor{\bsnm{Costello}, \binits{Z.}},
\bauthor{\bsnm{Barber}, \binits{K.W.}},
\bauthor{\bsnm{Wang}, \binits{W.}},
\bauthor{\bsnm{Ismail}, \binits{A.}},
\bauthor{\bsnm{Frappier}, \binits{V.}},
\bauthor{\bsnm{Lord}, \binits{D.M.}},
\bauthor{\bsnm{Ng-Thow-Hing}, \binits{C.}},
\bauthor{\bsnm{Van~Vlack}, \binits{E.R.}}, \betal:
\batitle{Illuminating protein space with a programmable generative model}.
\bjtitle{Nature}
\bvolume{623},
\bfpage{1070}--\blpage{1078}
(\byear{2023})
\end{barticle}
\endbibitem

%%% 37
\bibitem[\protect\citeauthoryear{Fasman}{2012}]{fasman2012prediction}
\begin{bbook}
\bauthor{\bsnm{Fasman}, \binits{G.D.}}:
\bbtitle{Prediction of Protein Structure and the Principles of Protein Conformation}.
\bpublisher{Springer}, \blocation{???}
(\byear{2012})
\end{bbook}
\endbibitem

%%% 38
\bibitem[\protect\citeauthoryear{Brooks and Case}{1993}]{brooks1993simulations}
\begin{barticle}
\bauthor{\bsnm{Brooks}, \binits{C.}},
\bauthor{\bsnm{Case}, \binits{D.A.}}:
\batitle{Simulations of peptide conformational dynamics and thermodynamics}.
\bjtitle{Chemical Reviews}
\bvolume{93},
\bfpage{2487}--\blpage{2502}
(\byear{1993})
\end{barticle}
\endbibitem

%%% 39
\bibitem[\protect\citeauthoryear{Abagyan and Argos}{1992}]{abagyan1992optimal}
\begin{barticle}
\bauthor{\bsnm{Abagyan}, \binits{R.}},
\bauthor{\bsnm{Argos}, \binits{P.}}:
\batitle{Optimal protocol and trajectory visualization for conformational searches of peptides and proteins}.
\bjtitle{Journal of molecular biology}
\bvolume{225},
\bfpage{519}--\blpage{532}
(\byear{1992})
\end{barticle}
\endbibitem

%%% 40
\bibitem[\protect\citeauthoryear{Long and Tycko}{1998}]{long1998biopolymer}
\begin{barticle}
\bauthor{\bsnm{Long}, \binits{H.W.}},
\bauthor{\bsnm{Tycko}, \binits{R.}}:
\batitle{Biopolymer conformational distributions from solid-state nmr: $\alpha$-helix and 310-helix contents of a helical peptide}.
\bjtitle{Journal of the American Chemical Society}
\bvolume{120},
\bfpage{7039}--\blpage{7048}
(\byear{1998})
\end{barticle}
\endbibitem

%%% 41
\bibitem[\protect\citeauthoryear{Lisanza et~al.}{2023}]{lisanza2023joint}
\begin{botherref}
\oauthor{\bsnm{Lisanza}, \binits{S.L.}},
\oauthor{\bsnm{Gershon}, \binits{J.M.}},
\oauthor{\bsnm{Tipps}, \binits{S.W.K.}},
\oauthor{\bsnm{Arnoldt}, \binits{L.}},
\oauthor{\bsnm{Hendel}, \binits{S.}},
\oauthor{\bsnm{Sims}, \binits{J.N.}},
\oauthor{\bsnm{Li}, \binits{X.}},
\oauthor{\bsnm{Baker}, \binits{D.}}:
Joint generation of protein sequence and structure with rosettafold sequence space diffusion.
bioRxiv,
2023--05
(2023)
\end{botherref}
\endbibitem

%%% 42
\bibitem[\protect\citeauthoryear{Sohl-Dickstein et~al.}{2015}]{sohl2015deep}
\begin{bchapter}
\bauthor{\bsnm{Sohl-Dickstein}, \binits{J.}},
\bauthor{\bsnm{Weiss}, \binits{E.}},
\bauthor{\bsnm{Maheswaranathan}, \binits{N.}},
\bauthor{\bsnm{Ganguli}, \binits{S.}}:
\bctitle{Deep unsupervised learning using nonequilibrium thermodynamics}.
In: \bbtitle{International Conference on Machine Learning},
pp. \bfpage{2256}--\blpage{2265}
(\byear{2015}).
\bcomment{PMLR}
\end{bchapter}
\endbibitem

%%% 43
\bibitem[\protect\citeauthoryear{Ho et~al.}{2020}]{ho2020denoising}
\begin{barticle}
\bauthor{\bsnm{Ho}, \binits{J.}},
\bauthor{\bsnm{Jain}, \binits{A.}},
\bauthor{\bsnm{Abbeel}, \binits{P.}}:
\batitle{Denoising diffusion probabilistic models}.
\bjtitle{Advances in neural information processing systems}
\bvolume{33},
\bfpage{6840}--\blpage{6851}
(\byear{2020})
\end{barticle}
\endbibitem

%%% 44
\bibitem[\protect\citeauthoryear{Song et~al.}{2020}]{song2020score}
\begin{botherref}
\oauthor{\bsnm{Song}, \binits{Y.}},
\oauthor{\bsnm{Sohl-Dickstein}, \binits{J.}},
\oauthor{\bsnm{Kingma}, \binits{D.P.}},
\oauthor{\bsnm{Kumar}, \binits{A.}},
\oauthor{\bsnm{Ermon}, \binits{S.}},
\oauthor{\bsnm{Poole}, \binits{B.}}:
Score-based generative modeling through stochastic differential equations.
arXiv preprint arXiv:2011.13456
(2020)
\end{botherref}
\endbibitem

%%% 45
\bibitem[\protect\citeauthoryear{Kosugi and Ohue}{2022}]{kosugi2022solubility}
\begin{barticle}
\bauthor{\bsnm{Kosugi}, \binits{T.}},
\bauthor{\bsnm{Ohue}, \binits{M.}}:
\batitle{Solubility-aware protein binding peptide design using alphafold}.
\bjtitle{Biomedicines}
\bvolume{10},
\bfpage{1626}
(\byear{2022})
\end{barticle}
\endbibitem

%%% 46
\bibitem[\protect\citeauthoryear{Yim et~al.}{2023}]{yim2023se}
\begin{botherref}
\oauthor{\bsnm{Yim}, \binits{J.}},
\oauthor{\bsnm{Trippe}, \binits{B.L.}},
\oauthor{\bsnm{De~Bortoli}, \binits{V.}},
\oauthor{\bsnm{Mathieu}, \binits{E.}},
\oauthor{\bsnm{Doucet}, \binits{A.}},
\oauthor{\bsnm{Barzilay}, \binits{R.}},
\oauthor{\bsnm{Jaakkola}, \binits{T.}}:
Se (3) diffusion model with application to protein backbone generation.
arXiv preprint arXiv:2302.02277
(2023)
\end{botherref}
\endbibitem

%%% 47
\bibitem[\protect\citeauthoryear{Wang et~al.}{2005}]{wang2005pdbbind}
\begin{barticle}
\bauthor{\bsnm{Wang}, \binits{R.}},
\bauthor{\bsnm{Fang}, \binits{X.}},
\bauthor{\bsnm{Lu}, \binits{Y.}},
\bauthor{\bsnm{Yang}, \binits{C.-Y.}},
\bauthor{\bsnm{Wang}, \binits{S.}}:
\batitle{The pdbbind database: methodologies and updates}.
\bjtitle{Journal of medicinal chemistry}
\bvolume{48},
\bfpage{4111}--\blpage{4119}
(\byear{2005})
\end{barticle}
\endbibitem

%%% 48
\bibitem[\protect\citeauthoryear{Martins et~al.}{2021}]{martins2021propedia}
\begin{barticle}
\bauthor{\bsnm{Martins}, \binits{P.M.}},
\bauthor{\bsnm{Santos}, \binits{L.H.}},
\bauthor{\bsnm{Mariano}, \binits{D.}},
\bauthor{\bsnm{Queiroz}, \binits{F.C.}},
\bauthor{\bsnm{Bastos}, \binits{L.L.}},
\bauthor{\bsnm{Gomes}, \binits{I.d.S.}},
\bauthor{\bsnm{Fischer}, \binits{P.H.}},
\bauthor{\bsnm{Rocha}, \binits{R.E.}},
\bauthor{\bsnm{Silveira}, \binits{S.A.}},
\bauthor{\bsnm{Lima}, \binits{L.H.}}, \betal:
\batitle{Propedia: a database for protein--peptide identification based on a hybrid clustering algorithm}.
\bjtitle{BMC bioinformatics}
\bvolume{22},
\bfpage{1}--\blpage{20}
(\byear{2021})
\end{barticle}
\endbibitem

%%% 49
\bibitem[\protect\citeauthoryear{Wen et~al.}{2019}]{wen2019pepbdb}
\begin{barticle}
\bauthor{\bsnm{Wen}, \binits{Z.}},
\bauthor{\bsnm{He}, \binits{J.}},
\bauthor{\bsnm{Tao}, \binits{H.}},
\bauthor{\bsnm{Huang}, \binits{S.-Y.}}:
\batitle{Pepbdb: a comprehensive structural database of biological peptide--protein interactions}.
\bjtitle{Bioinformatics}
\bvolume{35},
\bfpage{175}--\blpage{177}
(\byear{2019})
\end{barticle}
\endbibitem

%%% 50
\bibitem[\protect\citeauthoryear{Cock et~al.}{2009}]{cock2009biopython}
\begin{barticle}
\bauthor{\bsnm{Cock}, \binits{P.J.}},
\bauthor{\bsnm{Antao}, \binits{T.}},
\bauthor{\bsnm{Chang}, \binits{J.T.}},
\bauthor{\bsnm{Chapman}, \binits{B.A.}},
\bauthor{\bsnm{Cox}, \binits{C.J.}},
\bauthor{\bsnm{Dalke}, \binits{A.}},
\bauthor{\bsnm{Friedberg}, \binits{I.}},
\bauthor{\bsnm{Hamelryck}, \binits{T.}},
\bauthor{\bsnm{Kauff}, \binits{F.}},
\bauthor{\bsnm{Wilczynski}, \binits{B.}}, \betal:
\batitle{Biopython: freely available python tools for computational molecular biology and bioinformatics}.
\bjtitle{Bioinformatics}
\bvolume{25},
\bfpage{1422}
(\byear{2009})
\end{barticle}
\endbibitem

%%% 51
\bibitem[\protect\citeauthoryear{Zavala-Ruiz et~al.}{2004}]{zavala2004hairpin}
\begin{barticle}
\bauthor{\bsnm{Zavala-Ruiz}, \binits{Z.}},
\bauthor{\bsnm{Strug}, \binits{I.}},
\bauthor{\bsnm{Walker}, \binits{B.D.}},
\bauthor{\bsnm{Norris}, \binits{P.J.}},
\bauthor{\bsnm{Stern}, \binits{L.J.}}:
\batitle{A hairpin turn in a class ii mhc-bound peptide orients residues outside the binding groove for t cell recognition}.
\bjtitle{Proceedings of the National Academy of Sciences}
\bvolume{101},
\bfpage{13279}--\blpage{13284}
(\byear{2004})
\end{barticle}
\endbibitem

%%% 52
\bibitem[\protect\citeauthoryear{Martin et~al.}{2018}]{martin2018structure}
\begin{barticle}
\bauthor{\bsnm{Martin}, \binits{S.E.}},
\bauthor{\bsnm{Tan}, \binits{Z.-W.}},
\bauthor{\bsnm{Itkonen}, \binits{H.M.}},
\bauthor{\bsnm{Duveau}, \binits{D.Y.}},
\bauthor{\bsnm{Paulo}, \binits{J.A.}},
\bauthor{\bsnm{Janetzko}, \binits{J.}},
\bauthor{\bsnm{Boutz}, \binits{P.L.}},
\bauthor{\bsnm{T{\"o}rk}, \binits{L.}},
\bauthor{\bsnm{Moss}, \binits{F.A.}},
\bauthor{\bsnm{Thomas}, \binits{C.J.}}, \betal:
\batitle{Structure-based evolution of low nanomolar o-glcnac transferase inhibitors}.
\bjtitle{Journal of the American Chemical Society}
\bvolume{140},
\bfpage{13542}--\blpage{13545}
(\byear{2018})
\end{barticle}
\endbibitem

%%% 53
\bibitem[\protect\citeauthoryear{Pazgier et~al.}{2009}]{pazgier2009structural}
\begin{barticle}
\bauthor{\bsnm{Pazgier}, \binits{M.}},
\bauthor{\bsnm{Liu}, \binits{M.}},
\bauthor{\bsnm{Zou}, \binits{G.}},
\bauthor{\bsnm{Yuan}, \binits{W.}},
\bauthor{\bsnm{Li}, \binits{C.}},
\bauthor{\bsnm{Li}, \binits{C.}},
\bauthor{\bsnm{Li}, \binits{J.}},
\bauthor{\bsnm{Monbo}, \binits{J.}},
\bauthor{\bsnm{Zella}, \binits{D.}},
\bauthor{\bsnm{Tarasov}, \binits{S.G.}}, \betal:
\batitle{Structural basis for high-affinity peptide inhibition of p53 interactions with mdm2 and mdmx}.
\bjtitle{Proceedings of the National Academy of Sciences}
\bvolume{106},
\bfpage{4665}--\blpage{4670}
(\byear{2009})
\end{barticle}
\endbibitem

%%% 54
\bibitem[\protect\citeauthoryear{Guerlavais et~al.}{2023}]{guerlavais2023discovery}
\begin{barticle}
\bauthor{\bsnm{Guerlavais}, \binits{V.}},
\bauthor{\bsnm{Sawyer}, \binits{T.K.}},
\bauthor{\bsnm{Carvajal}, \binits{L.}},
\bauthor{\bsnm{Chang}, \binits{Y.S.}},
\bauthor{\bsnm{Graves}, \binits{B.}},
\bauthor{\bsnm{Ren}, \binits{J.-G.}},
\bauthor{\bsnm{Sutton}, \binits{D.}},
\bauthor{\bsnm{Olson}, \binits{K.A.}},
\bauthor{\bsnm{Packman}, \binits{K.}},
\bauthor{\bsnm{Darlak}, \binits{K.}}, \betal:
\batitle{Discovery of sulanemadlin (alrn-6924), the first cell-permeating, stabilized $\alpha$-helical peptide in clinical development}.
\bjtitle{Journal of Medicinal Chemistry}
\bvolume{66},
\bfpage{9401}--\blpage{9417}
(\byear{2023})
\end{barticle}
\endbibitem

%%% 55
\bibitem[\protect\citeauthoryear{B{\"o}ttger et~al.}{1997}]{bottger1997molecular}
\begin{barticle}
\bauthor{\bsnm{B{\"o}ttger}, \binits{A.}},
\bauthor{\bsnm{B{\"o}ttger}, \binits{V.}},
\bauthor{\bsnm{Garcia-Echeverria}, \binits{C.}},
\bauthor{\bsnm{Ch{\`e}ne}, \binits{P.}},
\bauthor{\bsnm{Hochkeppel}, \binits{H.-K.}},
\bauthor{\bsnm{Sampson}, \binits{W.}},
\bauthor{\bsnm{Ang}, \binits{K.}},
\bauthor{\bsnm{Howard}, \binits{S.F.}},
\bauthor{\bsnm{Picksley}, \binits{S.M.}},
\bauthor{\bsnm{Lane}, \binits{D.P.}}:
\batitle{Molecular characterization of the hdm2-p53 interaction}.
\bjtitle{Journal of molecular biology}
\bvolume{269},
\bfpage{744}--\blpage{756}
(\byear{1997})
\end{barticle}
\endbibitem

%%% 56
\bibitem[\protect\citeauthoryear{Klein and Vassilev}{2004}]{klein2004targeting}
\begin{barticle}
\bauthor{\bsnm{Klein}, \binits{C.}},
\bauthor{\bsnm{Vassilev}, \binits{L.}}:
\batitle{Targeting the p53--mdm2 interaction to treat cancer}.
\bjtitle{British journal of cancer}
\bvolume{91},
\bfpage{1415}--\blpage{1419}
(\byear{2004})
\end{barticle}
\endbibitem

%%% 57
\bibitem[\protect\citeauthoryear{Owen et~al.}{2021}]{owen2021oral}
\begin{barticle}
\bauthor{\bsnm{Owen}, \binits{D.R.}},
\bauthor{\bsnm{Allerton}, \binits{C.M.}},
\bauthor{\bsnm{Anderson}, \binits{A.S.}},
\bauthor{\bsnm{Aschenbrenner}, \binits{L.}},
\bauthor{\bsnm{Avery}, \binits{M.}},
\bauthor{\bsnm{Berritt}, \binits{S.}},
\bauthor{\bsnm{Boras}, \binits{B.}},
\bauthor{\bsnm{Cardin}, \binits{R.D.}},
\bauthor{\bsnm{Carlo}, \binits{A.}},
\bauthor{\bsnm{Coffman}, \binits{K.J.}}, \betal:
\batitle{An oral sars-cov-2 mpro inhibitor clinical candidate for the treatment of covid-19}.
\bjtitle{Science}
\bvolume{374},
\bfpage{1586}--\blpage{1593}
(\byear{2021})
\end{barticle}
\endbibitem

%%% 58
\bibitem[\protect\citeauthoryear{Mitchell et~al.}{2010}]{mitchell2010alk1}
\begin{barticle}
\bauthor{\bsnm{Mitchell}, \binits{D.}},
\bauthor{\bsnm{Pobre}, \binits{E.G.}},
\bauthor{\bsnm{Mulivor}, \binits{A.W.}},
\bauthor{\bsnm{Grinberg}, \binits{A.V.}},
\bauthor{\bsnm{Castonguay}, \binits{R.}},
\bauthor{\bsnm{Monnell}, \binits{T.E.}},
\bauthor{\bsnm{Solban}, \binits{N.}},
\bauthor{\bsnm{Ucran}, \binits{J.A.}},
\bauthor{\bsnm{Pearsall}, \binits{R.S.}},
\bauthor{\bsnm{Underwood}, \binits{K.W.}}, \betal:
\batitle{Alk1-fc inhibits multiple mediators of angiogenesis and suppresses tumor growth}.
\bjtitle{Molecular cancer therapeutics}
\bvolume{9},
\bfpage{379}--\blpage{388}
(\byear{2010})
\end{barticle}
\endbibitem

%%% 59
\bibitem[\protect\citeauthoryear{Bendell et~al.}{2014}]{bendell2014safety}
\begin{barticle}
\bauthor{\bsnm{Bendell}, \binits{J.C.}},
\bauthor{\bsnm{Gordon}, \binits{M.S.}},
\bauthor{\bsnm{Hurwitz}, \binits{H.I.}},
\bauthor{\bsnm{Jones}, \binits{S.F.}},
\bauthor{\bsnm{Mendelson}, \binits{D.S.}},
\bauthor{\bsnm{Blobe}, \binits{G.C.}},
\bauthor{\bsnm{Agarwal}, \binits{N.}},
\bauthor{\bsnm{Condon}, \binits{C.H.}},
\bauthor{\bsnm{Wilson}, \binits{D.}},
\bauthor{\bsnm{Pearsall}, \binits{A.E.}}, \betal:
\batitle{Safety, pharmacokinetics, pharmacodynamics, and antitumor activity of dalantercept, an activin receptor--like kinase-1 ligand trap, in patients with advanced cancer}.
\bjtitle{Clinical Cancer Research}
\bvolume{20},
\bfpage{480}--\blpage{489}
(\byear{2014})
\end{barticle}
\endbibitem

%%% 60
\bibitem[\protect\citeauthoryear{Jang et~al.}{2021}]{jang2021role}
\begin{barticle}
\bauthor{\bsnm{Jang}, \binits{D.-i.}},
\bauthor{\bsnm{Lee}, \binits{A.-H.}},
\bauthor{\bsnm{Shin}, \binits{H.-Y.}},
\bauthor{\bsnm{Song}, \binits{H.-R.}},
\bauthor{\bsnm{Park}, \binits{J.-H.}},
\bauthor{\bsnm{Kang}, \binits{T.-B.}},
\bauthor{\bsnm{Lee}, \binits{S.-R.}},
\bauthor{\bsnm{Yang}, \binits{S.-H.}}:
\batitle{The role of tumor necrosis factor alpha (tnf-$\alpha$) in autoimmune disease and current tnf-$\alpha$ inhibitors in therapeutics}.
\bjtitle{International journal of molecular sciences}
\bvolume{22},
\bfpage{2719}
(\byear{2021})
\end{barticle}
\endbibitem

%%% 61
\bibitem[\protect\citeauthoryear{Monaco et~al.}{2015}]{monaco2015anti}
\begin{barticle}
\bauthor{\bsnm{Monaco}, \binits{C.}},
\bauthor{\bsnm{Nanchahal}, \binits{J.}},
\bauthor{\bsnm{Taylor}, \binits{P.}},
\bauthor{\bsnm{Feldmann}, \binits{M.}}:
\batitle{Anti-tnf therapy: past, present and future}.
\bjtitle{International immunology}
\bvolume{27},
\bfpage{55}--\blpage{62}
(\byear{2015})
\end{barticle}
\endbibitem

%%% 62
\bibitem[\protect\citeauthoryear{Rau}{2002}]{rau2002adalimumab}
\begin{barticle}
\bauthor{\bsnm{Rau}, \binits{R.}}:
\batitle{Adalimumab (a fully human anti-tumour necrosis factor $\alpha$ monoclonal antibody) in the treatment of active rheumatoid arthritis: the initial results of five trials}.
\bjtitle{Annals of the rheumatic diseases}
\bvolume{61},
\bfpage{70}--\blpage{73}
(\byear{2002})
\end{barticle}
\endbibitem

%%% 63
\bibitem[\protect\citeauthoryear{Xu et~al.}{2018}]{xu2018cavityplus}
\begin{barticle}
\bauthor{\bsnm{Xu}, \binits{Y.}},
\bauthor{\bsnm{Wang}, \binits{S.}},
\bauthor{\bsnm{Hu}, \binits{Q.}},
\bauthor{\bsnm{Gao}, \binits{S.}},
\bauthor{\bsnm{Ma}, \binits{X.}},
\bauthor{\bsnm{Zhang}, \binits{W.}},
\bauthor{\bsnm{Shen}, \binits{Y.}},
\bauthor{\bsnm{Chen}, \binits{F.}},
\bauthor{\bsnm{Lai}, \binits{L.}},
\bauthor{\bsnm{Pei}, \binits{J.}}:
\batitle{Cavityplus: a web server for protein cavity detection with pharmacophore modelling, allosteric site identification and covalent ligand binding ability prediction}.
\bjtitle{Nucleic acids research}
\bvolume{46},
\bfpage{374}--\blpage{379}
(\byear{2018})
\end{barticle}
\endbibitem

%%% 64
\bibitem[\protect\citeauthoryear{Wang et~al.}{2023}]{wang2023cavityplus}
\begin{barticle}
\bauthor{\bsnm{Wang}, \binits{S.}},
\bauthor{\bsnm{Xie}, \binits{J.}},
\bauthor{\bsnm{Pei}, \binits{J.}},
\bauthor{\bsnm{Lai}, \binits{L.}}:
\batitle{Cavityplus 2022 update: an integrated platform for comprehensive protein cavity detection and property analyses with user-friendly tools and cavity databases}.
\bjtitle{Journal of Molecular Biology}
\bvolume{435},
\bfpage{168141}
(\byear{2023})
\end{barticle}
\endbibitem

%%% 65
\bibitem[\protect\citeauthoryear{Miura et~al.}{2023}]{miura2023vitro}
\begin{barticle}
\bauthor{\bsnm{Miura}, \binits{T.}},
\bauthor{\bsnm{Malla}, \binits{T.R.}},
\bauthor{\bsnm{Owen}, \binits{C.D.}},
\bauthor{\bsnm{Tumber}, \binits{A.}},
\bauthor{\bsnm{Brewitz}, \binits{L.}},
\bauthor{\bsnm{McDonough}, \binits{M.A.}},
\bauthor{\bsnm{Salah}, \binits{E.}},
\bauthor{\bsnm{Terasaka}, \binits{N.}},
\bauthor{\bsnm{Katoh}, \binits{T.}},
\bauthor{\bsnm{Lukacik}, \binits{P.}}, \betal:
\batitle{In vitro selection of macrocyclic peptide inhibitors containing cyclic $\gamma$2, 4-amino acids targeting the sars-cov-2 main protease}.
\bjtitle{Nature Chemistry}
\bvolume{15},
\bfpage{998}--\blpage{1005}
(\byear{2023})
\end{barticle}
\endbibitem

%%% 66
\bibitem[\protect\citeauthoryear{Salmon et~al.}{2020}]{salmon2020molecular}
\begin{barticle}
\bauthor{\bsnm{Salmon}, \binits{R.M.}},
\bauthor{\bsnm{Guo}, \binits{J.}},
\bauthor{\bsnm{Wood}, \binits{J.H.}},
\bauthor{\bsnm{Tong}, \binits{Z.}},
\bauthor{\bsnm{Beech}, \binits{J.S.}},
\bauthor{\bsnm{Lawera}, \binits{A.}},
\bauthor{\bsnm{Yu}, \binits{M.}},
\bauthor{\bsnm{Grainger}, \binits{D.J.}},
\bauthor{\bsnm{Reckless}, \binits{J.}},
\bauthor{\bsnm{Morrell}, \binits{N.W.}}, \betal:
\batitle{Molecular basis of alk1-mediated signalling by bmp9/bmp10 and their prodomain-bound forms}.
\bjtitle{Nature communications}
\bvolume{11},
\bfpage{1621}
(\byear{2020})
\end{barticle}
\endbibitem

%%% 67
\bibitem[\protect\citeauthoryear{Das and Baker}{2008}]{das2008macromolecular}
\begin{barticle}
\bauthor{\bsnm{Das}, \binits{R.}},
\bauthor{\bsnm{Baker}, \binits{D.}}:
\batitle{Macromolecular modeling with rosetta}.
\bjtitle{Annu. Rev. Biochem.}
\bvolume{77},
\bfpage{363}--\blpage{382}
(\byear{2008})
\end{barticle}
\endbibitem

%%% 68
\bibitem[\protect\citeauthoryear{McMillan et~al.}{2021}]{mcmillan2021structural}
\begin{barticle}
\bauthor{\bsnm{McMillan}, \binits{D.}},
\bauthor{\bsnm{Martinez-Fleites}, \binits{C.}},
\bauthor{\bsnm{Porter}, \binits{J.}},
\bauthor{\bsnm{Fox~3rd}, \binits{D.}},
\bauthor{\bsnm{Davis}, \binits{R.}},
\bauthor{\bsnm{Mori}, \binits{P.}},
\bauthor{\bsnm{Ceska}, \binits{T.}},
\bauthor{\bsnm{Carrington}, \binits{B.}},
\bauthor{\bsnm{Lawson}, \binits{A.}},
\bauthor{\bsnm{Bourne}, \binits{T.}}, \betal:
\batitle{Structural insights into the disruption of tnf-tnfr1 signalling by small molecules stabilising a distorted tnf}.
\bjtitle{Nature communications}
\bvolume{12},
\bfpage{582}
(\byear{2021})
\end{barticle}
\endbibitem

%%% 69
\bibitem[\protect\citeauthoryear{Al~Musaimi et~al.}{2022}]{al2022strategies}
\begin{barticle}
\bauthor{\bsnm{Al~Musaimi}, \binits{O.}},
\bauthor{\bsnm{Lombardi}, \binits{L.}},
\bauthor{\bsnm{Williams}, \binits{D.R.}},
\bauthor{\bsnm{Albericio}, \binits{F.}}:
\batitle{Strategies for improving peptide stability and delivery}.
\bjtitle{Pharmaceuticals}
\bvolume{15},
\bfpage{1283}
(\byear{2022})
\end{barticle}
\endbibitem

%%% 70
\bibitem[\protect\citeauthoryear{Vald{\'e}s-Tresanco et~al.}{2021}]{valdes2021gmx_mmpbsa}
\begin{barticle}
\bauthor{\bsnm{Vald{\'e}s-Tresanco}, \binits{M.S.}},
\bauthor{\bsnm{Vald{\'e}s-Tresanco}, \binits{M.E.}},
\bauthor{\bsnm{Valiente}, \binits{P.A.}},
\bauthor{\bsnm{Moreno}, \binits{E.}}:
\batitle{gmx\_mmpbsa: a new tool to perform end-state free energy calculations with gromacs}.
\bjtitle{Journal of chemical theory and computation}
\bvolume{17},
\bfpage{6281}--\blpage{6291}
(\byear{2021})
\end{barticle}
\endbibitem

%%% 71
\bibitem[\protect\citeauthoryear{Miller~III et~al.}{2012}]{miller2012mmpbsa}
\begin{barticle}
\bauthor{\bsnm{Miller~III}, \binits{B.R.}},
\bauthor{\bsnm{McGee~Jr}, \binits{T.D.}},
\bauthor{\bsnm{Swails}, \binits{J.M.}},
\bauthor{\bsnm{Homeyer}, \binits{N.}},
\bauthor{\bsnm{Gohlke}, \binits{H.}},
\bauthor{\bsnm{Roitberg}, \binits{A.E.}}:
\batitle{Mmpbsa. py: an efficient program for end-state free energy calculations}.
\bjtitle{Journal of chemical theory and computation}
\bvolume{8},
\bfpage{3314}--\blpage{3321}
(\byear{2012})
\end{barticle}
\endbibitem

%%% 72
\bibitem[\protect\citeauthoryear{Zhang et~al.}{2024}]{zhang2024revolutionizing}
\begin{barticle}
\bauthor{\bsnm{Zhang}, \binits{J.}},
\bauthor{\bsnm{Durham}, \binits{J.}},
\bauthor{\bsnm{Cong}, \binits{Q.}}:
\batitle{Revolutionizing protein--protein interaction prediction with deep learning}.
\bjtitle{Current Opinion in Structural Biology}
\bvolume{85},
\bfpage{102775}
(\byear{2024})
\end{barticle}
\endbibitem

%%% 73
\bibitem[\protect\citeauthoryear{Notin et~al.}{2024}]{notin2024machine}
\begin{barticle}
\bauthor{\bsnm{Notin}, \binits{P.}},
\bauthor{\bsnm{Rollins}, \binits{N.}},
\bauthor{\bsnm{Gal}, \binits{Y.}},
\bauthor{\bsnm{Sander}, \binits{C.}},
\bauthor{\bsnm{Marks}, \binits{D.}}:
\batitle{Machine learning for functional protein design}.
\bjtitle{Nature Biotechnology}
\bvolume{42},
\bfpage{216}--\blpage{228}
(\byear{2024})
\end{barticle}
\endbibitem

%%% 74
\bibitem[\protect\citeauthoryear{Gupta et~al.}{2022}]{gupta2022design}
\begin{botherref}
\oauthor{\bsnm{Gupta}, \binits{S.}},
\oauthor{\bsnm{Azadvari}, \binits{N.}},
\oauthor{\bsnm{Hosseinzadeh}, \binits{P.}}:
Design of protein segments and peptides for binding to protein targets.
BioDesign Research
\textbf{2022}
(2022)
\end{botherref}
\endbibitem

%%% 75
\bibitem[\protect\citeauthoryear{Wang et~al.}{2024}]{wang2024recent}
\begin{botherref}
\oauthor{\bsnm{Wang}, \binits{Y.-C.}},
\oauthor{\bsnm{Bai}, \binits{S.-C.}},
\oauthor{\bsnm{Ye}, \binits{W.-L.}},
\oauthor{\bsnm{Jiang}, \binits{J.}},
\oauthor{\bsnm{Li}, \binits{G.}}:
Recent progress in site-selective modification of peptides and proteins using macrocycles.
Bioconjugate Chemistry
(2024)
\end{botherref}
\endbibitem

%%% 76
\bibitem[\protect\citeauthoryear{Scantlebury et~al.}{2020}]{scantlebury2020data}
\begin{barticle}
\bauthor{\bsnm{Scantlebury}, \binits{J.}},
\bauthor{\bsnm{Brown}, \binits{N.}},
\bauthor{\bsnm{Von~Delft}, \binits{F.}},
\bauthor{\bsnm{Deane}, \binits{C.M.}}:
\batitle{Data set augmentation allows deep learning-based virtual screening to better generalize to unseen target classes and highlight important binding interactions}.
\bjtitle{Journal of chemical information and modeling}
\bvolume{60},
\bfpage{3722}--\blpage{3730}
(\byear{2020})
\end{barticle}
\endbibitem

%%% 77
\bibitem[\protect\citeauthoryear{Gao et~al.}{2023}]{gao2023self}
\begin{botherref}
\oauthor{\bsnm{Gao}, \binits{B.}},
\oauthor{\bsnm{Jia}, \binits{Y.}},
\oauthor{\bsnm{Mo}, \binits{Y.}},
\oauthor{\bsnm{Ni}, \binits{Y.}},
\oauthor{\bsnm{Ma}, \binits{W.}},
\oauthor{\bsnm{Ma}, \binits{Z.}},
\oauthor{\bsnm{Lan}, \binits{Y.}}:
Self-supervised pocket pretraining via protein fragment-surroundings alignment.
arXiv preprint arXiv:2310.07229
(2023)
\end{botherref}
\endbibitem

%%% 78
\bibitem[\protect\citeauthoryear{Huang et~al.}{2023}]{huang2023learning}
\begin{botherref}
\oauthor{\bsnm{Huang}, \binits{H.}},
\oauthor{\bsnm{Sun}, \binits{L.}},
\oauthor{\bsnm{Du}, \binits{B.}},
\oauthor{\bsnm{Lv}, \binits{W.}}:
Learning joint 2d \& 3d diffusion models for complete molecule generation.
arXiv preprint arXiv:2305.12347
(2023)
\end{botherref}
\endbibitem

%%% 79
\bibitem[\protect\citeauthoryear{Le et~al.}{2023}]{le2023navigating}
\begin{botherref}
\oauthor{\bsnm{Le}, \binits{T.}},
\oauthor{\bsnm{Cremer}, \binits{J.}},
\oauthor{\bsnm{No{\'e}}, \binits{F.}},
\oauthor{\bsnm{Clevert}, \binits{D.-A.}},
\oauthor{\bsnm{Sch{\"u}tt}, \binits{K.}}:
Navigating the design space of equivariant diffusion-based generative models for de novo 3d molecule generation.
arXiv preprint arXiv:2309.17296
(2023)
\end{botherref}
\endbibitem

%%% 80
\bibitem[\protect\citeauthoryear{Anand and Achim}{2022}]{anand2022protein}
\begin{botherref}
\oauthor{\bsnm{Anand}, \binits{N.}},
\oauthor{\bsnm{Achim}, \binits{T.}}:
Protein structure and sequence generation with equivariant denoising diffusion probabilistic models.
arXiv preprint arXiv:2205.15019
(2022)
\end{botherref}
\endbibitem

%%% 81
\bibitem[\protect\citeauthoryear{Campbell et~al.}{2024}]{campbell2024generative}
\begin{botherref}
\oauthor{\bsnm{Campbell}, \binits{A.}},
\oauthor{\bsnm{Yim}, \binits{J.}},
\oauthor{\bsnm{Barzilay}, \binits{R.}},
\oauthor{\bsnm{Rainforth}, \binits{T.}},
\oauthor{\bsnm{Jaakkola}, \binits{T.}}:
Generative flows on discrete state-spaces: Enabling multimodal flows with applications to protein co-design.
arXiv preprint arXiv:2402.04997
(2024)
\end{botherref}
\endbibitem

%%% 82
\bibitem[\protect\citeauthoryear{Lipman et~al.}{2022}]{lipman2022flow}
\begin{botherref}
\oauthor{\bsnm{Lipman}, \binits{Y.}},
\oauthor{\bsnm{Chen}, \binits{R.T.}},
\oauthor{\bsnm{Ben-Hamu}, \binits{H.}},
\oauthor{\bsnm{Nickel}, \binits{M.}},
\oauthor{\bsnm{Le}, \binits{M.}}:
Flow matching for generative modeling.
arXiv preprint arXiv:2210.02747
(2022)
\end{botherref}
\endbibitem

%%% 83
\bibitem[\protect\citeauthoryear{Chen and Lipman}{2023}]{chen2023riemannian}
\begin{botherref}
\oauthor{\bsnm{Chen}, \binits{R.T.}},
\oauthor{\bsnm{Lipman}, \binits{Y.}}:
Riemannian flow matching on general geometries.
arXiv preprint arXiv:2302.03660
(2023)
\end{botherref}
\endbibitem

%%% 84
\bibitem[\protect\citeauthoryear{Mannhold et~al.}{2006}]{mannhold2006pharmacophores}
\begin{bbook}
\bauthor{\bsnm{Mannhold}, \binits{R.}},
\bauthor{\bsnm{Kubinyi}, \binits{H.}},
\bauthor{\bsnm{Folkers}, \binits{G.}}:
\bbtitle{Pharmacophores and Pharmacophore Searches}.
\bpublisher{Wiley Online Library}, \blocation{???}
(\byear{2006})
\end{bbook}
\endbibitem

%%% 85
\bibitem[\protect\citeauthoryear{Weikl and Paul}{2014}]{weikl2014conformational}
\begin{barticle}
\bauthor{\bsnm{Weikl}, \binits{T.R.}},
\bauthor{\bsnm{Paul}, \binits{F.}}:
\batitle{Conformational selection in protein binding and function}.
\bjtitle{Protein Science}
\bvolume{23},
\bfpage{1508}--\blpage{1518}
(\byear{2014})
\end{barticle}
\endbibitem

%%% 86
\bibitem[\protect\citeauthoryear{Zarutskie et~al.}{1999}]{zarutskie1999conformational}
\begin{barticle}
\bauthor{\bsnm{Zarutskie}, \binits{J.A.}},
\bauthor{\bsnm{Sato}, \binits{A.K.}},
\bauthor{\bsnm{Rushe}, \binits{M.M.}},
\bauthor{\bsnm{Chan}, \binits{I.C.}},
\bauthor{\bsnm{Lomakin}, \binits{A.}},
\bauthor{\bsnm{Benedek}, \binits{G.B.}},
\bauthor{\bsnm{Stern}, \binits{L.J.}}:
\batitle{A conformational change in the human major histocompatibility complex protein hla-dr1 induced by peptide binding}.
\bjtitle{Biochemistry}
\bvolume{38},
\bfpage{5878}--\blpage{5887}
(\byear{1999})
\end{barticle}
\endbibitem

%%% 87
\bibitem[\protect\citeauthoryear{Springer et~al.}{1998}]{springer1998fast}
\begin{barticle}
\bauthor{\bsnm{Springer}, \binits{S.}},
\bauthor{\bsnm{D{\"o}ring}, \binits{K.}},
\bauthor{\bsnm{Skipper}, \binits{J.C.}},
\bauthor{\bsnm{Townsend}, \binits{A.R.}},
\bauthor{\bsnm{Cerundolo}, \binits{V.}}:
\batitle{Fast association rates suggest a conformational change in the mhc class i molecule h-2db upon peptide binding}.
\bjtitle{Biochemistry}
\bvolume{37},
\bfpage{3001}--\blpage{3012}
(\byear{1998})
\end{barticle}
\endbibitem

%%% 88
\bibitem[\protect\citeauthoryear{Armstrong et~al.}{2008}]{armstrong2008conformational}
\begin{barticle}
\bauthor{\bsnm{Armstrong}, \binits{K.M.}},
\bauthor{\bsnm{Piepenbrink}, \binits{K.H.}},
\bauthor{\bsnm{Baker}, \binits{B.M.}}:
\batitle{Conformational changes and flexibility in t-cell receptor recognition of peptide--mhc complexes}.
\bjtitle{Biochemical Journal}
\bvolume{415},
\bfpage{183}--\blpage{196}
(\byear{2008})
\end{barticle}
\endbibitem

%%% 89
\bibitem[\protect\citeauthoryear{Malde et~al.}{2019}]{malde2019crystal}
\begin{barticle}
\bauthor{\bsnm{Malde}, \binits{A.K.}},
\bauthor{\bsnm{Hill}, \binits{T.A.}},
\bauthor{\bsnm{Iyer}, \binits{A.}},
\bauthor{\bsnm{Fairlie}, \binits{D.P.}}:
\batitle{Crystal structures of protein-bound cyclic peptides}.
\bjtitle{Chemical reviews}
\bvolume{119},
\bfpage{9861}--\blpage{9914}
(\byear{2019})
\end{barticle}
\endbibitem

%%% 90
\bibitem[\protect\citeauthoryear{Nielsen et~al.}{2017}]{nielsen2017orally}
\begin{barticle}
\bauthor{\bsnm{Nielsen}, \binits{D.S.}},
\bauthor{\bsnm{Shepherd}, \binits{N.E.}},
\bauthor{\bsnm{Xu}, \binits{W.}},
\bauthor{\bsnm{Lucke}, \binits{A.J.}},
\bauthor{\bsnm{Stoermer}, \binits{M.J.}},
\bauthor{\bsnm{Fairlie}, \binits{D.P.}}:
\batitle{Orally absorbed cyclic peptides}.
\bjtitle{Chemical reviews}
\bvolume{117},
\bfpage{8094}--\blpage{8128}
(\byear{2017})
\end{barticle}
\endbibitem

%%% 91
\bibitem[\protect\citeauthoryear{Gavenonis et~al.}{2014}]{gavenonis2014comprehensive}
\begin{barticle}
\bauthor{\bsnm{Gavenonis}, \binits{J.}},
\bauthor{\bsnm{Sheneman}, \binits{B.A.}},
\bauthor{\bsnm{Siegert}, \binits{T.R.}},
\bauthor{\bsnm{Eshelman}, \binits{M.R.}},
\bauthor{\bsnm{Kritzer}, \binits{J.A.}}:
\batitle{Comprehensive analysis of loops at protein-protein interfaces for macrocycle design}.
\bjtitle{Nature chemical biology}
\bvolume{10},
\bfpage{716}--\blpage{722}
(\byear{2014})
\end{barticle}
\endbibitem

%%% 92
\bibitem[\protect\citeauthoryear{Demmer et~al.}{2009}]{demmer2009design}
\begin{botherref}
\oauthor{\bsnm{Demmer}, \binits{O.}},
\oauthor{\bsnm{Frank}, \binits{A.O.}},
\oauthor{\bsnm{Kessler}, \binits{H.}}:
Design of cyclic peptides.
Peptide and protein design for biopharmaceutical applications,
133--176
(2009)
\end{botherref}
\endbibitem

%%% 93
\bibitem[\protect\citeauthoryear{Eastman et~al.}{2017}]{eastman2017openmm}
\begin{barticle}
\bauthor{\bsnm{Eastman}, \binits{P.}},
\bauthor{\bsnm{Swails}, \binits{J.}},
\bauthor{\bsnm{Chodera}, \binits{J.D.}},
\bauthor{\bsnm{McGibbon}, \binits{R.T.}},
\bauthor{\bsnm{Zhao}, \binits{Y.}},
\bauthor{\bsnm{Beauchamp}, \binits{K.A.}},
\bauthor{\bsnm{Wang}, \binits{L.-P.}},
\bauthor{\bsnm{Simmonett}, \binits{A.C.}},
\bauthor{\bsnm{Harrigan}, \binits{M.P.}},
\bauthor{\bsnm{Stern}, \binits{C.D.}}, \betal:
\batitle{Openmm 7: Rapid development of high performance algorithms for molecular dynamics}.
\bjtitle{PLoS computational biology}
\bvolume{13},
\bfpage{1005659}
(\byear{2017})
\end{barticle}
\endbibitem

%%% 94
\bibitem[\protect\citeauthoryear{Fu et~al.}{2012}]{fu2012cd}
\begin{barticle}
\bauthor{\bsnm{Fu}, \binits{L.}},
\bauthor{\bsnm{Niu}, \binits{B.}},
\bauthor{\bsnm{Zhu}, \binits{Z.}},
\bauthor{\bsnm{Wu}, \binits{S.}},
\bauthor{\bsnm{Li}, \binits{W.}}:
\batitle{Cd-hit: accelerated for clustering the next-generation sequencing data}.
\bjtitle{Bioinformatics}
\bvolume{28},
\bfpage{3150}--\blpage{3152}
(\byear{2012})
\end{barticle}
\endbibitem

%%% 95
\bibitem[\protect\citeauthoryear{Li and Godzik}{2006}]{li2006cd}
\begin{barticle}
\bauthor{\bsnm{Li}, \binits{W.}},
\bauthor{\bsnm{Godzik}, \binits{A.}}:
\batitle{Cd-hit: a fast program for clustering and comparing large sets of protein or nucleotide sequences}.
\bjtitle{Bioinformatics}
\bvolume{22},
\bfpage{1658}--\blpage{1659}
(\byear{2006})
\end{barticle}
\endbibitem

%%% 96
\bibitem[\protect\citeauthoryear{Huang et~al.}{2010}]{huang2010cd}
\begin{barticle}
\bauthor{\bsnm{Huang}, \binits{Y.}},
\bauthor{\bsnm{Niu}, \binits{B.}},
\bauthor{\bsnm{Gao}, \binits{Y.}},
\bauthor{\bsnm{Fu}, \binits{L.}},
\bauthor{\bsnm{Li}, \binits{W.}}:
\batitle{Cd-hit suite: a web server for clustering and comparing biological sequences}.
\bjtitle{Bioinformatics}
\bvolume{26},
\bfpage{680}--\blpage{682}
(\byear{2010})
\end{barticle}
\endbibitem

%%% 97
\bibitem[\protect\citeauthoryear{K{\"o}hler et~al.}{2020}]{kohler2020equivariant}
\begin{bchapter}
\bauthor{\bsnm{K{\"o}hler}, \binits{J.}},
\bauthor{\bsnm{Klein}, \binits{L.}},
\bauthor{\bsnm{No{\'e}}, \binits{F.}}:
\bctitle{Equivariant flows: exact likelihood generative learning for symmetric densities}.
In: \bbtitle{International Conference on Machine Learning},
pp. \bfpage{5361}--\blpage{5370}
(\byear{2020}).
\bcomment{PMLR}
\end{bchapter}
\endbibitem

%%% 98
\bibitem[\protect\citeauthoryear{Xu et~al.}{2022}]{xu2022geodiff}
\begin{botherref}
\oauthor{\bsnm{Xu}, \binits{M.}},
\oauthor{\bsnm{Yu}, \binits{L.}},
\oauthor{\bsnm{Song}, \binits{Y.}},
\oauthor{\bsnm{Shi}, \binits{C.}},
\oauthor{\bsnm{Ermon}, \binits{S.}},
\oauthor{\bsnm{Tang}, \binits{J.}}:
Geodiff: A geometric diffusion model for molecular conformation generation.
arXiv preprint arXiv:2203.02923
(2022)
\end{botherref}
\endbibitem

%%% 99
\bibitem[\protect\citeauthoryear{De~Bortoli et~al.}{2022}]{de2022riemannian}
\begin{barticle}
\bauthor{\bsnm{De~Bortoli}, \binits{V.}},
\bauthor{\bsnm{Mathieu}, \binits{E.}},
\bauthor{\bsnm{Hutchinson}, \binits{M.}},
\bauthor{\bsnm{Thornton}, \binits{J.}},
\bauthor{\bsnm{Teh}, \binits{Y.W.}},
\bauthor{\bsnm{Doucet}, \binits{A.}}:
\batitle{Riemannian score-based generative modelling}.
\bjtitle{Advances in Neural Information Processing Systems}
\bvolume{35},
\bfpage{2406}--\blpage{2422}
(\byear{2022})
\end{barticle}
\endbibitem

%%% 100
\bibitem[\protect\citeauthoryear{Vaswani et~al.}{2017}]{vaswani2017attention}
\begin{botherref}
\oauthor{\bsnm{Vaswani}, \binits{A.}},
\oauthor{\bsnm{Shazeer}, \binits{N.}},
\oauthor{\bsnm{Parmar}, \binits{N.}},
\oauthor{\bsnm{Uszkoreit}, \binits{J.}},
\oauthor{\bsnm{Jones}, \binits{L.}},
\oauthor{\bsnm{Gomez}, \binits{A.N.}},
\oauthor{\bsnm{Kaiser}, \binits{{\L}.}},
\oauthor{\bsnm{Polosukhin}, \binits{I.}}:
Attention is all you need.
Advances in neural information processing systems
\textbf{30}
(2017)
\end{botherref}
\endbibitem

%%% 101
\bibitem[\protect\citeauthoryear{Lin et~al.}{2022}]{lin2022language}
\begin{barticle}
\bauthor{\bsnm{Lin}, \binits{Z.}},
\bauthor{\bsnm{Akin}, \binits{H.}},
\bauthor{\bsnm{Rao}, \binits{R.}},
\bauthor{\bsnm{Hie}, \binits{B.}},
\bauthor{\bsnm{Zhu}, \binits{Z.}},
\bauthor{\bsnm{Lu}, \binits{W.}},
\bauthor{\bsnm{Santos~Costa}, \binits{A.}},
\bauthor{\bsnm{Fazel-Zarandi}, \binits{M.}},
\bauthor{\bsnm{Sercu}, \binits{T.}},
\bauthor{\bsnm{Candido}, \binits{S.}}, \betal:
\batitle{Language models of protein sequences at the scale of evolution enable accurate structure prediction}.
\bjtitle{BioRxiv}
\bvolume{2022},
\bfpage{500902}
(\byear{2022})
\end{barticle}
\endbibitem

%%% 102
\bibitem[\protect\citeauthoryear{Chen et~al.}{2022}]{chen2022analog}
\begin{botherref}
\oauthor{\bsnm{Chen}, \binits{T.}},
\oauthor{\bsnm{Zhang}, \binits{R.}},
\oauthor{\bsnm{Hinton}, \binits{G.}}:
Analog bits: Generating discrete data using diffusion models with self-conditioning.
arXiv preprint arXiv:2208.04202
(2022)
\end{botherref}
\endbibitem

%%% 103
\bibitem[\protect\citeauthoryear{Ba et~al.}{2016}]{ba2016layer}
\begin{botherref}
\oauthor{\bsnm{Ba}, \binits{J.L.}},
\oauthor{\bsnm{Kiros}, \binits{J.R.}},
\oauthor{\bsnm{Hinton}, \binits{G.E.}}:
Layer normalization.
arXiv preprint arXiv:1607.06450
(2016)
\end{botherref}
\endbibitem

%%% 104
\bibitem[\protect\citeauthoryear{Conway et~al.}{2014}]{conway2014relaxation}
\begin{barticle}
\bauthor{\bsnm{Conway}, \binits{P.}},
\bauthor{\bsnm{Tyka}, \binits{M.D.}},
\bauthor{\bsnm{DiMaio}, \binits{F.}},
\bauthor{\bsnm{Konerding}, \binits{D.E.}},
\bauthor{\bsnm{Baker}, \binits{D.}}:
\batitle{Relaxation of backbone bond geometry improves protein energy landscape modeling}.
\bjtitle{Protein science}
\bvolume{23},
\bfpage{47}--\blpage{55}
(\byear{2014})
\end{barticle}
\endbibitem

%%% 105
\bibitem[\protect\citeauthoryear{Dauparas et~al.}{2023}]{dauparas2023atomic}
\begin{botherref}
\oauthor{\bsnm{Dauparas}, \binits{J.}},
\oauthor{\bsnm{Lee}, \binits{G.R.}},
\oauthor{\bsnm{Pecoraro}, \binits{R.}},
\oauthor{\bsnm{An}, \binits{L.}},
\oauthor{\bsnm{Anishchenko}, \binits{I.}},
\oauthor{\bsnm{Glasscock}, \binits{C.}},
\oauthor{\bsnm{Baker}, \binits{D.}}:
Atomic context-conditioned protein sequence design using ligandmpnn.
Biorxiv,
2023--12
(2023)
\end{botherref}
\endbibitem

%%% 106
\bibitem[\protect\citeauthoryear{Loshchilov and Hutter}{2017}]{loshchilov2017decoupled}
\begin{botherref}
\oauthor{\bsnm{Loshchilov}, \binits{I.}},
\oauthor{\bsnm{Hutter}, \binits{F.}}:
Decoupled weight decay regularization.
arXiv preprint arXiv:1711.05101
(2017)
\end{botherref}
\endbibitem

%%% 107
\bibitem[\protect\citeauthoryear{Paszke et~al.}{2019}]{paszke2019pytorch}
\begin{botherref}
\oauthor{\bsnm{Paszke}, \binits{A.}},
\oauthor{\bsnm{Gross}, \binits{S.}},
\oauthor{\bsnm{Massa}, \binits{F.}},
\oauthor{\bsnm{Lerer}, \binits{A.}},
\oauthor{\bsnm{Bradbury}, \binits{J.}},
\oauthor{\bsnm{Chanan}, \binits{G.}},
\oauthor{\bsnm{Killeen}, \binits{T.}},
\oauthor{\bsnm{Lin}, \binits{Z.}},
\oauthor{\bsnm{Gimelshein}, \binits{N.}},
\oauthor{\bsnm{Antiga}, \binits{L.}}, et al.:
Pytorch: An imperative style, high-performance deep learning library.
Advances in neural information processing systems
\textbf{32}
(2019)
\end{botherref}
\endbibitem

%%% 108
\bibitem[\protect\citeauthoryear{Bennett et~al.}{2023}]{bennett2023improving}
\begin{barticle}
\bauthor{\bsnm{Bennett}, \binits{N.R.}},
\bauthor{\bsnm{Coventry}, \binits{B.}},
\bauthor{\bsnm{Goreshnik}, \binits{I.}},
\bauthor{\bsnm{Huang}, \binits{B.}},
\bauthor{\bsnm{Allen}, \binits{A.}},
\bauthor{\bsnm{Vafeados}, \binits{D.}},
\bauthor{\bsnm{Peng}, \binits{Y.P.}},
\bauthor{\bsnm{Dauparas}, \binits{J.}},
\bauthor{\bsnm{Baek}, \binits{M.}},
\bauthor{\bsnm{Stewart}, \binits{L.}}, \betal:
\batitle{Improving de novo protein binder design with deep learning}.
\bjtitle{Nature Communications}
\bvolume{14},
\bfpage{2625}
(\byear{2023})
\end{barticle}
\endbibitem

\end{thebibliography}



\begin{thebibliography}{13}
\ifx \bisbn   \undefined \def \bisbn  #1{ISBN #1}\fi
\ifx \binits  \undefined \def \binits#1{#1}\fi
\ifx \bauthor  \undefined \def \bauthor#1{#1}\fi
\ifx \batitle  \undefined \def \batitle#1{#1}\fi
\ifx \bjtitle  \undefined \def \bjtitle#1{#1}\fi
\ifx \bvolume  \undefined \def \bvolume#1{\textit{#1}}\fi
\ifx \byear  \undefined \def \byear#1{#1}\fi
\ifx \bissue  \undefined \def \bissue#1{#1}\fi
\ifx \bfpage  \undefined \def \bfpage#1{#1}\fi
\ifx \blpage  \undefined \def \blpage #1{#1}\fi
\ifx \burl  \undefined \def \burl#1{\textsf{#1}}\fi
\ifx \doiurl  \undefined \def \doiurl#1{\url{https://doi.org/#1}}\fi
\ifx \betal  \undefined \def \betal{\textit{et al.}}\fi
\ifx \binstitute  \undefined \def \binstitute#1{#1}\fi
\ifx \binstitutionaled  \undefined \def \binstitutionaled#1{#1}\fi
\ifx \bctitle  \undefined \def \bctitle#1{#1}\fi
\ifx \beditor  \undefined \def \beditor#1{#1}\fi
\ifx \bpublisher  \undefined \def \bpublisher#1{#1}\fi
\ifx \bbtitle  \undefined \def \bbtitle#1{#1}\fi
\ifx \bedition  \undefined \def \bedition#1{#1}\fi
\ifx \bseriesno  \undefined \def \bseriesno#1{#1}\fi
\ifx \blocation  \undefined \def \blocation#1{#1}\fi
\ifx \bsertitle  \undefined \def \bsertitle#1{#1}\fi
\ifx \bsnm \undefined \def \bsnm#1{#1}\fi
\ifx \bsuffix \undefined \def \bsuffix#1{#1}\fi
\ifx \bparticle \undefined \def \bparticle#1{#1}\fi
\ifx \barticle \undefined \def \barticle#1{#1}\fi
\bibcommenthead
\ifx \bconfdate \undefined \def \bconfdate #1{#1}\fi
\ifx \botherref \undefined \def \botherref #1{#1}\fi
\ifx \url \undefined \def \url#1{\textsf{#1}}\fi
\ifx \bchapter \undefined \def \bchapter#1{#1}\fi
\ifx \bbook \undefined \def \bbook#1{#1}\fi
\ifx \bcomment \undefined \def \bcomment#1{#1}\fi
\ifx \oauthor \undefined \def \oauthor#1{#1}\fi
\ifx \citeauthoryear \undefined \def \citeauthoryear#1{#1}\fi
\ifx \endbibitem  \undefined \def \endbibitem {}\fi
\ifx \bconflocation  \undefined \def \bconflocation#1{#1}\fi
\ifx \arxivurl  \undefined \def \arxivurl#1{\textsf{#1}}\fi
\csname PreBibitemsHook\endcsname

%%% 1
\bibitem[\protect\citeauthoryear{Fu et~al.}{2012}]{fu2012cd}
\begin{barticle}
\bauthor{\bsnm{Fu}, \binits{L.}},
\bauthor{\bsnm{Niu}, \binits{B.}},
\bauthor{\bsnm{Zhu}, \binits{Z.}},
\bauthor{\bsnm{Wu}, \binits{S.}},
\bauthor{\bsnm{Li}, \binits{W.}}:
\batitle{Cd-hit: accelerated for clustering the next-generation sequencing data}.
\bjtitle{Bioinformatics}
\bvolume{28},
\bfpage{3150}--\blpage{3152}
(\byear{2012})
\end{barticle}
\endbibitem

%%% 2
\bibitem[\protect\citeauthoryear{Li and Godzik}{2006}]{li2006cd}
\begin{barticle}
\bauthor{\bsnm{Li}, \binits{W.}},
\bauthor{\bsnm{Godzik}, \binits{A.}}:
\batitle{Cd-hit: a fast program for clustering and comparing large sets of protein or nucleotide sequences}.
\bjtitle{Bioinformatics}
\bvolume{22},
\bfpage{1658}--\blpage{1659}
(\byear{2006})
\end{barticle}
\endbibitem

%%% 3
\bibitem[\protect\citeauthoryear{Huang et~al.}{2010}]{huang2010cd}
\begin{barticle}
\bauthor{\bsnm{Huang}, \binits{Y.}},
\bauthor{\bsnm{Niu}, \binits{B.}},
\bauthor{\bsnm{Gao}, \binits{Y.}},
\bauthor{\bsnm{Fu}, \binits{L.}},
\bauthor{\bsnm{Li}, \binits{W.}}:
\batitle{Cd-hit suite: a web server for clustering and comparing biological sequences}.
\bjtitle{Bioinformatics}
\bvolume{26},
\bfpage{680}--\blpage{682}
(\byear{2010})
\end{barticle}
\endbibitem

%%% 4
\bibitem[\protect\citeauthoryear{Wang et~al.}{2005}]{wang2005pdbbind}
\begin{barticle}
\bauthor{\bsnm{Wang}, \binits{R.}},
\bauthor{\bsnm{Fang}, \binits{X.}},
\bauthor{\bsnm{Lu}, \binits{Y.}},
\bauthor{\bsnm{Yang}, \binits{C.-Y.}},
\bauthor{\bsnm{Wang}, \binits{S.}}:
\batitle{The pdbbind database: methodologies and updates}.
\bjtitle{Journal of medicinal chemistry}
\bvolume{48},
\bfpage{4111}--\blpage{4119}
(\byear{2005})
\end{barticle}
\endbibitem

%%% 5
\bibitem[\protect\citeauthoryear{Van Der~Spoel et~al.}{2005}]{van2005gromacs}
\begin{barticle}
\bauthor{\bsnm{Van Der~Spoel}, \binits{D.}},
\bauthor{\bsnm{Lindahl}, \binits{E.}},
\bauthor{\bsnm{Hess}, \binits{B.}},
\bauthor{\bsnm{Groenhof}, \binits{G.}},
\bauthor{\bsnm{Mark}, \binits{A.E.}},
\bauthor{\bsnm{Berendsen}, \binits{H.J.}}:
\batitle{Gromacs: fast, flexible, and free}.
\bjtitle{Journal of computational chemistry}
\bvolume{26},
\bfpage{1701}--\blpage{1718}
(\byear{2005})
\end{barticle}
\endbibitem

%%% 6
\bibitem[\protect\citeauthoryear{Huang et~al.}{2017}]{huang2017charmm36}
\begin{barticle}
\bauthor{\bsnm{Huang}, \binits{J.}},
\bauthor{\bsnm{Rauscher}, \binits{S.}},
\bauthor{\bsnm{Nawrocki}, \binits{G.}},
\bauthor{\bsnm{Ran}, \binits{T.}},
\bauthor{\bsnm{Feig}, \binits{M.}},
\bauthor{\bsnm{Groot}, \binits{B.L.}},
\bauthor{\bsnm{Grubm{\"u}ller}, \binits{H.}},
\bauthor{\bsnm{MacKerell}, \binits{A.D.}}:
\batitle{Charmm36: An improved force field for folded and intrinsically disordered proteins}.
\bjtitle{Biophysical Journal}
\bvolume{112},
\bfpage{175}--\blpage{176}
(\byear{2017})
\end{barticle}
\endbibitem

%%% 7
\bibitem[\protect\citeauthoryear{Price and Brooks~III}{2004}]{price2004modified}
\begin{barticle}
\bauthor{\bsnm{Price}, \binits{D.J.}},
\bauthor{\bsnm{Brooks~III}, \binits{C.L.}}:
\batitle{A modified tip3p water potential for simulation with ewald summation}.
\bjtitle{The Journal of chemical physics}
\bvolume{121},
\bfpage{10096}--\blpage{10103}
(\byear{2004})
\end{barticle}
\endbibitem

%%% 8
\bibitem[\protect\citeauthoryear{Berendsen et~al.}{1984}]{berendsen1984molecular}
\begin{barticle}
\bauthor{\bsnm{Berendsen}, \binits{H.J.}},
\bauthor{\bsnm{Postma}, \binits{J.v.}},
\bauthor{\bsnm{Van~Gunsteren}, \binits{W.F.}},
\bauthor{\bsnm{DiNola}, \binits{A.}},
\bauthor{\bsnm{Haak}, \binits{J.R.}}:
\batitle{Molecular dynamics with coupling to an external bath}.
\bjtitle{The Journal of chemical physics}
\bvolume{81},
\bfpage{3684}--\blpage{3690}
(\byear{1984})
\end{barticle}
\endbibitem

%%% 9
\bibitem[\protect\citeauthoryear{Bussi et~al.}{2007}]{bussi2007canonical}
\begin{botherref}
\oauthor{\bsnm{Bussi}, \binits{G.}},
\oauthor{\bsnm{Donadio}, \binits{D.}},
\oauthor{\bsnm{Parrinello}, \binits{M.}}:
Canonical sampling through velocity rescaling.
The Journal of chemical physics
\textbf{126}
(2007)
\end{botherref}
\endbibitem

%%% 10
\bibitem[\protect\citeauthoryear{Essmann et~al.}{1995}]{essmann1995smooth}
\begin{barticle}
\bauthor{\bsnm{Essmann}, \binits{U.}},
\bauthor{\bsnm{Perera}, \binits{L.}},
\bauthor{\bsnm{Berkowitz}, \binits{M.L.}},
\bauthor{\bsnm{Darden}, \binits{T.}},
\bauthor{\bsnm{Lee}, \binits{H.}},
\bauthor{\bsnm{Pedersen}, \binits{L.G.}}:
\batitle{A smooth particle mesh ewald method}.
\bjtitle{The Journal of chemical physics}
\bvolume{103},
\bfpage{8577}--\blpage{8593}
(\byear{1995})
\end{barticle}
\endbibitem

%%% 11
\bibitem[\protect\citeauthoryear{P{\'a}ll and Hess}{2013}]{pall2013flexible}
\begin{barticle}
\bauthor{\bsnm{P{\'a}ll}, \binits{S.}},
\bauthor{\bsnm{Hess}, \binits{B.}}:
\batitle{A flexible algorithm for calculating pair interactions on simd architectures}.
\bjtitle{Computer Physics Communications}
\bvolume{184},
\bfpage{2641}--\blpage{2650}
(\byear{2013})
\end{barticle}
\endbibitem

%%% 12
\bibitem[\protect\citeauthoryear{Hess et~al.}{1997}]{hess1997lincs}
\begin{barticle}
\bauthor{\bsnm{Hess}, \binits{B.}},
\bauthor{\bsnm{Bekker}, \binits{H.}},
\bauthor{\bsnm{Berendsen}, \binits{H.J.}},
\bauthor{\bsnm{Fraaije}, \binits{J.G.}}:
\batitle{Lincs: A linear constraint solver for molecular simulations}.
\bjtitle{Journal of computational chemistry}
\bvolume{18},
\bfpage{1463}--\blpage{1472}
(\byear{1997})
\end{barticle}
\endbibitem

%%% 13
\bibitem[\protect\citeauthoryear{Vald{\'e}s-Tresanco et~al.}{2021}]{valdes2021gmx_mmpbsa}
\begin{barticle}
\bauthor{\bsnm{Vald{\'e}s-Tresanco}, \binits{M.S.}},
\bauthor{\bsnm{Vald{\'e}s-Tresanco}, \binits{M.E.}},
\bauthor{\bsnm{Valiente}, \binits{P.A.}},
\bauthor{\bsnm{Moreno}, \binits{E.}}:
\batitle{gmx\_mmpbsa: a new tool to perform end-state free energy calculations with gromacs}.
\bjtitle{Journal of chemical theory and computation}
\bvolume{17},
\bfpage{6281}--\blpage{6291}
(\byear{2021})
\end{barticle}
\endbibitem

\end{thebibliography}

% \begin{figure*}[htb]
% \centering
% \includegraphics[width=0.98\linewidth]{figs_si/diffpepbuilder_toc.pdf}
% \caption{\textbf{\textbar}  TOC Figure }\label{toc}
% \end{figure*}

\end{document}

\title[Article Title]{Supplementary Information of Target-Specific \textit{De Novo} Peptide Binder Design with DiffPepbuilder}


\author[1]{\fnm{Fanhao} \sur{Wang}}
\equalcont{These authors contributed equally}

\author[1]{\fnm{Yuzhe} \sur{Wang}}
\equalcont{These authors contributed equally}

\author[3]{\fnm{Laiyi} \sur{Feng}}

\author*[2]{\fnm{Changsheng} \sur{Zhang} \email{changshengzhang@pku.edu.cn}}
%\altaffiliation{A shared footnote}

\author*[1,2,3]{\fnm{Luhua} \sur{Lai} \email{lhlai@pku.edu.cn}}

\affil[1]{\orgname{Center for Quantitative Biology, Academy for Advanced Interdisciplinary Studies}, \orgdiv{Peking University}, \orgaddress{\city{Beijing}, \postcode{100871},  \country{China}}}

\affil[2]{\orgname{BNLMS, College of Chemistry and Molecular Engineering}, \orgdiv{Peking University}, \orgaddress{\city{Beijing}, \postcode{100871}, \country{China}}}

\affil[3]{\orgname{Center for Life Sciences, Academy for Advanced Interdisciplinary Studies}, \orgdiv{Peking University}, \orgaddress{\city{Beijing}, \postcode{100871},  \country{China}}}




\maketitle

\newpage

% PepPC-ts22 similarity distribution figure
% Original complex structure
% Cross update block

\section{Additional information of dataset curation}\label{secs1}

Structures were selected from the Protein Data Bank (https://www.rcsb.org/) based on the following criteria: composed solely of proteins, comprising more than two chains, and resolution lower than 2.5  \AA. Complex structures are classified into different groups based on the length of the shortest chain, and other chains interact with the shortest chain were output paired with the shortest chain. Structures with shortest chain length between 8 and 30 are classified into peptide-protein complex (5,364 in total), more than 30 are defined as protein-protein complex (11,469 in total). Symmetric structures were completed using biounit information, and missing atoms were rectified with pdbfixer. Membrane proteins in PepPC-F were removed based on Uniprot information (UniProt: the Universal Protein Knowledgebase in 2023), if the information contained lines starting with 'FT TRANSMEM', which indicated that the information related to the transmembrane region, the structure was removed. Structures contained chains linked by disulfide bonds were also excluded. CD-hit\cite{fu2012cd,li2006cd,huang2010cd} was used to cluster, with 90\% cutoff and word length = 5. After clustering, 117 clusters were generated from 275 helical peptides and 1,021 clusters for loop entries in PepPC, while 2,588 clusters were generated from 16,553 helical complexes and 3,386 clusters for 18,065 loop complexes in PepPC-F. \\

\begin{figure}[H]
\centering
\includegraphics[width=1\linewidth]{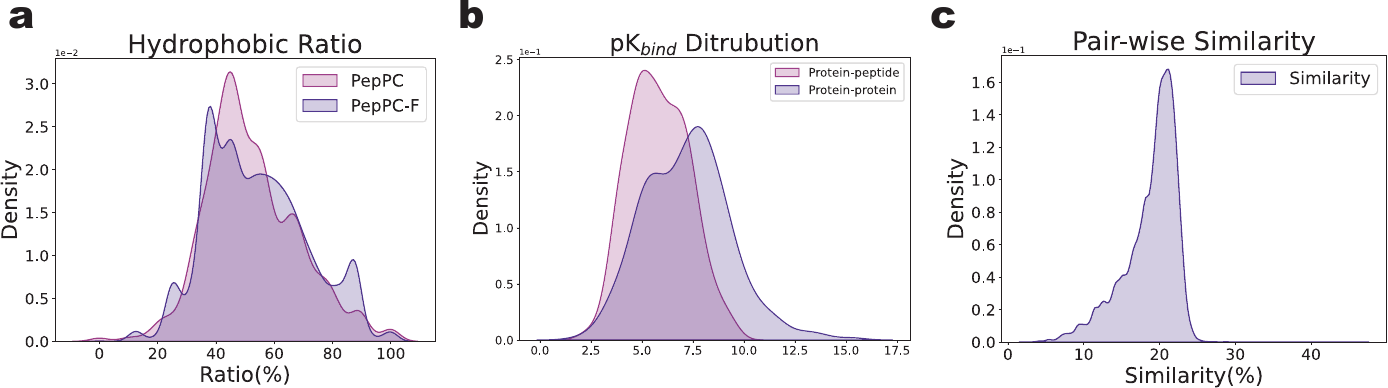}
\caption{\textbf{\textbar} \textbf{a.} Hydrophobic ratio distribution of PepPC and PepPC-F. \textbf{b.}  Experimental binding affinity distribution. The pIC$_{50}$, p$K_d$, p$K_i$ distribution of protein-protein binding data (in pink) and protein-peptide binding data (in purple) in PDBbind2020. \textbf{c.} Pairwise sequence similarity distribution between the training set and the regeneration test set.}\label{figs0}
\end{figure}

We have additionally analyzed the distribution of peptide hydrophobicity (the number of hydrophobic residues divided by the total number of residues) in PepPC-F and PepPC, as illustrated in Fig. \ref{figs0}a. There is a slightly higher peak for PepPC-F at a larger hydrophobic ratio, suggesting a modest increase in the frequency of more hydrophobic peptides in PepPC-F. However the overall shapes of the distributions are quite similar, indicating that the hydrophobicity profiles of PepPC and PepPC-F do not differ significantly.  \\

\section{Additional information of regeneration test set and result analysis}\label{secs2}
PDB ID, bioactivity, max and average similarity of dataset used in regeneration task with PepPC-F are shown in Table \ref{tab:bioactivity}. We ensured that the average sequence similarity of the receptors was controlled at approximately 30\%, and removed entries from the test set with a similarity greater than 40\%. The specific similarity distribution is illustrated in Fig. \ref{figs0}c.\\

We selected all protein-ligand data from PDBbind2020\cite{wang2005pdbbind} that intersect with the PepPC database, retaining those cases with binding activity data referred to protein-peptide interactions (some activity data may correspond to other interaction interfaces within the structure). This resulted in a total of 384 entries. We also included all 641 entries of protein-protein interaction data. The distribution of their activity data, including pIC$_{50}$, p$K_d$, and p$K_i$, is depicted in the Fig.\ref{figs0}b.\\

\begin{figure}[H]
\centering
\includegraphics[width=0.6\linewidth]{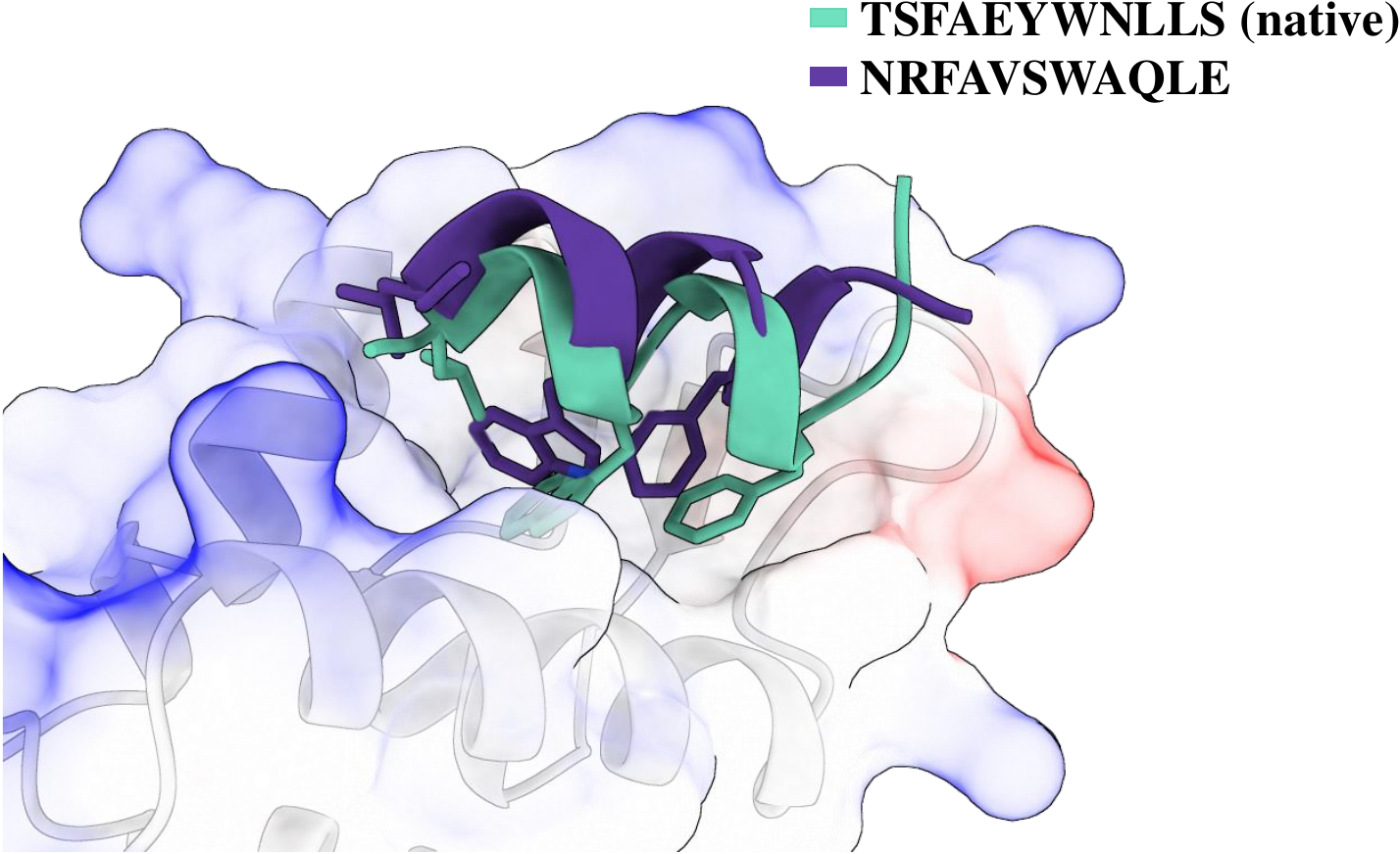}
\caption{\textbf{\textbar\  Additional Result of a DiffPepBuilder-Generated Peptide on MDM2.} Structure of the generated peptide (in violet) and the original peptide (in cyan) in the hydrophobic pocket of MDM2 (colored by electrostatic potential). The sequences are displayed as legends, respectively. The side chains of three critical conserved residues are illustrated for comparison.}\label{3eqs}
\end{figure}

\begin{longtable}{>{\centering\arraybackslash}p{1cm} 
                  >{\centering\arraybackslash}p{2cm} 
                  >{\centering\arraybackslash}p{2cm} 
                  >{\centering\arraybackslash}p{1cm} 
                  >{\centering\arraybackslash}p{1.5cm} 
                  p{6.5cm}}
\caption{Dataset information in regeneration task} \label{tab:bioactivity}\\
\toprule
\textbf{pdbid} & \textbf{Bioactivity Type} & \textbf{Bioactivity Value} & \textbf{Max} & \textbf{Average} & \textbf{Description} \\
\midrule
\endfirsthead

\multicolumn{6}{c}%
{{\bfseries Table \thetable\ continued from previous page}} \\
\toprule
\textbf{pdbid} & \textbf{Bioactivity Type} & \textbf{Bioactivity Value} & \textbf{Max} & \textbf{Average} & \textbf{Description} \\
\midrule
\endhead

\bottomrule \multicolumn{6}{r}{{Continued on next page}} \\
\endfoot

\bottomrule
\endlastfoot

1BJR & IC$_{50}$ & 21 nM & 31.9 & 20.3 & Proteolytically generated Lactoferrin fragment and proteinase K \\
1J7Z & $K_d$ & 0.02 $\mu$M & 24.0 & 16.4 & Ribonuclease in complex with Osmolyte \\
1RJK & $K_d$ & 0.1 nM & 29.4 & 20.3 & Rat vitamin D receptor ligand binding domain complexed with 2MD and a synthetic peptide containing the NR2 box of DRIP 205 \\
1SJH & $K_d$ & 9 nM & 25.4 & 19.8 & HLA-DR1 complexed with a 13 residue HIV capsid peptide \\
2A4R & IC$_{50}$ & 15 nM & 26.3 & 20.2 & HCV NS3 protease domain with a ketoamide inhibitor \\
2AQ9 & IC$_{50}$ & 50 nM & 25.2 & 21.0 & E. coli LpxA with a bound peptide \\
2BBA & $K_d$ & 70 nM & 24.5 & 19.9 & EphB4 receptor in complex with an ephrin-B2 antagonist peptide \\
2FTS & $K_d$ & 90 nM & 24.8 & 20.3 & Glycine receptor-gephyrin complex \\
2IZX & $K_d$ & 0.4 nM & 26.0 & 11.5 & AKAP and PKA regulatory subunits \\
3EQS & $K_d$ & 3.3 nM & 26.7 & 16.1 & Human MDM2 in complex with a 12-mer peptide inhibitor \\
3H0A & IC$_{50}$ & 33 nM & 26.8 & 20.9 & Peroxisome proliferator-activated receptor gamma (PPAR$\gamma$) and retinoic acid receptor alpha (RXR$\alpha$) in complex with 9-cis retinoic acid, co-activator peptide, and a partial agonist \\
3ZQI & $K_d$ & 64 nM & 40.4 & 20.9 & Tetracycline repressor in complex with inducer peptide- TIP2 \\
4ERZ & $K_d$ & 88 nM & 25.4 & 20.8 & WDR5-MLL4 Win motif peptide binary complex \\
4GQ6 & $K_d$ & 53 nM & 24.2 & 19.8 & Human menin in complex with MLL peptide \\
4K0U & $K_d$ & 55 nM & 25.7 & 17.4 & Pilotin/secretin peptide complex \\
4P6X & IC$_{50}$ & 91 nM & 26.0 & 20.3 & Cortisol-bound glucocorticoid receptor ligand binding domain \\
4QJR & $K_d$ & 80 nM & 27.4 & 20.3 & Human nuclear receptor SF-1 (nr5a1) bound to its hormone PIP3 \\
4R1E & $K_d$ & 52 nM & 32.4 & 19.1 & MTIP from Plasmodium falciparum in complex with a peptide-fragment chimera \\
4RRV & $K_d$ & 40 nM & 29.0 & 21.0 & PDK1 in complex with ATP and PIFtide \\
5IZU & $K_d$ & 0.054 $\mu$M & 25.1 & 18.5 & SHANK PDZ domain and SAPAP \\
5LY1 & IC$_{50}$ & 29.8 nM & 24.7 & 20.4 & JMJD2A/KDM4A complexed with Ni(II) and Macrocyclic peptide Inhibitor CP2 \\
5LY3 & $K_d$ & 31 nM & 26.0 & 20.4 & P. calidifontis crenactin in complex with arcadin-2 C-terminal peptide \\
5N8B & $K_d$ & 0.043 $\mu$M & 27.0 & 18.9 & Streptavidin with peptide \\
5UL6 & $K_d$ & 0.006 $\mu$M & 24.7 & 13.4 & nSH3CrkII–PRMNS1 complex \\
5V1Y & IC$_{50}$ & 7.1 nM & 24.4 & 17.6 & Ternary RPN13 PRU-RPN2 (940-953)-ubiquitin complex \\
5WUK & $K_d$ & 14 nM & 23.9 & 19.9 & EED [G255D] in complex with EZH2 peptide \\
6H7B & $K_d$ & 33 nM & 28.6 & 16.0 & Leishmania PABP1 (domain J) complexed with a peptide containing the PAM2 motif of eIF4E4 \\
6JJZ & $K_d$ & 2.3 nM & 25.8 & 15.9 & MAGI2 and Dendrin complex \\
6MA3 & $K_d$ & 0.005 $\mu$M & 23.7 & 18.0 & Human O-GlcNAc transferase bound to a peptide from HCF-1 pro-repeat 2 (11-26) and inhibitor 2a \\
6S07 & $K_d$ & 40 nM & 25.3 & 20.0 & Formylglycine-generating enzyme in complex with copper and a substrate \\
\end{longtable}

\section{Cross update mudule architecture}\label{secs3}

\begin{figure}[H]
\centering
\includegraphics[width=1.0\linewidth]{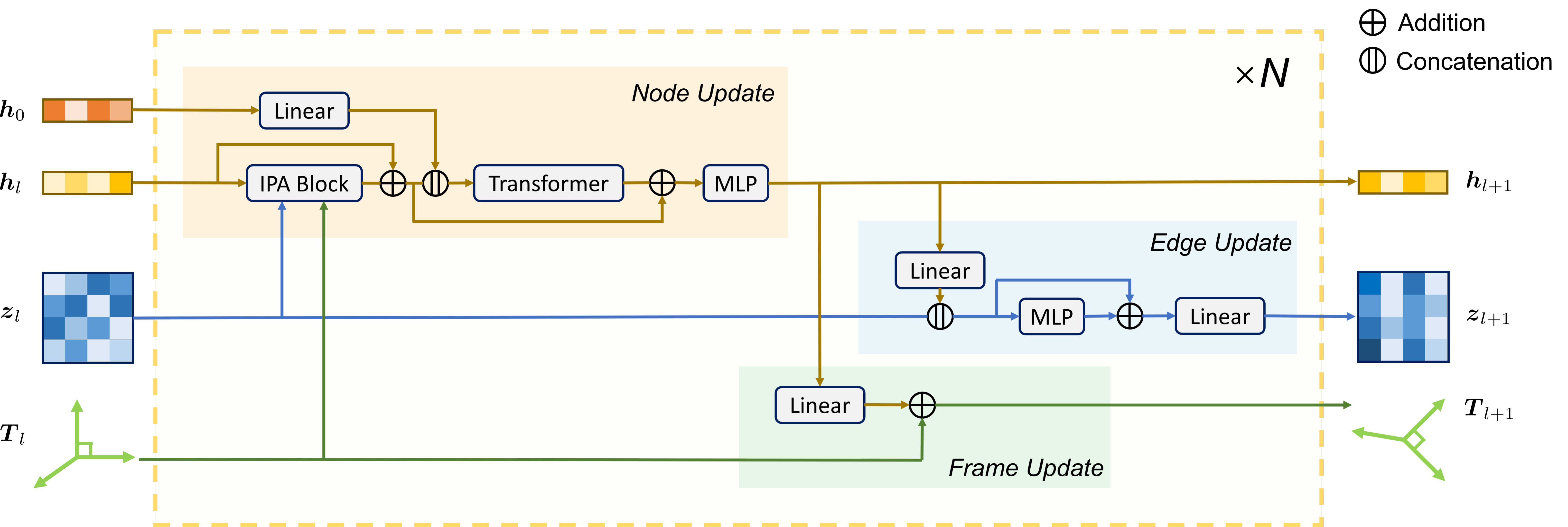}
\caption{\textbf{\textbar\  Architecture of the Cross Update Module.} Each layer $l$ accepts the initial node embedding $\textbf{h}_0$, the current node embedding $\textbf{h}_l$, the edge embedding $\textbf{z}_l$, and the frame representations $\textbf{T}_l$ as inputs. In the Node Update Module, the node embedding is updated with IPA and then using a vanilla transformer. This updated node embedding is subsequently utilized to update the edge embedding and frame representations. }\label{figs1}
\end{figure}

\section{Analysis on the effect of hotspot identification cutoff}

\begin{figure}[H]
\centering
\includegraphics[width=0.5\linewidth]{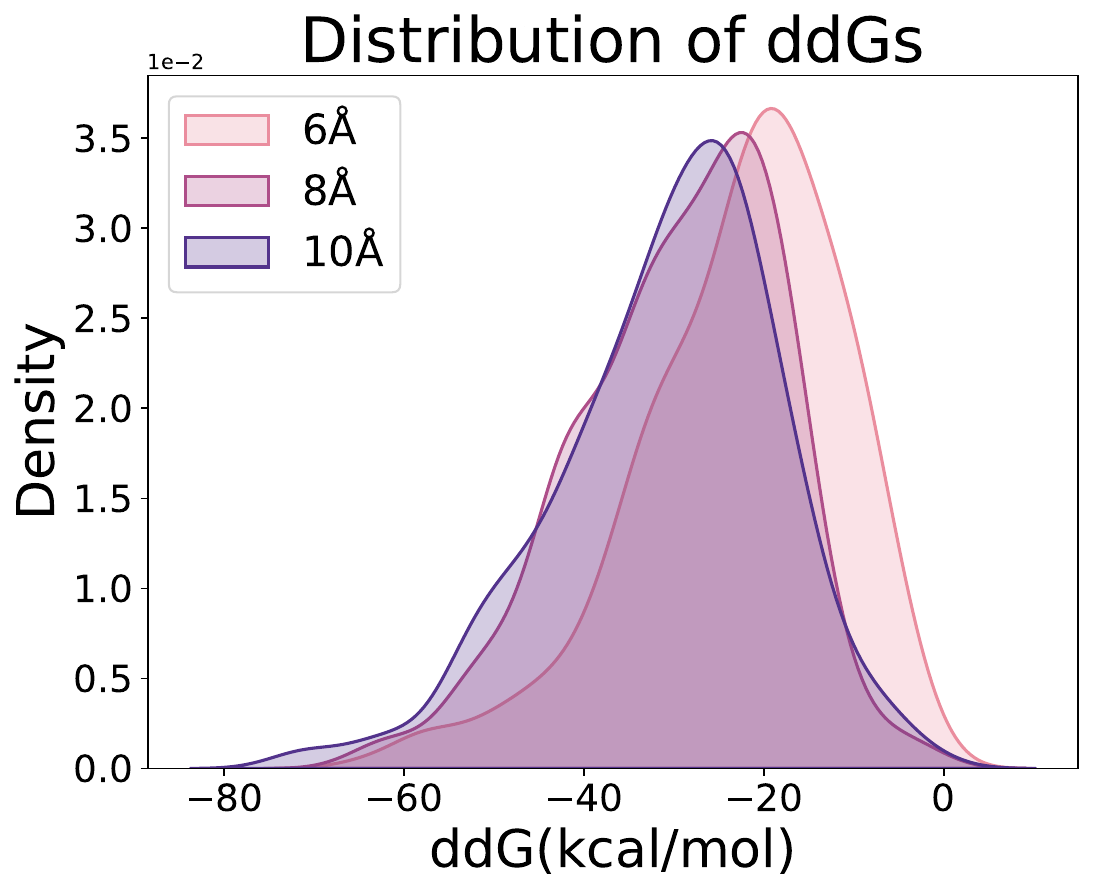}  
\caption{\textbf{\textbar\ Effects of hotspot identification cutoff.} Distribution of binding free energy of regenerated peptide binders under three different cutoff distances: 6 \AA, 8 \AA, and 10 \AA. }\label{hotspot}
\end{figure}

We have elucidated the impact of the cutoff (default 8 \AA) used for identifying hotspots in the regeneration test. Empirically, we found that this cutoff maintains an optimal protein motif size. Specifically, we tested three different cutoff distances (6 \AA, 8 \AA, and 10 \AA) using our program on PepPC-HF, and the results are shown in Fig.\ref{hotspot} and Table \ref{tab:hotspot}. 

\begin{table*}[htb]
\centering
\caption{Regeneration results under different hotspot identification cutoff distances. For 6 \AA\ cutoff, only 25 out of 30 cases were found to contain valid hotspots.}\label{tab:hotspot}
\begin{tabular*}{0.7\textwidth}{@{\extracolsep\fill}ccc@{}}
\toprule
\multirow{2}{*}{Cutoff (\AA)} & \multicolumn{2}{c}{Metric} \\ 
\cmidrule(l){2-3} &  ddG (kcal/mol, $\downarrow$) & Validity \% ($\uparrow$) \\
\midrule
6 & -22.71 & 0.9876  \\
8  & -29.39 & 0.9687  \\
10 & -30.70 & 0.8889  \\
\bottomrule                        
\end{tabular*}
\end{table*}

Our observations indicate that a smaller cutoff (6 \AA) results in some targets lacking hotspots, leading to unsatisfactory binding ddG. Conversely, a larger cutoff (10 \AA) may slightly enhance the binding strengths of the generated peptide binders by promoting more interactions. However, it also causes a significant drop in the valid rate due to steric clashes with the receptors.\\

\section{Additional analysis of \textit{de nove} generation tests}\label{secs4}

\begin{figure}[H]
\centering
\includegraphics[width=0.7\linewidth]{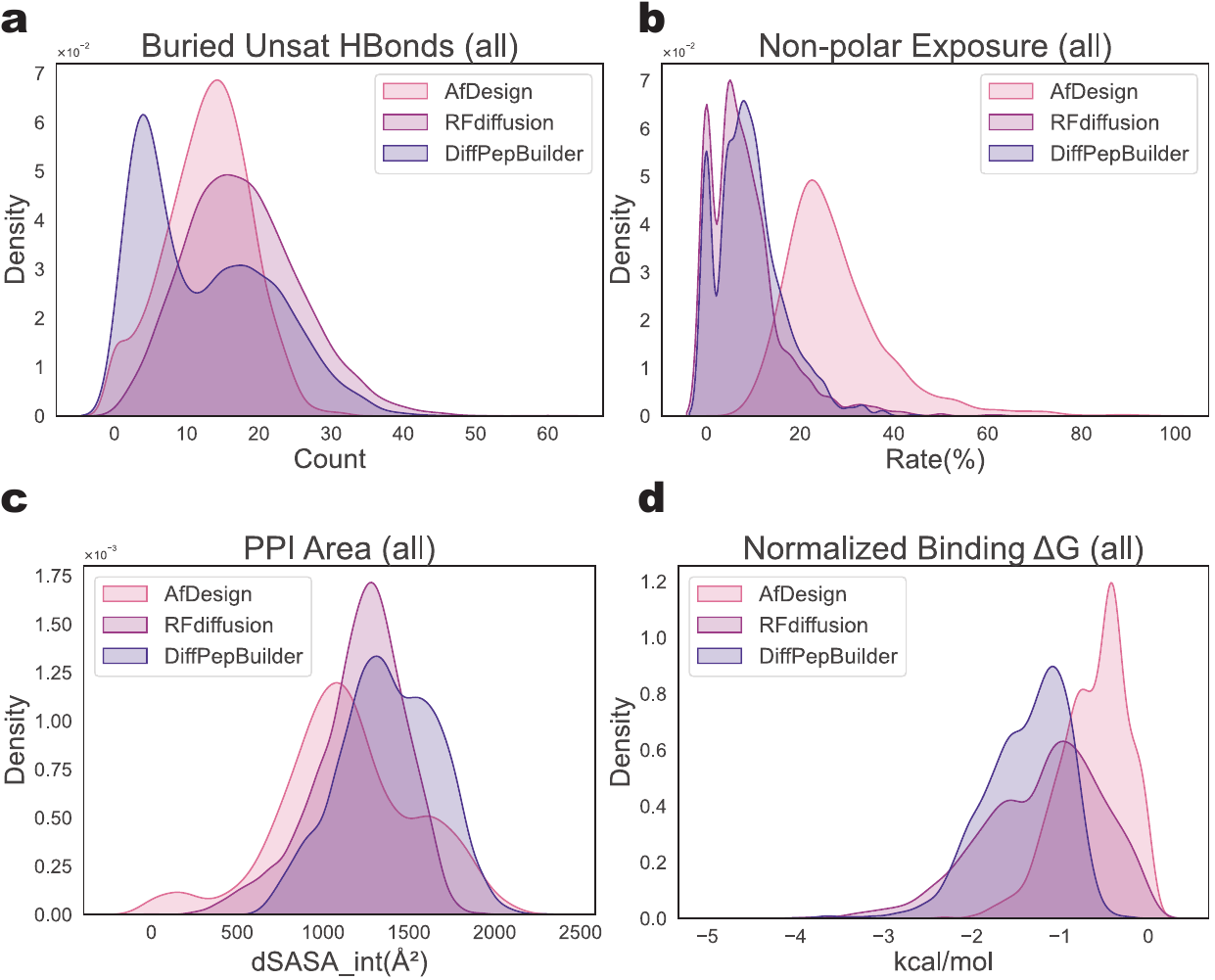}
\caption{\textbf{\textbar\  Additional interface features of \textit{de novo} generation test across all cases.} \textbf{a.} Statistical analysis of unsaturated hydrogen bonds at the interface. \textbf{b.} Comparative analysis of the proportion of exposed hydrophobic residues. \textbf{c.} Visualization of the interface areas between peptides and proteins, it was discerned that DiffPepBuilder facilitated the creation of notably more expansive interface regions. \textbf{d.} Evaluation of normalized binding energies for ligands produced by various software tools, with normalization adjusted according to the count of residues, DiffPepBuilder exhibited superior overall performance. }\label{figs2}
\end{figure}

\begin{figure}[H]
\centering
\includegraphics[width=0.7\linewidth]{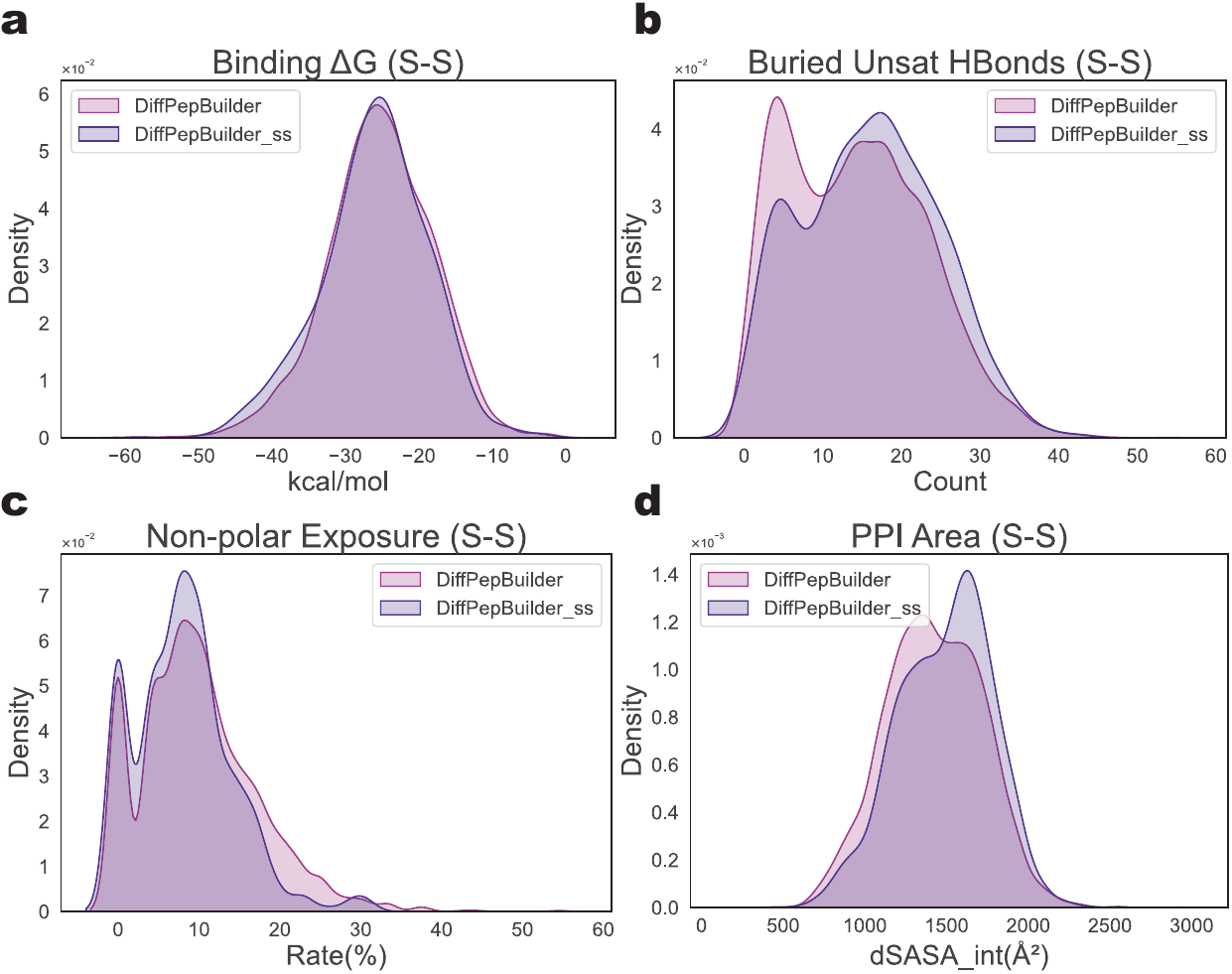}
\caption{\textbf{\textbar\ Interface features of \textit{de novo} generation tests with SSbuilder across all cases.} \textbf{a.} Comparison of the total binding free energy between ligands with (DiffPepBuilder\_ss, in purple) and without (DiffPepBuilder, in pink) disulfide bonds. The incorporation of disulfide bonds had minimal impact on the binding energy. \textbf{b.} Analysis of unsaturated hydrogen bonds at the interface of generated ligands. The introduction of disulfide bonds led to an slight increase in unsaturated hydrogen bonds at the interface. \textbf{c.} Statistics of exposed hydrophobic residues. Incorporating disulfide bonds resulted in a slight reduction in the number of unsaturated hydrogen bonds at the interface. \textbf{d.} Statistical chart of interface areas. Structures incorporating disulfide bonds were shown to help enlarge the interface area. }\label{figs4}
\end{figure}

\begin{figure}[H]
\centering
\includegraphics[width=1.0\linewidth]{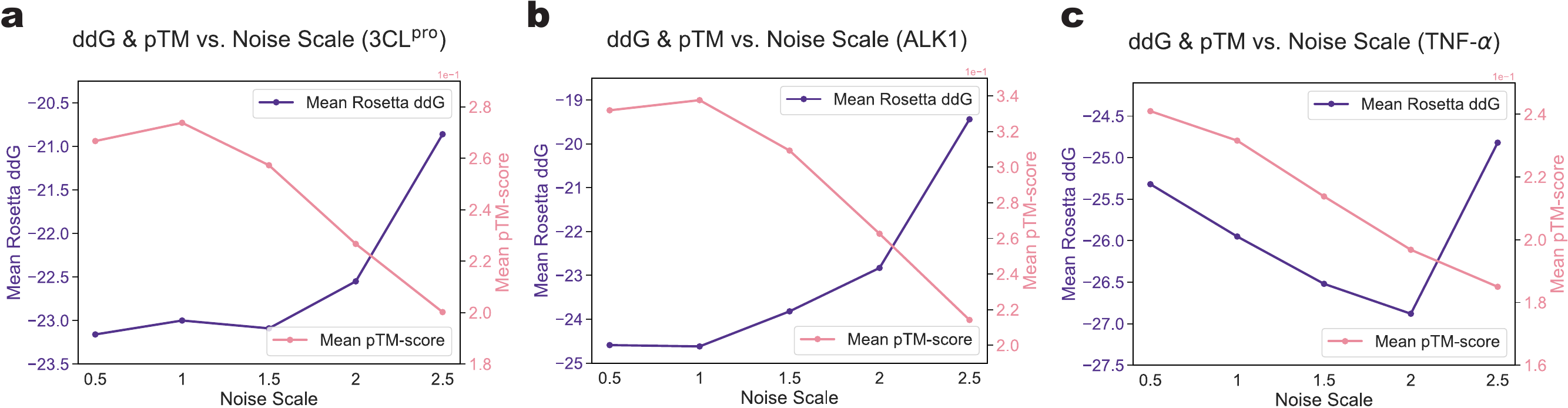}
\caption{\textbf{\textbar\ Effects of noise scales on \textit{de novo} generation tasks.}  \textbf{a}-\textbf{c.} Effects of noise scale on 3CL$^{\rm pro}$, ALK1 and TNF-$\alpha$, respectively. }\label{figs5}
\end{figure}

It was noted that the addition of disulfide bonds enlarges the interface area, reduces the exposure of hydrophobic residues in the ligand peptide, and essentially leaves the binding free energy unaffected. (Fig.\ref{figs4}). Additionally, as the noise scale increases, as shown in Fig.\ref{figs5}, the diversity of the scaffolds generated by the model becomes more pronounced (as indicated by the decreased pTM-score), yet there is a slight decline in the binding free energy.\\

% \section{Ablation of DiffPepBuilder on 3CL$^{\rm pro}$ and ALK1}\label{sec4}

% \begin{figure}[H]
% \centering
% \includegraphics[width=1.0\linewidth]{figs_si/SI_ablation.pdf}
% \caption{\textbf{\textbar\ Ablation studies on 3CL$^{\rm pro}$ and ALK1}  \textbf{a,b.} Ablation studies on 3CL$^{\rm pro}$ and ALK1 respectively, with ddG performance of different models.}\label{figs5}
% \end{figure}

%Vanilla DiffPepBuilder performed optimally in the examples of TNF-$\alpha$ and ALK1. However, in the case of 3CL$^{\rm pro}$, we found that models trained on the PepPC dataset performed better. This may be related to the prevalence of loop structures in natural peptides, as the substrate binding pocket of 3CL$^{\rm pro}$ is more conducive to binding with loop structures.\\

\section{MD \& MMPBSA settings and additional results }\label{secs5}

Molecular Dynamics simulations were performed using GROMACS 2023\cite{van2005gromacs} with the CHARMM36 forcefield\cite{huang2017charmm36}. Each peptide or complex was solvated in cubic box of explicit TIP3P waters\cite{price2004modified} and neutralized with either sodium or chloride ions. The solvated systems were energy-minimized using the steepest descent minimization method. Next, the system was equilibrated for 10 ns under the NVT ensemble with position restraints (1000 kJ/mol$^{-1}$nm$^{-1}$) applied on all the heavy atoms of the peptide. During this equilibration, pressure coupling to 1 atm was performed with the Berendsen barostat\cite{berendsen1984molecular}, and temperature coupling to 310.5 K using the velocity-rescaling thermostat\cite{bussi2007canonical}. From each equilibrated system, simulations of 100 ns were performed in the NPT ensemble. The systems were simulated using periodic boundary conditions. A cutoff at 10 \AA \ was used for van der Waals and short-range electrostatic interactions. The Particle-Mesh Ewald (PME) summation method was used for the long-range electrostatic interactions\cite{essmann1995smooth}. The Verlet cutoff scheme was used\cite{pall2013flexible}. All chemical bonds were constrained using the LINCS algorithm\cite{hess1997lincs}. The integration time-step was 2 fs, and simulations were analyzed using GROMACS tools. We calculated the root-mean-square deviation (RMSD) of the position of the C$\alpha$ atoms of the peptides, compared to the initial conformation, using \textit{gmx rms}.\\

The MMPBSA analysis were established by gmx\_MMPBSA (version 1.6.1)\cite{valdes2021gmx_mmpbsa}. The molecular mechanics/Poisson-Boltzmann surface area method (recommended for the CHARMM force field) was used. Single-term total non-polar solvation free energy($inp=1$) was used. The molecular surface was used for cavity term calculation ($use\_uav=0$). The charmm\_radii ($PBRadii = 7$) was used to build amber topology files. Default parameters were applied for other terms. \\

\begin{figure}[H]
\centering
\includegraphics[width=1.0\linewidth]{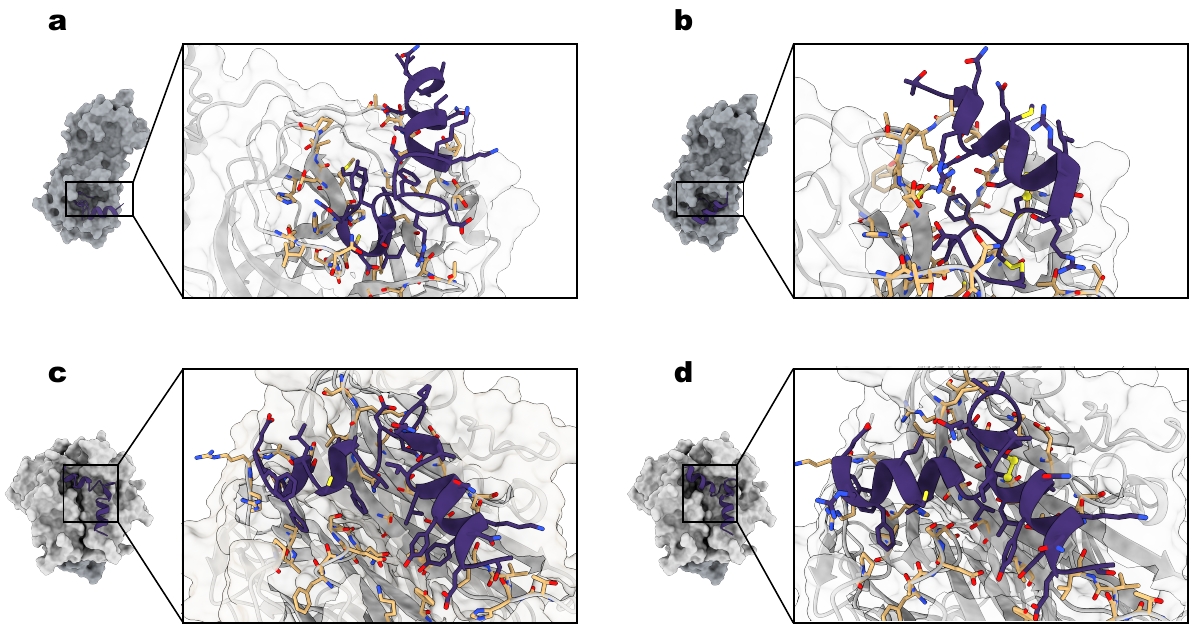}
\caption{\textbf{\textbar\ Structures of the \textit{de novo} generation candidates on the targets of 3CL$\rm ^{pro}$ and TNF-$\alpha$ after MD simulations.}  \textbf{a,b.} Complex structures of 3CL$^{\rm pro}$ before and after disulfide bond construction, respectively. \textbf{c,d.} Complex structures of TNF-$\alpha$ before and after disulfide bond construction, respectively. }\label{figs6}
\end{figure}

\section{SSbuilder details}\label{secs6}

\begin{figure}[H]
\centering
\includegraphics[width=0.9\linewidth]{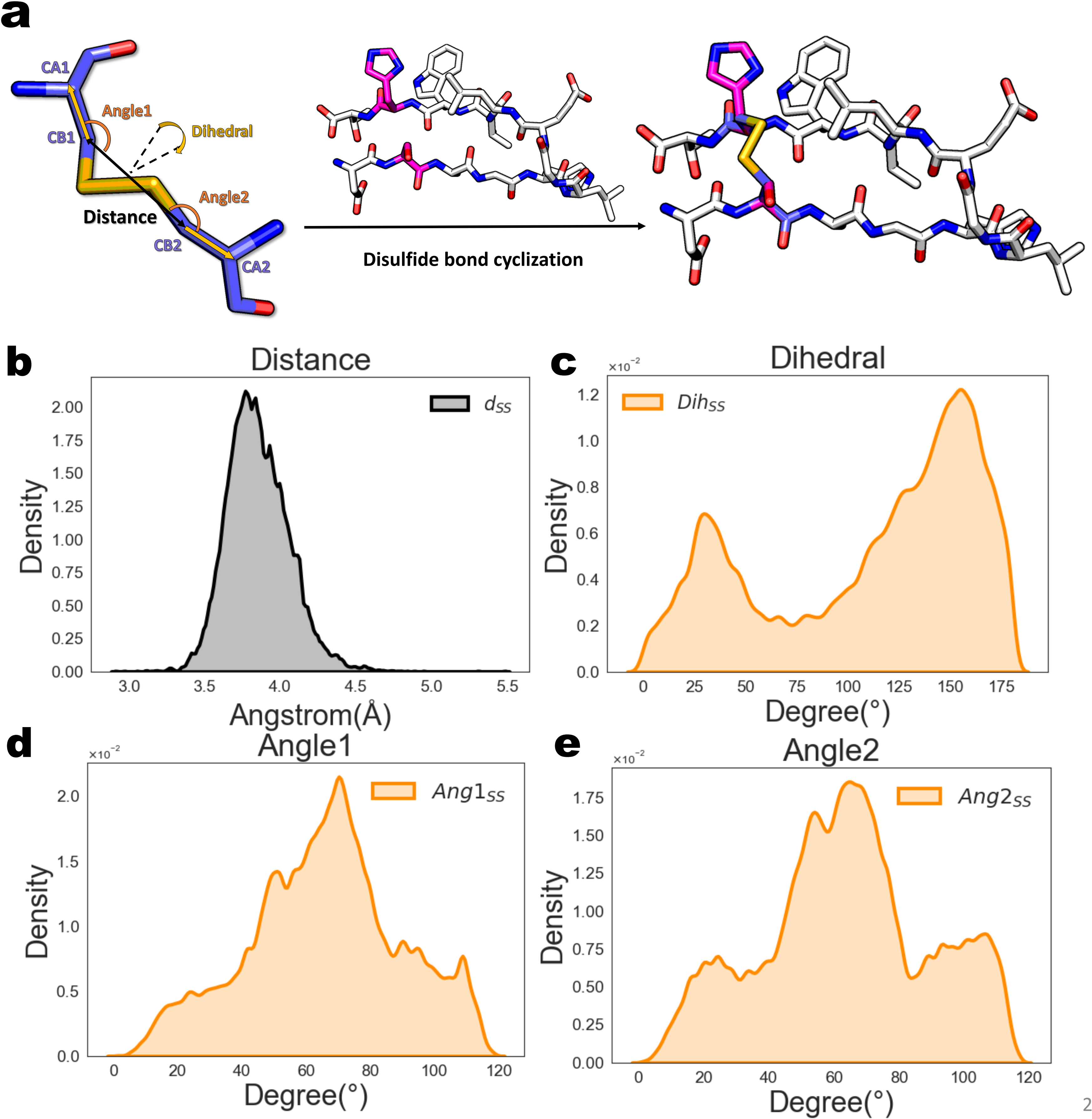}
\caption{\textbf{\textbar\ SSbuilder process schematic and distributions of geometric data.} \textbf{a.} Left: the matched disulfide bond building block. Middle: the substructure containing the potential disulfide sites (in pink). Right: a schematic overlay of the cyclized structure with the atomic structure. "Dehedral" represents the dihedral angle formed by $\rm C\alpha1-C\beta1-C\beta2-C\alpha2$, "Angle1" represents the planar angle formed by $\rm C\alpha1-C\beta1-C\beta2$, and "Angle2" represents the angle formed by $\rm C\beta1-C\beta2-C\alpha2$. \textbf{b.} Distribution diagram of the "Distance" parameter for the disulfide bond building block. \textbf{c.} Distribution diagram of the "Dehedral" parameter. \textbf{d.} Distribution diagram of the "Angle1" parameter. \textbf{e.} Distribution diagram of the 'Angle2' parameter.}\label{figs7}
\end{figure}

\newpage
\bibliography{sn-bibliography}